\documentclass[aps,prb,twocolumn,amsmath,amssymb,groupedaddress]{revtex4-2}

\usepackage{graphicx}
\usepackage{dcolumn}
\usepackage{bm}
\usepackage{multirow}
\usepackage{braket}
\usepackage{color}
\usepackage{ulem}
\usepackage{booktabs}

\begin{document}
\title{
Charge density waves in multiple-$Q$ spin states
}
\author{Satoru~Hayami and Yukitoshi~Motome}
\affiliation{Department of Applied Physics, the University of Tokyo, Tokyo 113-8656, Japan}

\begin{abstract}
Coupling between spin and charge degrees of freedom in electrons is a source of various electronic and magnetic properties of solids. 
We theoretically study charge density waves induced by the spin-charge coupling in the presence of magnetic orderings in itinerant magnets. 
By performing a perturbative calculation in the weak-coupling limit of the Kondo lattice model, we derive a useful formula for the relationship between charge and spin density waves, which can be applied to any magnetic orderings, including noncollinear and noncoplanar ones composed of multiple spin density waves called multiple-$Q$ magnetic orderings. 
We demonstrate the predictive power for single-$Q$ and double-$Q$ states including 
skyrmion and meron-antimeron crystals on a square lattice, in comparison with the numerical calculations. 
Moreover, we show that the charge density waves contain richer information than the spin density waves, and are indeed useful in distinguishing the spin textures with similar spin structure factors. 
We discuss the relation to bond modulation in terms of the kinetic bond energy and the vector spin chirality. 
We also perform numerical calculations beyond the perturbative regime and find that the charge density waves can be enhanced when the electron filling is commensurate. 
Furthermore, we investigate the effect of the spin-orbit coupling, which 
can lead to additional charge density waves owing to effective anisotropic magnetic interactions in momentum space. 
Our result will provide a way to identify complex magnetic orderings and their origins from the charge modulations. 
\end{abstract}

\maketitle

\section{Introduction}

Itinerant magnets consisting of itinerant electrons and localized spins have long been studied in condensed matter physics~\cite{Stewart_RevModPhys.56.755,Stewart_RevModPhys.73.797}. 
The key concept is spin-charge entanglement that arises from the exchange coupling between the itinerant electron spins and the localized spins. 
The interplay between the spin and charge degrees of freedom results in a variety of magnetic, transport, and optical properties. 
For example, it stabilizes helical magnetic orderings through an effective magnetic interaction via the kinetic motion of itinerant electrons, which is called the Ruderman-Kittel-Kasuya-Yosida interaction~\cite{Ruderman,Kasuya,Yosida1957}. 
Besides, various types of noncollinear and noncoplanar magnetic orderings consisting of multiple spin density waves (SDWs) dubbed multiple-$Q$ magnetic orderings are also induced by effective multiple-spin interactions arising from the spin-charge entanglement~\cite{Shindou_PhysRevLett.87.116801,Martin_PhysRevLett.101.156402,Akagi_JPSJ.79.083711,Hayami_PhysRevB.90.060402,batista2016frustration,Ozawa_PhysRevLett.118.147205,hayami2021topological}. 
Conversely, the magnetic structures of the localized spins affect the electronic properties of itinerant electrons, such as the colossal magnetoresistance~\cite{baibich1988giant,ramirez1997colossal,tokura1999colossal,Tokura200005,dagotto2001colossal}, the topological Hall effect~\cite{Ye_PhysRevLett.83.3737,Ohgushi_PhysRevB.62.R6065,tatara2002chirality,Nagaosa_RevModPhys.82.1539,nagaosa2013topological,Lee_PhysRevLett.102.186601,Neubauer_PhysRevLett.102.186602}, the magnetoelectric effect~\cite{Hayami_PhysRevB.90.024432,thole2018magnetoelectric,Gao_PhysRevB.97.134423,Shitade_PhysRevB.98.020407}, and nonreciprocal transport~\cite{tokura2018nonreciprocal,Gao_PhysRevB.98.060402,ishizuka2020anomalous,Hayami_PhysRevB.102.144441}. 

Among rich spin-charge coupled physics, we focus on the charge density wave (CDW) in itinerant magnets.
The spin-charge interplay brings about the possibility of spontaneous formation of the CDW, without relying on repulsive Coulomb interactions or electron-phonon interactions~\cite{Hirsch_PhysRevB.30.5383}.  
Indeed, it was shown that a CDW appears in the Kondo lattice model in one~\cite{Huang_PhysRevB.99.195109}, two~\cite{Misawa_PhysRevLett.110.246401}, and infinite dimensions~\cite{Otsuki_JPSJ.78.034719,Peters_PhysRevB.87.165133}. 
Similar instability was also discussed for the periodic Anderson model~\cite{Zadeh_PhysRevB.55.R3332,Majidi_2007charge}. 
These CDWs are attributed to a quantum many-body effect via the Kondo singlet formation, and thus do not necessarily require magnetic orderings.  
Meanwhile, CDWs can occur in the presence of magnetic ordering. 
In this case, the internal field from the magnetic moments affects the charge degree of freedom via the spin-charge coupling. 
The typical examples were found in a partial magnetic disorder~\cite{Motome2010,Hayami2011,Ishizuka_PhysRevLett.108.257205,Hayami2012,Aulbach_PhysRevB.92.235131,Keler_PhysRevB.102.235125} and a ferrimagnetic order~\cite{Ishizuka_PhysRevLett.109.237207,Akagi_PhysRevB.91.155132} on a triangular lattice, and a noncoplanar triple-$Q$ order on a cubic lattice~\cite{Hayami_PhysRevB.89.085124,hayami2014charge}. 
Interestingly, it was recently shown that such a CDW appears in more complex spin textures, such as a skyrmion crystal (SkX)~\cite{heinze2011spontaneous,Bergmann_PhysRevB.86.134422,von2015influence,hanneken2015electrical,Ozawa_PhysRevLett.118.147205,Kubetzka_PhysRevB.95.104433,Kathyat_PhysRevB.103.035111}. 

In the spin-charge coupled systems, the identification of the CDW provides us important information on the magnetic state and its microscopic origin. 
A successful example has recently been achieved in the centrosymmetric 4$f$-electron material GdRu$_2$Si$_2$~\cite{khanh2020nanometric,Yasui2020imaging,Hayami_PhysRevB.103.024439}, which hosts three multiple-$Q$ magnetic states in an external magnetic field~\cite{khanh2020nanometric}. 
Although one of the three, the square SkX in the intermediate field region, has been identified by a magnetic probe of the Lorentz transmission electron microscopy~\cite{khanh2020nanometric}, the magnetic structures of the other two were indirectly resolved by an electric probe of the spectroscopic-imaging scanning tunneling microscopy measurement~\cite{Yasui2020imaging}. 
The observed CDW modulations were well reproduced based on the Kondo lattice model, which indicates that the concomitant SDWs and CDWs in GdRu$_2$Si$_2$ are a consequence of the spin-charge coupling inherent to itinerant magnets~\cite{Hayami_PhysRevB.103.024439}. 

In the present study, we investigate the CDW formation in the presence of SDWs in a broader context, in order to understand their relationship in details. 
On the basis of the perturbation in terms of the spin-charge coupling, we derive a compact formula for predicting the CDW modulation induced by SDW ordering, which can be applied to arbitrary complex multiple-$Q$ states on any lattices. 
We test the formula for several magnetic orderings including the double-$Q$ (2$Q$) SkX and the $2Q$ meron-antimeron crystal (MAX) on a square lattice, and show that the associated CDW patterns obtained by the numerical diagonalization are well accounted for by the formula. 
In particular, we find that the $2Q$ coplanar state and the MAX are clearly distinguished in terms of the CDWs, although they look similar in terms of the spin structure factor. 
We comment on the relationship between the CDW formation and the bond modulation by the underlying spin texture, from the results for the kinetic bond energy and the vector spin chirality. 
We also show that, in the strong spin-charge coupling region beyond the perturbative regime, the CDW can be enhanced and modulated when the electron filling is commensurate and the electronic band structure tends to be gapped. 
Moreover, we discuss the role of the spin-orbit coupling in the CDW formation, which also induces additional CDWs through effective anisotropic magnetic interactions arising from the spin-orbit coupling. 
Our result indicates the relevance of the CDW modulations to the identification of complex magnetic orderings and their microscopic origins. 

The rest of the paper is organized as follows. 
After introducing the Kondo lattice model in Sec.~\ref{sec:Kondo Lattice model}, we present the results obtained by the perturbative calculation and the numerical diagonalization in a complementary way in Sec.~\ref{sec:Results}. 
Section~\ref{sec:Summary} is devoted to the summary.

\section{Model}
\label{sec:Kondo Lattice model}

We consider the Kondo lattice model, which is one of the prototypical models for itinerant magnets. 
The model consists of itinerant electrons and localized spins, which are coupled via the exchange coupling. 
The Hamiltonian is given by 
\begin{align}
\label{eq:Ham_KLM}
\mathcal{H}= -\sum_{i, j,  \sigma} t_{ij} c^{\dagger}_{i\sigma}c_{j \sigma}
+J \sum_{i, \sigma, \sigma'} c^{\dagger}_{i\sigma} \bm{\sigma}_{\sigma \sigma'} c_{i \sigma'}
\cdot \bm{S}_i, 
\end{align}
where $c^{\dagger}_{i\sigma}$ ($c_{i \sigma}$) is a creation (annihilation) operator of an itinerant electron at site $i$ and spin $\sigma$, and $\bm{S}_i$ is a localized spin at site $i$. 
The first term in Eq.~(\ref{eq:Ham_KLM}) represents the kinetic energy of the itinerant electrons with the hopping parameter $t_{ij}$ between sites $i$ and $j$. 
The second term stands for the onsite exchange (spin-charge) coupling between the itinerant electron spins $\bm{s}_i=(1/2)\sum_{\sigma,\sigma'}c^{\dagger}_{i\sigma} \bm{\sigma}_{\sigma \sigma'} c_{i \sigma'}$ and $\bm{S}_i$, where $\bm{\sigma}=(\sigma^x,\sigma^y,\sigma^z)$ is the vector of Pauli matrices. 
The coupling constant is denoted as $J$. 
We regard $\bm{S}_i$ as the classical spin with $|\bm{S}_i|=1$; the sign of $J$ is irrelevant and no Kondo screening occurs.

The Kondo lattice model in Eq.~(\ref{eq:Ham_KLM}) is an appropriate model to examine the relation between SDW and CDW, since it exhibits a variety of multiple-$Q$ magnetic states, such as SkXs~\cite{Ozawa_PhysRevLett.118.147205,Hayami_PhysRevB.99.094420,Mohanta_PhysRevB.100.064429,Wang_PhysRevLett.124.207201,Kathyat_PhysRevB.103.035111}, chirality density waves~\cite{Solenov_PhysRevLett.108.096403,Hayami_PhysRevB.94.024424,Ozawa_doi:10.7566/JPSJ.85.103703}, and vortex crystals~\cite{hayami2020phase}, in spite of the simple two constituents in the Hamiltonian. 
This is attributed to effective spin interactions that arise from the spin-charge coupling combined with the kinetic motion of itinerant electrons.  
Indeed, it was shown that the Kondo lattice model generally includes effective multiple-spin interactions described in momentum space, e.g., $(\bm{S}_{\bm{q}_1}\cdot \bm{S}_{\bm{q}_2})(\bm{S}_{\bm{q}_3}\cdot \bm{S}_{\bm{q}_4})$, by the perturbation in terms of $J$, where $\bm{S}_{\bm{q}}$ is the Fourier transform of $\bm{S}_i$~\cite{Hayami_PhysRevB.95.224424}. 

In the following, we discuss the CDW under the SDW by postulating magnetic textures for $\bm{S}_i$ in the ground state. 
As will be discussed in Sec.~\ref{sec:Perturbative Analysis}, 
the charge modulation is caused by multiple spin scatterings of itinerant electrons, which are formulated by effective spin interactions in momentum space arising from the kinetic motion of itinerant electrons [see Eq.~(\ref{eq:perturbation_nq})]. 
Thereby, the following results can also be applied to other itinerant electron models, such as the Hubbard model and the periodic Anderson model in the region where the effective interactions are described by the multiple spin interactions in momentum space.

\section{Results}
\label{sec:Results}

In this section, we study the CDW modulation in the presence of SDW for the Kondo lattice model in Eq.~(\ref{eq:Ham_KLM}). 
First, we derive a general formula for the CDW induced by the SDW on the basis of the perturbation in terms of the spin-charge coupling in Sec.~\ref{sec:Perturbative Analysis}. 
Then, we test the formula for representative examples, in comparison with the numerical results by the direct exact diagonalization. 
Specifically, we take single-$Q$ (1$Q$) and 2$Q$ magnetic orderings on a square lattice. 
While the formula holds for generic cases, for simplicity we take into account only the nearest-neighbor hopping, $t=1$, in the model in Eq.~(\ref{eq:Ham_KLM}) on the square lattice in the calculations.  
In Sec.~\ref{sec:Charge density waves in single-$Q$ sinusoidal orderings}, we discuss the CDW in the 1$Q$ sinusoidal orderings with and without a net magnetization. 
We further investigate the CDWs in the presence of the $2Q$ magnetic orderings composed of superpositions of two SDWs in Sec.~\ref{sec:Charge density waves in multiple-$Q$ orderings}. 
In Sec.~\ref{sec:Relation to other quantities}, we show the relationship between the CDW and the other spin and charge related quantities, the vector chirality and the kinetic bond energy. 
We discuss the behavior in a wider parameter region beyond the perturbative regime in Sec.~\ref{sec:Parameter dependence}. 
We also discuss the effect of the antisymmetric spin-orbit interaction on the CDW for a polar tetragonal system in Sec.~\ref{sec:Effect of spin-orbit coupling}.

\subsection{Perturbative analysis}
\label{sec:Perturbative Analysis}

\begin{figure}[htb!]
\begin{center}
\includegraphics[width=0.7 \hsize]{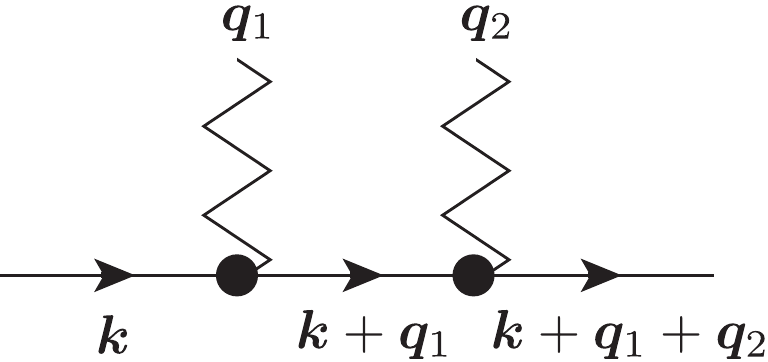} 
\caption{
\label{Fig:diagram}
Feynman diagram for the lowest-order contribution to the charge density $n_{\bm{q}}$ in the perturbation expansion in terms of the spin-charge coupling $J$; see Eq.~(\ref{eq:perturbation_nq}).  
The vertices with wavy lines denote the scattering of the itinerant electrons by the localized spins, and the solid lines with arrows represent the bare propagators of itinerant electrons. 
}
\end{center}
\end{figure}

For the preparation of the perturbative argument, we perform the Fourier transform of the model in Eq.~(\ref{eq:Ham_KLM}) as 
\begin{align}
\label{eq:Ham_kspace}
\mathcal{H}=\sum_{\bm{k},\sigma} \varepsilon_{\bm{k}} c^{\dagger}_{\bm{k}\sigma}c_{\bm{k}\sigma} +
\frac{J}{\sqrt{N}} \sum_{\bm{k},\bm{q},\sigma, \sigma'} c^{\dagger}_{\bm{k}\sigma}\bm{\sigma}_{\sigma \sigma'} c_{\bm{k}+\bm{q}\sigma'} \cdot \bm{S}_{\bm{q}}, 
\end{align}
where $\varepsilon_{\bm{k}}$ is the energy dispersion and $N$ is the number of sites; $c_{\bm{k} \sigma}^{\dagger}$ and $c_{\bm{k}\sigma}$ are the Fourier transform of $c_{i\sigma}^{\dagger}$ and $c_{i\sigma}$, respectively. 
The second term denotes the scattering of itinerant electrons by the localized spins with momentum transfer $\bm{q}$.

In the presence of SDW, the CDW modulation of itinerant electrons is caused by the scattering by the localized spins $\bm{S}_{\bm{q}}$ in the second term of Eq.~(\ref{eq:Ham_kspace}). 
When the spin-charge coupling $J$ is sufficiently small compared to the bare bandwidth of itinerant electrons, we can estimate the charge density with wave vector $\bm{q}$, $n_{\bm{q}}=(1/N)\sum_{\sigma}c^{\dagger}_{\bm{k}+\bm{q}\sigma}c_{\bm{k}\sigma}$, by using the perturbative expansion in terms of $J$. 
The lowest-order contribution comes from the second-order process, as represented by the Feynman diagram in Fig.~\ref{Fig:diagram}. 
The diagram consists of two scattering vertices by the localized spins and three bare propagators of the itinerant electrons, which is explicitly given by 
\begin{align}
\label{eq:perturbation_nq}
n_{\bm{q}}=\frac{2J^2}{N^2}T\sum_{\bm{k},\bm{q}_1,\bm{q}_2}\sum_{\omega_n}&G^0_{\bm{k}}G^0_{\bm{k}+\bm{q}_1}G^0_{\bm{k}+\bm{q}_1+\bm{q}_2} \nonumber \\
&\times (\bm{S}_{\bm{q}_1}\cdot \bm{S}_{\bm{q}_2}) \delta_{\bm{q}_1+\bm{q}_2,\bm{q}+l\bm{G}},
\end{align}
where $G^0_{\bm{k}}(i \omega_p)=[i \omega_p-(\varepsilon_{\bm{k}}-\mu)]^{-1}$ is noninteracting Green's function, $T$ is the temperature (the Boltzmann constant is taken as unity), $\omega_p$ is the Matsubara frequency, $\mu$ is the chemical potential, $\delta$ is the Kronecker delta, and $\bm{G}$ is the reciprocal lattice vector ($l$ is an integer).  
We drop the spin index for Green's function because the kinetic term in the Hamiltonian is spin independent (we will discuss a spin-dependent case in Sec.~\ref{sec:Effect of spin-orbit coupling}). 
We note that the summation with respect to the Matsubara frequency is analytically taken as 
\begin{align}
&T\sum_{\omega_n}G^0_{\bm{k}}G^0_{\bm{k}'}G^0_{\bm{k}''} = \nonumber \\
&\frac{f(\varepsilon_{\bm{k}})(\varepsilon_{\bm{k}'}-\varepsilon_{\bm{k}''})
+f(\varepsilon_{\bm{k}'})(\varepsilon_{\bm{k}''}-\varepsilon_{\bm{k}})
+
f(\varepsilon_{\bm{k}''})(\varepsilon_{\bm{k}}-\varepsilon_{\bm{k}'})}
{(\varepsilon_{\bm{k}}-\varepsilon_{\bm{k}'})(\varepsilon_{\bm{k}}-\varepsilon_{\bm{k}''})(\varepsilon_{\bm{k}'}-\varepsilon_{\bm{k}''})}. 
\end{align}

The expression in Eq.~(\ref{eq:perturbation_nq}) indicates that the CDW modulation with wave vector $\bm{q}$ is predominantly induced by the two-spin scattering in the form of $\bm{S}_{\bm{q}_1}\cdot \bm{S}_{\bm{q}_2}$ which satisfies the momentum conservation $\bm{q}=\bm{q}_1+\bm{q}_2$. 
In other words, only the magnetic correlations with nonzero $\bm{S}_{\bm{q}_1}\cdot \bm{S}_{\bm{q}-\bm{q}_1}$ contribute to the CDW modulation with wave vector $\bm{q}$. 
In addition, the magnitude of the CDW modulation is strongly affected by the band structure and the electron filling through Green's functions in Eq.~(\ref{eq:perturbation_nq}). 

We can also obtain higher-order contributions to $n_{\bm{q}}$ by higher-order expansions in terms of $J$. 
Note that no odd-order terms in $J$ appear in the presence of time-reversal symmetry. 
As expected from Eq.~(\ref{eq:perturbation_nq}), the $2n$th-order contribution is given in the form of $(\bm{S}_{\bm{q}_1}\cdot \bm{S}_{\bm{q}_2}) (\bm{S}_{\bm{q}_3}\cdot \bm{S}_{\bm{q}_4})\cdots (\bm{S}_{\bm{q}_{2n-1}}\cdot \bm{S}_{\bm{q}_{2n}})$: for example, the fourth-order contribution is proportional to $(\bm{S}_{\bm{q}_1}\cdot \bm{S}_{\bm{q}_2}) (\bm{S}_{\bm{q}_3}\cdot \bm{S}_{\bm{q}_4})$ with $\bm{q}=\bm{q}_1+\bm{q}_2+\bm{q}_3+\bm{q}_4 +l\bm{G}$.

\subsection{Single-$Q$ spin states}
\label{sec:Charge density waves in single-$Q$ sinusoidal orderings}

To test the perturbative argument in Sec.~\ref{sec:Perturbative Analysis}, we first numerically evaluate Eq.~(\ref{eq:perturbation_nq}) for $1Q$ states on a square lattice as the simplest example, and compare the results with those by the direct diagonalization of the Kondo lattice Hamiltonian in Eq.~(\ref{eq:Ham_KLM}). 
We discuss the CDW in the $1Q$ sinusoidal SDW without a net magnetization in Sec.~\ref{sec:Without a net magnetization} and that with a nonzero magnetization in Sec.~\ref{sec:With a net magnetization}.

\begin{figure*}[htb!]
\begin{center}
\includegraphics[width=1.0 \hsize]{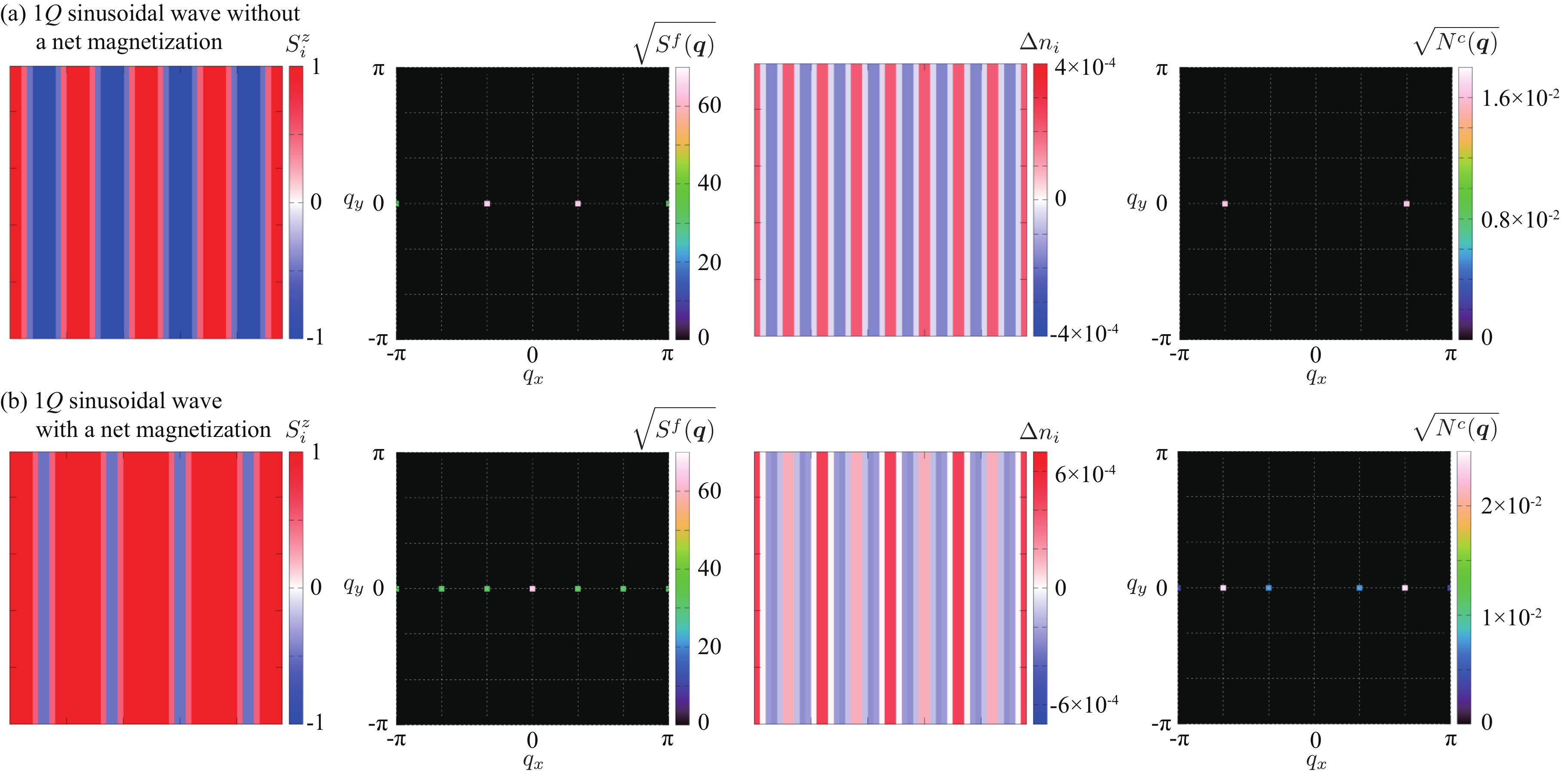} 
\caption{
\label{Fig:noSOC_1Q}
Left: Real-space spin configurations of the 1$Q$ sinusoidal spin state with $\bm{Q}_1=(\pi/3,0)$, (a) without a net magnetization in Eq.~(\ref{eq:1Qsinusoidal_zero}) and (b) with a nonzero magnetization $\tilde{M}^z = 0.7$ in Eq.~(\ref{eq:1Qsinusoidal_nonzero}). 
The contour shows the $z$ component of the spin moment. 
Note that the spin states are collinear: (a) up-up-up-down-down-down and (b) up-up-up-up-up-down. 
Middle left: The square root of the spin structure factor for the localized spins in the first Brillouin zone.
Middle right:  Real-space distributions of the local charge density measured from the average density. 
Right: The square root of the charge structure factor. 
The data for the CDW modulations are obtained by the direct diagnalization of the Hamiltonian in Eq.~(\ref{eq:Ham_KLM}) at $J=0.05$ and $\mu=3$ for the system size with $N=96^2$ under the periodic boundary conditions.
}
\end{center}
\end{figure*}

\begin{figure}[htb!]
\begin{center}
\includegraphics[width=1.0 \hsize]{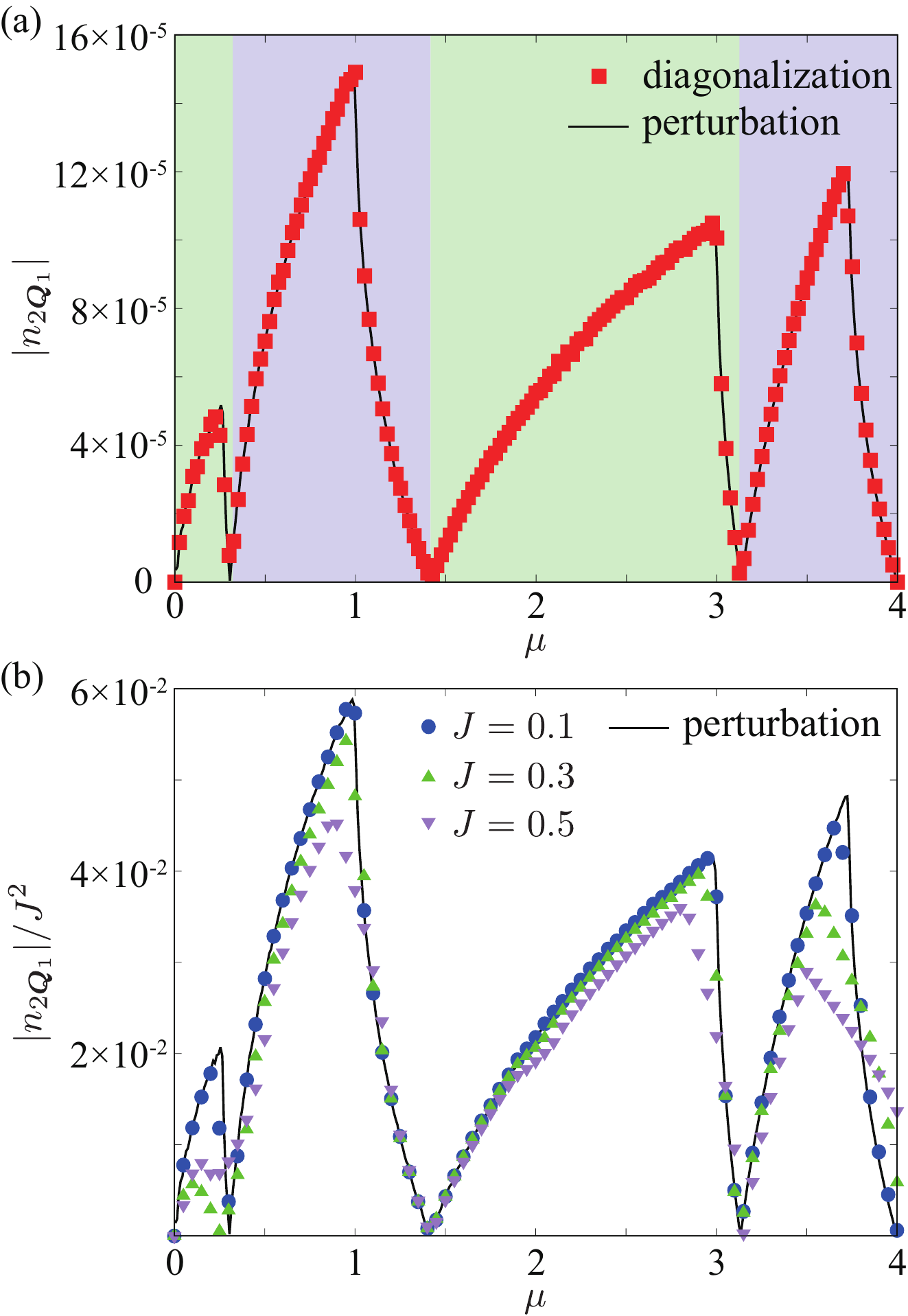} 
\caption{
\label{Fig:perturbation}
(a) $|n_{2\bm{Q}_1}|$ under the 1$Q$ sinusoidal spin ordering with $\bm{Q}_1=(\pi/3,0)$ in Eq.~(\ref{eq:1Qsinusoidal_zero}) as a function of the chemical potential $\mu$ for $J=0.05$ and $N=1440^2$. 
The squares show the results obtained by the direct diagonalization of Eq.~(\ref{eq:Ham_KLM}) and the solid lines show those by the perturbative formula in Eq.~(\ref{eq:perturbation_nq}). 
The blue (green) shaded regions represent $n_{2\bm{Q}_1}>0$ ($n_{2\bm{Q}_1}<0$). 
(b) The same plot as (a) for $J=0.1$, $0.3$, and $0.5$, where $|n_{2\bm{Q}_1}|$ is renormalized by $J^2$ for comparison. 
}
\end{center}
\end{figure}

\subsubsection{Without a net magnetization}
\label{sec:Without a net magnetization}

The spin texture characterized by the 1$Q$ sinusoidal wave is represented by 
\begin{align}
\label{eq:1Qsinusoidal_zero}
\bm{S}_i =\mathcal{N}_i (0,0, \cos \bm{Q}_1\cdot \bm{r}_i), 
\end{align}
where $\bm{r}_i$ is the position vector at site $i$ and $\mathcal{N}_i$ represents the normalization to satisfy $|\bm{S}_i|=1$ at each site. 
Here we take the ordering vector $\bm{Q}_1=(\pi/3,0)$ (the lattice constant is set to unity). 
The spin configuration is shown in the left panel of Fig.~\ref{Fig:noSOC_1Q}(a), where the spins are aligned in an up-up-up-down-down-down way along the $x$ direction, and hence, there is no net magnetization. 
This spin state shows dominant Bragg peaks at $\pm \bm{Q}_1$ in the spin structure factor, and in addition, higher harmonics at $\pm 3\bm{Q}_1$, as shown in the middle left panel of Fig.~\ref{Fig:noSOC_1Q}(a). 
Here, the spin structure factor for the localized spins is calculated as 
\begin{align}
S^f(\bm{q})=\frac{1}{N} \sum_{ij}\bm{S}_i \cdot  \bm{S}_j e^{i \bm{q}\cdot (\bm{r}_i-\bm{r}_j)}.
\end{align}

In this situation, we can predict the CDW modulation by using Eq.~(\ref{eq:perturbation_nq}). 
Specifically, by plugging $\pm \bm{Q}_1$ and $\pm 3\bm{Q}_1$ into $\bm{q}_1$ and $\bm{q}_2$ in Eq.~(\ref{eq:perturbation_nq}), we find that only the $\bm{q}=\pm 2\bm{Q}_1$ components of $n_{\bm{q}}$ become nonzero in the first Brillouin zone. 
Note that there are three combinations that contribute to $n_{2\bm{Q}_1}$: 
$(\bm{q}_1, \bm{q}_2)=(\bm{Q}_1, \bm{Q}_1)$, $(-\bm{Q}_1,3\bm{Q}_1)$, and $(3\bm{Q}_1, -\bm{Q}_1)$ ($\pm 3\bm{Q}_1$ are equivalent as they are on the zone boundary). 
In Fig.~\ref{Fig:perturbation}(a), the solid lines show the chemical potential $\mu$ dependence of $|n_{2\bm{Q}_1}|$ calculated by Eq.~(\ref{eq:perturbation_nq}) for the $\bm{k}$-space mesh of $1440 \times 1440$ ($N=1440^2$). 
We choose a small value of $J=0.05$, as the expression in Eq.~(\ref{eq:perturbation_nq}) is derived in the limit of the weak spin-charge coupling. 
The result for $\mu<0$ is obtained from that for $\mu>0$ owing to particle-hole symmetry of the model; 
$\mu=0$ corresponds to the half filling. 
As shown in Fig.~\ref{Fig:perturbation}(a), $|n_{2\bm{Q}_1}|$ shows a nonmonotonic behavior against $\mu$. 
The sign of $n_{2\bm{Q}_1}$, which corresponds to the phase of the CDW, changes at particular values of $\mu$; $n_{2\bm{Q}_1}>0$ for $0.35\lesssim \mu \lesssim 1.42$ and $3.13 \lesssim \mu \lesssim 4$, while $n_{2\bm{Q}_1}<0$ for the other regions. 

Independently, we can evaluate the CDW modulation by the direct numerical diagonalization of the Kondo lattice model in Eq.~(\ref{eq:Ham_KLM}). 
Substituting the $1Q$ sinusoidal state in Eq.~(\ref{eq:1Qsinusoidal_zero}) into Eq.~(\ref{eq:Ham_KLM}), the local charge density modulation is calculated as $\Delta n_i = \sum_\sigma \langle  c^{\dagger}_{i\sigma}c_{i\sigma} \rangle -n^{\rm ave}$, where $n^{\rm ave}$ is the average charge density. 
The real-space distribution of $\Delta n_i$ at $\mu=3$ is shown in the middle right panel of Fig.~\ref{Fig:noSOC_1Q}(a). 
The data are calculated for the system size $N=96^2$ (the figure shows a part of the whole system with $24^2$ sites).  
The characteristic wave vectors of this CDW is extracted from the charge structure factor defined by 
\begin{align}
N^c(\bm{q})=\frac{1}{N}\sum_{i,j} \Delta n_i \Delta n_j e^{i \bm{q}\cdot (\bm{r}_i-\bm{r}_j)}. 
\end{align}
Note that $|n_{\bm{q}}|=\sqrt{N^c(\bm{q})/N}$. 
The result is plotted in the right panel of Fig.~\ref{Fig:noSOC_1Q}(a). 
We find that the Bragg peaks appear only at $\pm 2\bm{Q}_1$, as predicted by the perturbative formula in Eq.~(\ref{eq:perturbation_nq}). 

In order to quantitatively test the perturbative argument, we compare the $\mu$ dependences of $|n_{2\bm{Q}_1}|$ between the results by Eq.~(\ref{eq:perturbation_nq}) and the direct diagonalization at $J=0.05$ in Fig.~\ref{Fig:perturbation}(a). 
We find good agreement in the whole range of $\mu$ including the sign change of $n_{2\bm{Q}_1}$, which validates the perturbative argument. 
Figure~\ref{Fig:perturbation}(b) shows further comparison for larger $J$. 
The results by the diagonalization gradually deviate from that by the perturbative calculations while increasing $J$, whereas we do not find any additional CDW modulations at other $\bm{q}$ within this range (see Sec.~\ref{sec:Parameter dependence} for larger $J$). 

The $2\bm{Q}_1$ CDW modulation for the $\bm{Q}_1$ sinusoidal SDW is also intuitively understood from the real-space picture. 
When the itinerant electrons move on the up-up-up-down-down-down spin texture, the effective hopping amplitude is modulated in a different way for neighboring parallel (up-up or down-down) spins and antiparallel (up-down or down-up) spins. 
This modulation of the kinetic energy results in the CDW modulation. 
As the period of the modulation of the kinetic energy is a half of that of the SDW, the period of the charge modulation is also halved, which results in the $2\bm{Q}_1$ CDW. 
(Detailed analysis will be given for the $2Q$ cases in Sec.~\ref{sec:Relation to other quantities}.) 

Similar results will apply to the $1Q$ SDW states with spin spirals, at least, when the spirals are elliptically modulated as 
\begin{align}
\label{eq:1Qelliptical}
\bm{S}_i =\mathcal{N}_i
(0,a_y \sin \bm{Q}_1\cdot \bm{r}_i, a_z\cos \bm{Q}_1\cdot \bm{r}_i), 
\end{align}
with $a_y \neq a_z$; note that Eq.~(\ref{eq:1Qelliptical}) with $a_y=0$ and $a_z\neq 0$ reduces to the 1$Q$ sinusoidal spin state in Eq.~(\ref{eq:1Qsinusoidal_zero}). 
Meanwhile, the CDW modulation vanishes for the 1$Q$ circular spiral state with $a_y=a_z$, since $n_{2\bm{Q}_1}$ in Eq.~(\ref{eq:perturbation_nq}) becomes zero owing to the cancellation between $S^y_{\bm{Q}_1} S^y_{\bm{Q}_1}$ and $S^z_{\bm{Q}_1} S^z_{\bm{Q}_1}$. 
This is reasonable from the real-space picture: the effective hopping is renormalized but remains spatially uniform because of the uniform twist of spins.

\subsubsection{With a net magnetization}
\label{sec:With a net magnetization}

We further perform the comparison for the 1$Q$ sinusoidal spin configuration with a net magnetization given by 
\begin{align}
\label{eq:1Qsinusoidal_nonzero}
\bm{S}_i =\mathcal{N}_i (0,0, \tilde{M}^z+\cos \bm{Q}_1\cdot \bm{r}_i). 
\end{align}
We take $\tilde{M}^z=0.7$ so that the real-space spin configuration becomes the up-up-up-up-up-down spin configuration, as shown in the left panel of Fig.~\ref{Fig:noSOC_1Q}(b). 
Note that $\tilde{M}^z$ is not the actual value of the net magnetization. 
In contrast to the case without the magnetization, this spin state shows additional peaks at $\bm{q}=\bm{0}$ and $\pm 2\bm{Q}_1$ in the spin structure factor, as shown in the middle left panel of Fig.~\ref{Fig:noSOC_1Q}(b). 
Reflecting the additional Fourier components, the perturbative formula in Eq.~(\ref{eq:perturbation_nq}) predicts additional CDW modulations at $\bm{q}=\pm \bm{Q}_1$ and $\pm 3\bm{Q}_1$ owing to nonzero spin products of $S^z_{\pm\bm{Q}_1} S^z_{\bm{0}}$ and $S^z_{\pm\bm{Q}_1} S^z_{\pm 2\bm{Q}_1}$. 
This is indeed confirmed by the direct diagonalization, as shown in the real-space charge modulation and the charge structure factor in the middle right and right panels of Fig.~\ref{Fig:noSOC_1Q}(b), respectively.

\subsection{Multiple-$Q$ spin states}
\label{sec:Charge density waves in multiple-$Q$ orderings}

The above analysis can be straightforwardly applied to more complex spin configurations. 
In this section, we demonstrate it for four types of the 2$Q$ spin textures with the ordering vectors $\bm{Q}_1=(\pi/3,0)$ and $\bm{Q}_2=(0,\pi/3)$: 
the 2$Q$ chiral stripe state in Sec.~\ref{sec:2$Q$ chiral stripe}, the 2$Q$ coplanar state in Sec.~\ref{sec:2$Q$ coplanar}, the 2$Q$ MAX in Sec.~\ref{sec:2$Q$ MAX}, and the 2$Q$ SkX in Sec.~\ref{sec:2$Q$ SkX}. 
We show that while all these spin states are described by the superpositions of two SDWs, the characteristic wave vectors of the CDW are different from each other, which are in good agreement with the predictions from the perturbative argument. 
Throughout this section, the numerical calculations by the direct diagonalization are done for $J=0.05$ and $\mu=3$, and the system size with 
$N=96^2$. 

\begin{figure*}[htb!]
\begin{center}
\includegraphics[width=1.0 \hsize]{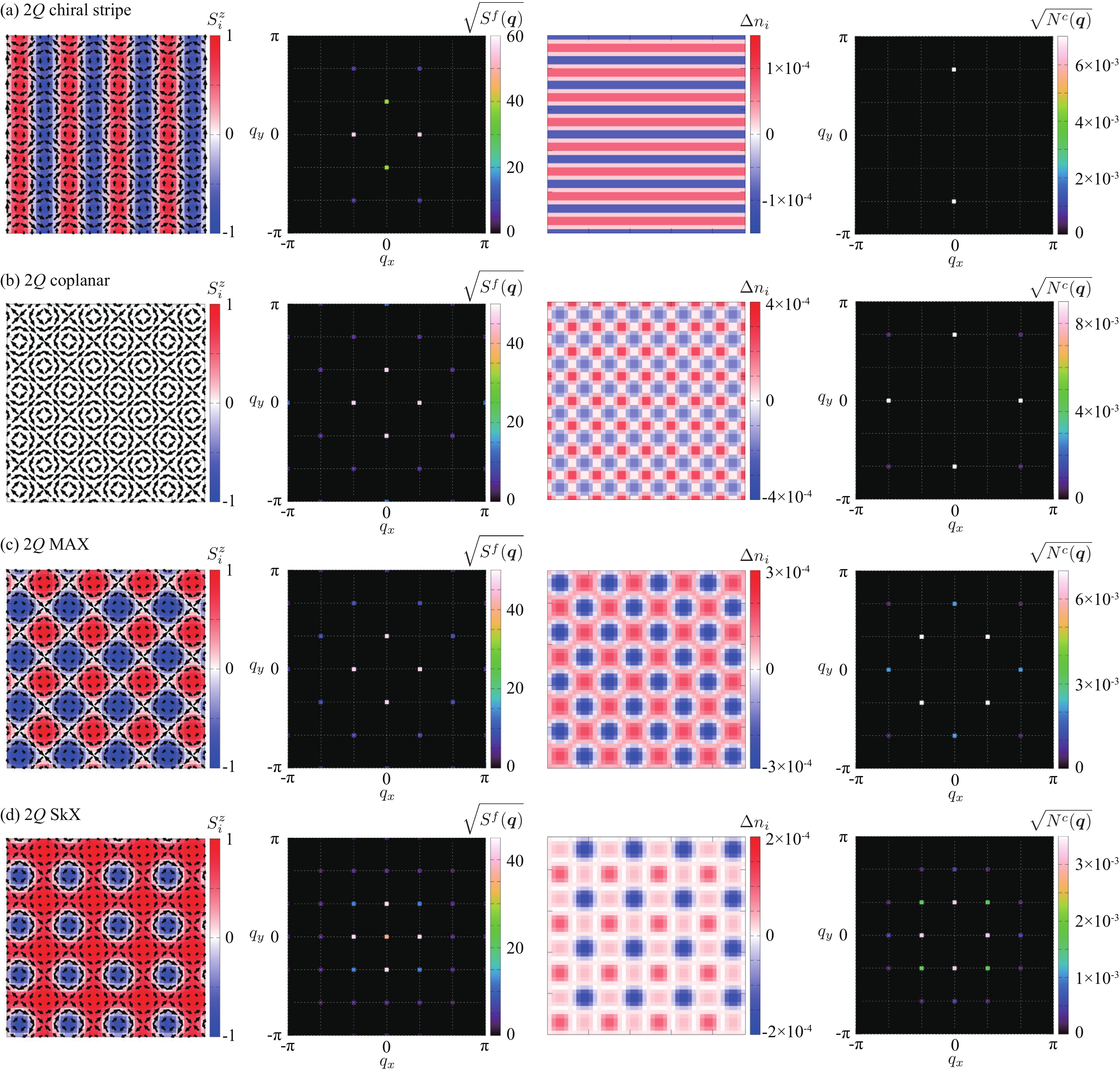} 
\caption{
\label{Fig:noSOC_multiQ}
Left: Real-space spin configurations of (a) the 2$Q$ chiral stripe state in Eq.~(\ref{eq:2QCS}), (b) the 2$Q$ coplanar state in Eq.~(\ref{eq:2Qcoplanar}), (c) the 2$Q$ MAX in Eq.~(\ref{eq:2QMAX}), and (d) the 2$Q$ SkX in Eq.~(\ref{eq:2QSkX}). 
The arrows and the contour show the $xy$ and $z$ components of the localized spins, respectively. 
Middle left: The square root of the spin structure factor for the localized spins in the first Brillouin zone.
Middle right:  Real-space distributions of the local charge density measured from the average density. 
Right: The square root of the charge structure factor. 
}
\end{center}
\end{figure*}

\subsubsection{2$Q$ chiral stripe}
\label{sec:2$Q$ chiral stripe}

We begin with a superposition of the 1$Q$ spiral wave along the $\bm{Q}_1$ direction and the 1$Q$ sinusoidal wave along the $\bm{Q}_2$ direction. 
This is called the $2Q$ chiral stripe state, whose spin configuration is given by~\cite{Ozawa_doi:10.7566/JPSJ.85.103703} 
\begin{align}
\label{eq:2QCS}
\bm{S}_i =\mathcal{N}_i
\left(
    \begin{array}{c}
    b \sin  \bm{Q}_2 \cdot \bm{r}_i  \\
     \sqrt{1-b^2 \sin^2  \bm{Q}_2 \cdot \bm{r}_i }\cos \bm{Q}_1\cdot \bm{r}_i \\
    \sqrt{1-b^2 \sin^2  \bm{Q}_2 \cdot \bm{r}_i }\sin \bm{Q}_1\cdot \bm{r}_i
          \end{array}
  \right)^{\rm T},
\end{align}
where $b$ is a parameter that controls the mixture of the second component with $\bm{Q}_2$; we here take $b=0.8$. 
The real-space spin configuration is shown in the left panel of Fig.~\ref{Fig:noSOC_multiQ}(a), which consists of a periodic array of vortices and antivortices in the $xy$-spin components and the stripe modulation in the $z$-spin component. 
Owing to the noncoplanar spin texture, this state accompanies a density wave of the scalar spin chirality defined by the triple product of three neighboring spins, $\bm{S}_i \cdot (\bm{S}_j \times \bm{S}_k)$, along the $\bm{Q}_2$ direction, which is the reason why this spin state is called the $2Q$ chiral stripe. 
There is no net scalar chirality, and hence, no topological Hall effect occurs in this state. 
The spin structure factor exhibits the Bragg peaks at $\pm \bm{Q}_1$ and $\pm \bm{Q}_2$ with different intensities and at the higher harmonics at $\pm \bm{Q}_1 \pm 2 \bm{Q}_2$, as shown in the middle left panel of Fig.~\ref{Fig:noSOC_multiQ}(a). 
The nonzero intensities at $\pm \bm{Q}_1 \pm 2 \bm{Q}_2$ are attributed to the factor $\sqrt{1-b^2 \sin^2 \bm{Q}_2 \cdot  \bm{r}_i}$ in Eq.~(\ref{eq:2QCS}), which includes the contribution of $ \cos  2\bm{Q}_2 \cdot  \bm{r}_i$. 
This 2$Q$ chiral stripe has been widely found in itinerant magnets on various lattices, e.g., the Kondo lattice model on the square~\cite{Ozawa_doi:10.7566/JPSJ.85.103703,Hayami_PhysRevB.95.224424,Hayami_PhysRevB.103.024439}, triangular~\cite{Hayami_PhysRevB.95.224424,hayami2020multiple,Hayami_PhysRevB.103.054422}, and cubic~\cite{Okumura_PhysRevB.101.144416} lattices, and the $d$-$p$ model on the square lattice~\cite{yambe2020double}. 

The right two panels of Fig.~\ref{Fig:noSOC_multiQ}(a) represent the real-space charge distribution and the charge structure factor in the 2$Q$ chiral stripe state. 
The local charge density oscillates only along the $\bm{Q}_2$ direction, and indeed, the charge structure factor exhibits the peaks only at $\pm \bm{Q}_2$. 
At first glance, the result appears to contradict with the perturbative argument for the given peaks in the spin structure factor at $\pm \bm{Q}_1$, $\pm \bm{Q}_2$, and $\pm \bm{Q}_1 \pm 2 \bm{Q}_2$, but it is understood as follows. 
Equation~(\ref{eq:2QCS}) is represented by a superposition of the circular spiral wave with $|S^y_{\bm{Q}_1}|=|S^z_{\bm{Q}_1}|$ along the $\bm{Q}_1$ direction and the sinusoidal wave along the $\bm{Q}_2$ direction. 
The latter gives rise to nonzero $n_{2\bm{Q}_2}$, whereas the former does not lead to any CDW, as discussed in Sec.~\ref{sec:Without a net magnetization}. 
Thus, the CDW with $\pm \bm{Q}_2$ is compatible with the perturbative formula in Eq.~(\ref{eq:perturbation_nq}).

\subsubsection{2$Q$ coplanar}
\label{sec:2$Q$ coplanar}

Next, we consider the 2$Q$ coplanar state, which is characterized by a superposition of two sinusoidal waves. 
The spin configuration is given by 
\begin{align}
\label{eq:2Qcoplanar}
\bm{S}_i =\mathcal{N}_i
\left(
    -\cos  \bm{Q}_2 \cdot \bm{r}_i,
     \cos  \bm{Q}_1 \cdot \bm{r}_i,
  0
  \right).
\end{align}
The real-space picture of the localized spins is shown in the left panel of Fig.~\ref{Fig:noSOC_multiQ}(b). 
This state also consists of a periodic array of vortices and antivortices like the 2$Q$ chiral stripe state, but all the spins are coplanar with no $z$-spin component. 
Owing to the normalization condition of $|\bm{S}_i|=1$, the spin structure factor shows the peaks at higher harmonics $\pm \bm{Q}_\nu \pm 2\bm{Q}_\nu$ and $\pm 2\bm{Q}_\nu \pm 3\bm{Q}_\nu$ ($\nu=1,2$), in addition to $\pm \bm{Q}_\nu$ and $\pm 3\bm{Q}_\nu$ expected for each sinusoidal wave, as shown in the middle left panel of Fig.~\ref{Fig:noSOC_multiQ}(b). 
This state has been discussed in itinerant magnets at zero field~\cite{hayami2018multiple} and in frustrated and itinerant magnets under an external magnetic field~\cite{Hayami_PhysRevB.95.224424,Hayami_PhysRevB.103.024439,Utesov_PhysRevB.103.064414,Wang_PhysRevB.103.104408}. 

In the 2$Q$ coplanar state, the local charge density is modulated in both the $x$ and $y$ directions unlike the $2Q$ chiral stripe, as shown in the middle right panel of Fig.~\ref{Fig:noSOC_multiQ}(b). 
The charge structure factor exhibits the peaks at $\pm 2\bm{Q}_1\pm 2\bm{Q}_2$, in addition to $\pm 2\bm{Q}_1$ and $\pm 2\bm{Q}_2$ which are expected from each sinusoidal wave. 
This is because there are nonzero contributions from, e.g., 
$\bm{S}_{\bm{Q}_1}\cdot \bm{S}_{\bm{Q}_1+2\bm{Q}_2}$ and $\bm{S}_{3\bm{Q}_2}\cdot \bm{S}_{2\bm{Q}_1-\bm{Q}_2}$, to $n_{\bm{q}}$ with $\bm{q}=2\bm{Q}_1+2\bm{Q}_2$ in Eq.~(\ref{eq:perturbation_nq}).

\subsubsection{2$Q$ MAX}
\label{sec:2$Q$ MAX}

The 2$Q$ MAX is obtained as a modulation of the $2Q$ coplanar state by adding a nonzero $z$-spin component as 
\begin{align}
\label{eq:2QMAX}
\bm{S}_i =\mathcal{N}_i
\left(
    \begin{array}{c}
    -\cos  \bm{Q}_2 \cdot \bm{r}_i  \\
     \cos  \bm{Q}_1 \cdot \bm{r}_i  \\
 -\sin  \bm{Q}_1 \cdot \bm{r}_i  -\sin  \bm{Q}_2 \cdot \bm{r}_i 
          \end{array}
  \right)^{\rm T}.
\end{align}
This is regarded as a superposition of two spin helices: one is in the $yz$-spin component with wave vector $\bm{Q}_1$, and the other is in the $xz$-spin component with $\bm{Q}_2$. 
The real-space spin configuration is shown in the left panel of Fig.~\ref{Fig:noSOC_multiQ}(c). 
Reflecting the modulation in the $z$-spin component, the 2$Q$ MAX consists of a periodic array of meron and antimeron characterized by a half skyrmion number each with opposite sign~\cite{brey1996skyrme,yu2018transformation,kurumaji2019skyrmion}. 
The cancellation of the skyrmion number between the merons and antimerons leads to no topological Hall effect. 
The peak positions of the spin structure factor are the same as those in the 2$Q$ coplanar state, as shown in the middle left panel of Fig.~\ref{Fig:noSOC_multiQ}(c). 
This state has been discussed in chiral magnets~\cite{Lin_PhysRevB.91.224407}, frustrated magnets~\cite{Wang_PhysRevB.103.104408}, and polar itinerant magnets~\cite{Hayami_PhysRevLett.121.137202,hayami2021meron}. 

In spite of the same set of the Bragg peaks in the spin structure factor, the CDW modulations are qualitatively different between the 2$Q$ coplanar state and the 2$Q$ MAX, as shown in the right two panels of Figs.~\ref{Fig:noSOC_multiQ}(b) and \ref{Fig:noSOC_multiQ}(c), respectively. 
The primary difference is found in the $\pm\bm{Q}_1 \pm \bm{Q}_2$ components in the charge structure factor; the 2$Q$ MAX has nonzero $n_{\pm \bm{Q}_1 \pm \bm{Q}_2}$, while the 2$Q$ coplanar state does not. 
This is because the 2$Q$ structure in the $z$-spin component in the 2$Q$ MAX gives a contribution of $S^z_{\pm \bm{Q}_1}S^z_{\pm\bm{Q}_2}$ to $n_{\bm{q}}$ with $\bm{q}=\pm\bm{Q}_1 \pm \bm{Q}_2$ in Eq.~(\ref{eq:perturbation_nq}). 
Furthermore, it is worthwhile mentioning that the 2$Q$ MAX state exhibits the CDW modulations, although the constituent waves are the circular spiral waves rather than the sinusoidal waves.  
This appears to contradict with the observations in the $1Q$ case in Sec.~\ref{sec:Without a net magnetization}, but can be understood from the different amplitudes between the $xy$- and $z$-spin components; the former has the 1$Q$ component, whereas the latter has 2$Q$, which effectively make the $xy$ and $z$ spins inequivalent. 
Thus, the multiple-$Q$ spiral states may be accompanied by different CDW modulation from the 1$Q$ cases.

\subsubsection{2$Q$ SkX}
\label{sec:2$Q$ SkX}

Lastly, we discuss the CDW in the 2$Q$ SkX. 
The spin configuration is obtained by adding a uniform $z$-spin component to the 2$Q$ MAX in Eq.~(\ref{eq:2QMAX}), which is given by 
\begin{align}
\label{eq:2QSkX}
\bm{S}_i =\mathcal{N}_i
\left(
    \begin{array}{c}
    -\cos  \bm{Q}_2 \cdot \bm{r}_i  \\
     \cos  \bm{Q}_1 \cdot \bm{r}_i  \\
\tilde{M}^z   -\sin  \bm{Q}_1 \cdot \bm{r}_i  -\sin  \bm{Q}_2 \cdot \bm{r}_i 
          \end{array}
  \right)^{\rm T}. 
\end{align}
We here set $\tilde{M}^z=0.7$. 
As shown in the left panel of Fig.~\ref{Fig:noSOC_multiQ}(d), the skyrmion cores defined by $S^z_i \simeq -1$ form a square lattice, and hence, this state is called the square SkX. 
Different from the 2$Q$ MAX, this spin configuration shows a nonzero net value of the scalar spin chirality, which results in the topological Hall effect. 
Owing to the nonzero magnetization, there are additional peaks in the spin structure factor at $\bm{q}=\bm{0}$, $\pm \bm{Q}_1 \pm \bm{Q}_2$, $\pm 2\bm{Q}_1$, $\pm 2\bm{Q}_2$, and $\pm 2\bm{Q}_1 \pm 2\bm{Q}_2$, as shown in the middle left panel of Fig.~\ref{Fig:noSOC_multiQ}(d) (we note that there are also peaks with weak intensity on the Brillouin zone boundary, e.g., at $3\bm{Q}_1+\bm{Q}_2$). 
This state has been widely discussed in itinerant magnets~\cite{Hayami_PhysRevB.103.024439} and in localized magnets~\cite{Utesov_PhysRevB.103.064414,Wang_PhysRevB.103.104408}.

The additional $\bm{q}$ components of spins induce additional charge modulations, as shown in the right two panels of Fig.~\ref{Fig:noSOC_multiQ}(d). 
The intensities become nonzero at multiples of $\bm{Q}_1$ and $\bm{Q}_2$, i.e., $m_1 \bm{Q}_1+m_2 \bm{Q}_2$ where $m_1$ and $m_2$ are integers, all of which are accounted for by the perturbative formula in Eq.~(\ref{eq:perturbation_nq}). 

We note that a similar CDW modulation is expected to appear in the collinear 2$Q$ bubble crystal without the $xy$ spin components in Eq.~(\ref{eq:2QSkX}), which might be realized in CeAuSb$_2$~\cite{Marcus_PhysRevLett.120.097201,Seo_PhysRevX.10.011035,seo2021spin}.

\subsection{Relation to bond modulation}
\label{sec:Relation to other quantities}

\begin{figure}[htb!]
\begin{center}
\includegraphics[width=1.0 \hsize]{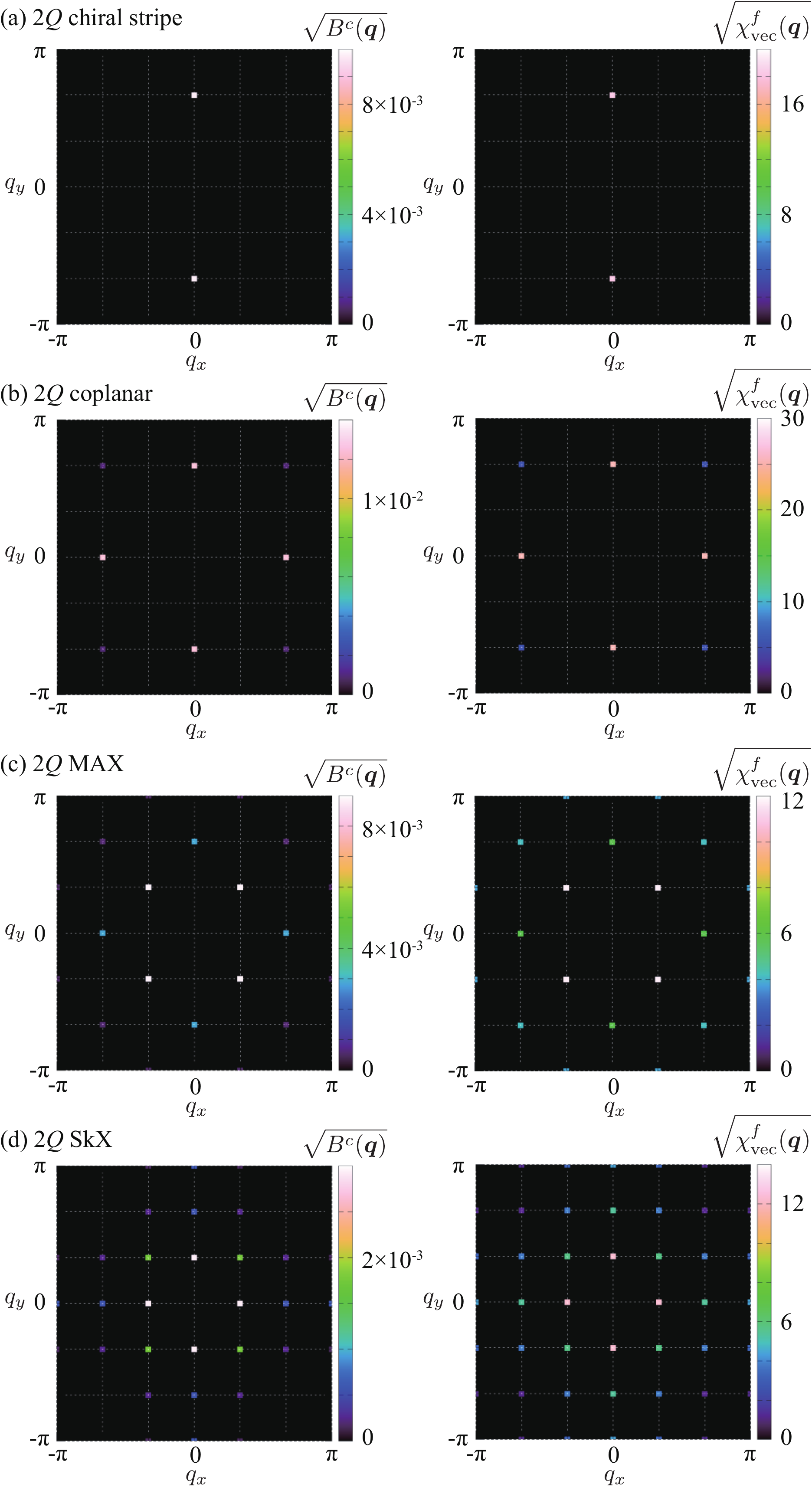} 
\caption{
\label{Fig:Other} 
The square root of the structure factor for the kinetic bond energy of the itinerant electrons (left) and the vector chirality of the localized spins (right) for (a) the 2$Q$ chiral stripe state in Fig.~\ref{Fig:noSOC_multiQ}(a), (b) the 2$Q$ coplanar state in Fig.~\ref{Fig:noSOC_multiQ}(b), (c) the 2$Q$ MAX in Fig.~\ref{Fig:noSOC_multiQ}(c), and (d) the 2$Q$ SkX in Fig.~\ref{Fig:noSOC_multiQ}(d). 
}
\end{center}
\end{figure}

As discussed for the $1Q$ case in Sec.~\ref{sec:Without a net magnetization}, the CDW formation in the $2Q$ cases can be understood from the modulation of the effective hopping amplitude on the underlying spin texture. 
This is demonstrated in the left panels of Fig.~\ref{Fig:Other} for (a) the 2$Q$ chiral stripe state in Eq.~(\ref{eq:2QCS}), (b) the 2$Q$ coplanar state in Eq.~(\ref{eq:2Qcoplanar}), (c) the 2$Q$ MAX in Eq.~(\ref{eq:2QMAX}), and (d) the 2$Q$ SkX in Eq.~(\ref{eq:2QSkX}). 
Here, we plot the structure factor for the kinetic bond energy defined by
\begin{align}
B^c(\bm{q})=\frac{1}{N} \sum_{ij\delta} K_{i\delta} \cdot  K_{j\delta} e^{i \bm{q}\cdot (\bm{r}_i-\bm{r}_j)}, 
\end{align}
where $K_{i\delta}$ is the local kinetic energy between site $i$ and $i+\delta$ as
\begin{align}
K_{i\delta}=\sum_{\sigma}\langle c^\dagger_{i\sigma}c_{i+\delta\sigma} + {\rm H.c.} \rangle.
\end{align}
Here, $\delta=\hat{x}$ and $\hat{y}$ denote the shift with lattice constant along the $x$ and $y$ directions, respectively.
We find that $B^c(\bm{q})$ shows the peaks at the same positions as the charge structure factor $N^c (\bm{q})$ in the right panels of Figs.~\ref{Fig:noSOC_multiQ}(a)-\ref{Fig:noSOC_multiQ}(d). 
This is because the hopping of itinerant electrons is modulated depending on the relative angle of the localized spins, as discussed in Sec.~\ref{sec:Without a net magnetization}.

While the above quantity is related to the inner product of the localized spins $\bm{S}_i\cdot \bm{S}_j$, we find that the outer product $\bm{S}_i \times \bm{S}_j$, which is called the vector spin chirality, also has a correlation with the CDW. 
In the right panels of Fig.~\ref{Fig:Other}, we plot the structure factor for the vector spin chirality defined by
\begin{align}
\chi_{\rm vec}^f(\bm{q})=\frac{1}{N} \sum_{ij\delta} \bm{\chi}_{i\delta} \cdot  \bm{\chi}_{j\delta} e^{i \bm{q}\cdot (\bm{r}_i-\bm{r}_j)}, 
\end{align}
where $\bm{\chi}_{i\delta}=\bm{S}_i \times \bm{S}_{i+\delta}$. 
The result shows that $\chi_{\rm vec}^f(\bm{q})$ exhibits the peaks at the same positions as $N^c (\bm{q})$ as well as $B^c(\bm{q})$. 
The origin of the correspondence is not clear but it might be attributed to the relation between the vector spin chirality and the local electronic polarization as $p_{ij} \propto \hat{\bm{r}}_{ij} \times (\bm{S}_i \times \bm{S}_j)$, where $\hat{\bm{r}}_{ij}$ is the unit vector from the site $i$ to $j$~\cite{Katsura_PhysRevLett.95.057205,Mostovoy_PhysRevLett.96.067601,tokura2014multiferroics}. 
Although this relation holds for the insulating systems, we speculate that the CDW in our metallic system is also affected by the vector spin chirality through a similar relationship between spin and charge. 
We note, however, that the correspondence does not hold for the collinear magnetic orderings.

\subsection{Parameter dependence}
\label{sec:Parameter dependence}

\begin{figure}[htb!]
\begin{center}
\includegraphics[width=0.95 \hsize]{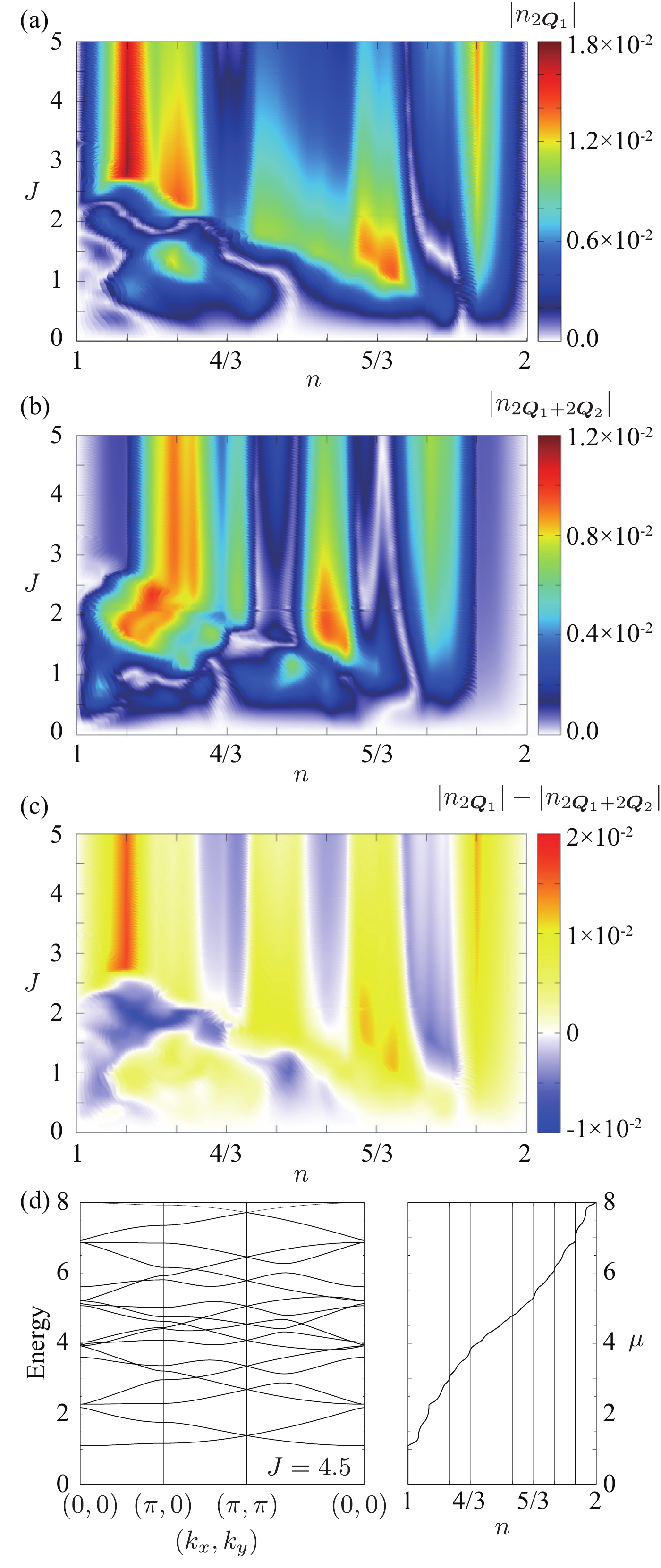} 
\caption{
\label{Fig:Jmu_dep}
Contour plot of (a) $|n_{2\bm{Q}_1}|$, (b) $|n_{2\bm{Q}_1+2\bm{Q}_2}|$, and (c) $|n_{2\bm{Q}_1}|-|n_{2\bm{Q}_1+2\bm{Q}_2}|$ in the $n$-$J$ plane in the case of the 2$Q$ coplanar state. 
(d) (left) Energy dispersion of the 2$Q$ coplanar state at $J=4.5$. 
(right) Chemical potential $\mu$ as a function of $n$ corresponding to the left panel. 
}
\end{center}
\end{figure}

In the previous sections, we confirmed that the perturbative formula in Eq.~(\ref{eq:perturbation_nq}) well explains the CDW formation in the model in Eq.~(\ref{eq:Ham_KLM}) for weak $J$. 
In this section, we examine how the CDW evolves while increasing $J$ beyond the perturbative regime. 
We take the 2$Q$ coplanar state in Eq.~(\ref{eq:2Qcoplanar}) as an example. 
Figure~\ref{Fig:Jmu_dep}(a) displays the intensity of $|n_{2\bm{Q}_1}|$ while varying 
$J$ and the electron filling $n$. 
Again, the results for less than half filling are obtained by using the particle-hole symmetry. 
In the weak $J$ region, $|n_{2\bm{Q}_1}|$ increases while increasing $J$ owing to the factor of $J^2$ in Eq.~(\ref{eq:perturbation_nq}) [see also Fig.~\ref{Fig:perturbation}(b)]. 
Meanwhile, the behavior of $|n_{2\bm{Q}_1}|$ deviates largely from Eq.~(\ref{eq:perturbation_nq}) for $J \gtrsim 0.5$ and depends on $J$ and $n$ in a complicated manner, as shown in Fig.~\ref{Fig:Jmu_dep}(a). 
We find that the value of $|n_{2\bm{Q}_1}|$ tends to be enhanced around some commensurate fillings to a multiple of 1/9, e.g., $n=10/9$, $11/9$, and $17/9$ in the large $J$ region. 
This is attributed to the tendency of the gap opening in the band structure of the itinerant electrons at the commensurate fillings. 
We demonstrate this for $J=4.5$ in Fig.~\ref{Fig:Jmu_dep}(d), where the system opens an energy gap at $n=10/9$ and $17/9$, and remains gapless but semimetallic at $n=11/9$.  
Thus, the strong spin-charge coupling provides the possibility to enhance the CDW modulation through the significant modification of the band structure. 

Besides, we find that, depending on $J$ and $n$, the dominant CDW modulation appears in a different wave vector from the prediction by the perturbative formula in Eq.~(\ref{eq:perturbation_nq}). 
This is demonstrated in Fig.~\ref{Fig:Jmu_dep}(b) which plots $|n_{2\bm{Q}_1+2\bm{Q}_2}|$. 
In the weak $J$ region, $|n_{2\bm{Q}_1+2\bm{Q}_2}|$ is smaller than $|n_{2\bm{Q}_1}|$ owing to the factor of $\bm{S}_{\bm{q}_1}\cdot \bm{S}_{\bm{q}_2}$ in $n_{\bm{q}}$; 
the dominant contribution of $|n_{2\bm{Q}_1+2\bm{Q}_2}|$ is $\bm{S}_{\bm{Q}_1}\cdot \bm{S}_{\bm{Q}_1+2\bm{Q}_2}$, which is smaller than that of $|n_{2\bm{Q}_1}|$, $\bm{S}_{\bm{Q}_1}\cdot \bm{S}_{\bm{Q}_1}$, since $\bm{S}_{\bm{Q}_1} > \bm{S}_{\bm{Q}_1+2\bm{Q}_2}$ [see the middle left panel of Fig.~\ref{Fig:noSOC_multiQ}(b)]. 
Meanwhile, there are regions for $|n_{2\bm{Q}_1+2\bm{Q}_2}|>|n_{2\bm{Q}_1}|$ beyond the perturbation regime, which is clearly shown in the plot of the difference $|n_{2\bm{Q}_1}|-|n_{2\bm{Q}_1+2\bm{Q}_2}|$ in Fig.~\ref{Fig:Jmu_dep}(c). 
The results indicate that the strong spin-charge coupling leads to not only the enhancement but also the modulation of CDW.

\subsection{Effect of spin-orbit coupling}
\label{sec:Effect of spin-orbit coupling}

Thus far, we have examined the CDW induced by the SDW via the isotropic interaction in spin space like $\bm{S}_{\bm{q}_1} \cdot \bm{S}_{\bm{q}_2}$ in Eq.~(\ref{eq:perturbation_nq}). 
On the other hand, an anisotropic interaction can arise from the cooperation between the spin-orbit coupling and the crystalline electric field. 
In this section, we consider the effect of such anisotropic interactions on the CDW modulation. 

We introduce an antisymmetric spin-orbit interaction by supposing mirror symmetry breaking with respect to the two-dimensional plane of the square lattice and inversion symmetry breaking, whose contribution is given by 
\begin{align}
\label{eq:ASOC}
\mathcal{H}^{\rm ASOC}= \sum_{\bm{k},\sigma,\sigma'} \bm{g}_{\bm{k}} \cdot c^{\dagger}_{\bm{k}\sigma}\bm{\sigma}_{\sigma\sigma'}c_{\bm{k}\sigma'},
\end{align}
where $\bm{g}_{\bm{k}}$ is an antisymmetric vector with respect to $\bm{k}$: 
\begin{align}
\bm{g}_{\bm{k}}=(g^x_{\bm{k}},g^y_{\bm{k}},g^z_{\bm{k}})= \tilde{\alpha} (\sin k_y, -\sin k_x, 0)=-\bm{g}_{-\bm{k}}.
\end{align}  
The antisymmetric spin-orbit interaction in Eq.~(\ref{eq:ASOC}) corresponds to the Rashba-type antisymmetric spin-orbit interaction~\cite{rashba1960properties,bychkov1984oscillatory}. 
By taking into account $\mathcal{H}^{\rm ASOC}$ in addition to $\mathcal{H}$ in Eq.~(\ref{eq:Ham_kspace}), the spin space in the total Hamiltonian becomes anisotropic, which results in effective long-range anisotropic magnetic interactions by tracing out the itinerant electron degree of freedom~\cite{Hayami_PhysRevLett.121.137202,lounis2020multiple}. 
There are mainly two types of effective anisotropic interactions in the lowest order in terms of $J$: One is the antisymmetric interaction in the form of $\bm{S}_{\bm{q}} \times \bm{S}_{-\bm{q}}$ and the other is the symmetric anisotropic interaction in the form of $S^x_{\bm{q}}S^x_{-\bm{q}}$ and $S^x_{\bm{q}}S^y_{-\bm{q}}$~\cite{shibuya2016magnetic,Hayami_PhysRevLett.121.137202,yambe2021skyrmion}. 
Note that the Hamiltonian $\mathcal{H}+\mathcal{H}^{\rm ASOC}$ including the other type of spin-orbit coupling was shown to exhibit a variety of multiple-$Q$ magnetic states~\cite{Okada_PhysRevB.98.224406}.

In a similar procedure to that in Sec.~\ref{sec:Perturbative Analysis}, we derive the lowest order contribution to the charge modulation in terms of $J$ for the Hamiltonian $\mathcal{H}+\mathcal{H}^{\rm ASOC}$, which is given by 
\begin{align}
\label{eq:perturbation_nq_SOC}
n_{\bm{q}}=&\frac{J^2}{4N^2}T\sum_{\bm{k},\bm{q}_1,\bm{q}_2}\sum_{\sigma,\sigma',\sigma''}\sum_{\alpha\beta}\sum_{\omega_n}G^0_{\bm{k}\sigma}G^0_{\bm{k}+\bm{q}_1 \sigma'}G^0_{\bm{k}+\bm{q}_1+\bm{q}_2\sigma''} \nonumber \\
&\times \Pi^{\alpha \beta}_{(\bm{k}\bm{q}_1\bm{q}_2)(\sigma\sigma'\sigma'')} S^\alpha_{\bm{q}_1} S^\beta_{\bm{q}_2} \delta_{\bm{q}_1+\bm{q}_2,\bm{q}+l\bm{G}}, 
\end{align}
where $\Pi^{\alpha \beta}_{(\bm{k}\bm{q}_1\bm{q}_2)(\sigma\sigma'\sigma'')}$ represents the form factor, and Green's functions depend on the spin index $\sigma$ reflecting the spin anisotropy. 
For example, $\Pi^{\alpha \beta}_{(\bm{k}\bm{q}_1\bm{q}_2)(\uparrow\uparrow\uparrow)}$ is explicitly given by 
\begin{widetext}
\begin{align}
\label{eq:Pixx}
\Pi^{xx}_{(\bm{k}\bm{q}_1\bm{q}_2)(\uparrow\uparrow\uparrow)}&=
1+
\tilde{g}^x_{\bm{k}}\tilde{g}^x_{\bm{k}+\bm{q}_1}+\tilde{g}^x_{\bm{k}+\bm{q}_1}\tilde{g}^x_{\bm{k}+\bm{q}_1+\bm{q}_2}+\tilde{g}^x_{\bm{k}+\bm{q}_1+\bm{q}_2}\tilde{g}^x_{\bm{k}}
-
\tilde{g}^y_{\bm{k}}\tilde{g}^y_{\bm{k}+\bm{q}_1}-\tilde{g}^y_{\bm{k}+\bm{q}_1}\tilde{g}^y_{\bm{k}+\bm{q}_1+\bm{q}_2}+\tilde{g}^y_{\bm{k}+\bm{q}_1+\bm{q}_2}\tilde{g}^y_{\bm{k}}, \\
\label{eq:Piyy}
\Pi^{yy}_{(\bm{k}\bm{q}_1\bm{q}_2)(\uparrow\uparrow\uparrow)}&=
1-
\tilde{g}^x_{\bm{k}}\tilde{g}^x_{\bm{k}+\bm{q}_1}-\tilde{g}^x_{\bm{k}+\bm{q}_1}\tilde{g}^x_{\bm{k}+\bm{q}_1+\bm{q}_2}+\tilde{g}^x_{\bm{k}+\bm{q}_1+\bm{q}_2}\tilde{g}^x_{\bm{k}}
+
\tilde{g}^y_{\bm{k}}\tilde{g}^y_{\bm{k}+\bm{q}_1}+\tilde{g}^y_{\bm{k}+\bm{q}_1}\tilde{g}^y_{\bm{k}+\bm{q}_1+\bm{q}_2}+\tilde{g}^y_{\bm{k}+\bm{q}_1+\bm{q}_2}\tilde{g}^y_{\bm{k}}, \\
\label{eq:Pizz}
\Pi^{zz}_{(\bm{k}\bm{q}_1\bm{q}_2)(\uparrow\uparrow\uparrow)}&=
1-
\tilde{g}^x_{\bm{k}}\tilde{g}^x_{\bm{k}+\bm{q}_1}-\tilde{g}^x_{\bm{k}+\bm{q}_1}\tilde{g}^x_{\bm{k}+\bm{q}_1+\bm{q}_2}+\tilde{g}^x_{\bm{k}+\bm{q}_1+\bm{q}_2}\tilde{g}^x_{\bm{k}}
-
\tilde{g}^y_{\bm{k}}\tilde{g}^y_{\bm{k}+\bm{q}_1}-\tilde{g}^y_{\bm{k}+\bm{q}_1}\tilde{g}^y_{\bm{k}+\bm{q}_1+\bm{q}_2}+\tilde{g}^y_{\bm{k}+\bm{q}_1+\bm{q}_2}\tilde{g}^y_{\bm{k}}, \\
\label{eq:Pixy}
\Pi^{xy}_{(\bm{k}\bm{q}_1\bm{q}_2)(\uparrow\uparrow\uparrow)}&=
\tilde{g}^x_{\bm{k}}\tilde{g}^y_{\bm{k}+\bm{q}_1}+\tilde{g}^x_{\bm{k}+\bm{q}_1}\tilde{g}^y_{\bm{k}+\bm{q}_1+\bm{q}_2}+\tilde{g}^x_{\bm{k}+\bm{q}_1+\bm{q}_2}\tilde{g}^y_{\bm{k}}
+
\tilde{g}^y_{\bm{k}}\tilde{g}^x_{\bm{k}+\bm{q}_1}+\tilde{g}^y_{\bm{k}+\bm{q}_1}\tilde{g}^x_{\bm{k}+\bm{q}_1+\bm{q}_2}-\tilde{g}^y_{\bm{k}+\bm{q}_1+\bm{q}_2}\tilde{g}^x_{\bm{k}}, \\
\label{eq:Piyx}
\Pi^{yx}_{(\bm{k}\bm{q}_1\bm{q}_2)(\uparrow\uparrow\uparrow)}&=
\tilde{g}^x_{\bm{k}}\tilde{g}^y_{\bm{k}+\bm{q}_1}+\tilde{g}^x_{\bm{k}+\bm{q}_1}\tilde{g}^y_{\bm{k}+\bm{q}_1+\bm{q}_2}-\tilde{g}^x_{\bm{k}+\bm{q}_1+\bm{q}_2}\tilde{g}^y_{\bm{k}}
+
\tilde{g}^y_{\bm{k}}\tilde{g}^x_{\bm{k}+\bm{q}_1}+\tilde{g}^y_{\bm{k}+\bm{q}_1}\tilde{g}^x_{\bm{k}+\bm{q}_1+\bm{q}_2}+\tilde{g}^y_{\bm{k}+\bm{q}_1+\bm{q}_2}\tilde{g}^x_{\bm{k}}
, \\
\label{eq:Piyz}
\Pi^{yz}_{(\bm{k}\bm{q}_1\bm{q}_2)(\uparrow\uparrow\uparrow)}& 
=-i(
\tilde{g}^x_{\bm{k}}-\tilde{g}^x_{\bm{k}+\bm{q}_1}+\tilde{g}^x_{\bm{k}+\bm{q}_1+\bm{q}_2}\nonumber\\
&\qquad\quad-
\tilde{g}^x_{\bm{k}}\tilde{g}^x_{\bm{k}+\bm{q}_1}\tilde{g}^x_{\bm{k}+\bm{q}_1+\bm{q}_2}
-
\tilde{g}^x_{\bm{k}}\tilde{g}^y_{\bm{k}+\bm{q}_1}\tilde{g}^y_{\bm{k}+\bm{q}_1+\bm{q}_2}
-
\tilde{g}^y_{\bm{k}}\tilde{g}^x_{\bm{k}+\bm{q}_1}\tilde{g}^y_{\bm{k}+\bm{q}_1+\bm{q}_2}
+
\tilde{g}^y_{\bm{k}}\tilde{g}^y_{\bm{k}+\bm{q}_1}\tilde{g}^x_{\bm{k}+\bm{q}_1+\bm{q}_2}
), \\
\label{eq:Pizy}
\Pi^{zy}_{(\bm{k}\bm{q}_1\bm{q}_2)(\uparrow\uparrow\uparrow)}& 
=
i(
\tilde{g}^x_{\bm{k}}-\tilde{g}^x_{\bm{k}+\bm{q}_1}+\tilde{g}^x_{\bm{k}+\bm{q}_1+\bm{q}_2}\nonumber\\
&\qquad-
\tilde{g}^x_{\bm{k}}\tilde{g}^x_{\bm{k}+\bm{q}_1}\tilde{g}^x_{\bm{k}+\bm{q}_1+\bm{q}_2}
+
\tilde{g}^x_{\bm{k}}\tilde{g}^y_{\bm{k}+\bm{q}_1}\tilde{g}^y_{\bm{k}+\bm{q}_1+\bm{q}_2}
-
\tilde{g}^y_{\bm{k}}\tilde{g}^x_{\bm{k}+\bm{q}_1}\tilde{g}^y_{\bm{k}+\bm{q}_1+\bm{q}_2}
-
\tilde{g}^y_{\bm{k}}\tilde{g}^y_{\bm{k}+\bm{q}_1}\tilde{g}^x_{\bm{k}+\bm{q}_1+\bm{q}_2}
), \\
\label{eq:Pixz}
\Pi^{xz}_{(\bm{k}\bm{q}_1\bm{q}_2)(\uparrow\uparrow\uparrow)}& 
=
i(
\tilde{g}^y_{\bm{k}}-\tilde{g}^y_{\bm{k}+\bm{q}_1}+\tilde{g}^y_{\bm{k}+\bm{q}_1+\bm{q}_2}\nonumber\\
&\qquad-
\tilde{g}^y_{\bm{k}}\tilde{g}^y_{\bm{k}+\bm{q}_1}\tilde{g}^y_{\bm{k}+\bm{q}_1+\bm{q}_2}
-
\tilde{g}^y_{\bm{k}}\tilde{g}^x_{\bm{k}+\bm{q}_1}\tilde{g}^x_{\bm{k}+\bm{q}_1+\bm{q}_2}
-
\tilde{g}^x_{\bm{k}}\tilde{g}^y_{\bm{k}+\bm{q}_1}\tilde{g}^x_{\bm{k}+\bm{q}_1+\bm{q}_2}
+
\tilde{g}^x_{\bm{k}}\tilde{g}^x_{\bm{k}+\bm{q}_1}\tilde{g}^y_{\bm{k}+\bm{q}_1+\bm{q}_2}
), \\
\label{eq:Pizx}
\Pi^{zx}_{(\bm{k}\bm{q}_1\bm{q}_2)(\uparrow\uparrow\uparrow)}& 
=
-i(
\tilde{g}^y_{\bm{k}}-\tilde{g}^y_{\bm{k}+\bm{q}_1}+\tilde{g}^y_{\bm{k}+\bm{q}_1+\bm{q}_2}\nonumber\\
&\qquad\quad-
\tilde{g}^y_{\bm{k}}\tilde{g}^y_{\bm{k}+\bm{q}_1}\tilde{g}^y_{\bm{k}+\bm{q}_1+\bm{q}_2}
+
\tilde{g}^y_{\bm{k}}\tilde{g}^x_{\bm{k}+\bm{q}_1}\tilde{g}^x_{\bm{k}+\bm{q}_1+\bm{q}_2}
-
\tilde{g}^x_{\bm{k}}\tilde{g}^y_{\bm{k}+\bm{q}_1}\tilde{g}^x_{\bm{k}+\bm{q}_1+\bm{q}_2}
-
\tilde{g}^x_{\bm{k}}\tilde{g}^x_{\bm{k}+\bm{q}_1}\tilde{g}^y_{\bm{k}+\bm{q}_1+\bm{q}_2}
), 
\end{align}
\end{widetext}
where $\tilde{g}^\alpha_{\bm{k}}=g^{\alpha}_{\bm{k}}/|\bm{g}_{\bm{k}}|$ for $\alpha=x$ and 
$y$.
The diagonal components of $\Pi^{\alpha \beta}_{(\bm{k}\bm{q}_1\bm{q}_2)(\uparrow\uparrow\uparrow)}$ in Eqs.~(\ref{eq:Pixx})-(\ref{eq:Piyx}) represent the contributions from the symmetric spin interaction including even power of the spin-orbit coupling, while the off-diagonal ones in Eqs.~(\ref{eq:Piyz})-(\ref{eq:Pizx}) represent the contributions from the antisymmetric spin interaction including odd power of the spin-orbit coupling. 
One can obtain the expression of $\Pi^{\alpha \beta}_{(\bm{k}\bm{q}_1\bm{q}_2)(\sigma\sigma'\sigma'')}$ for the other spin components in a similar manner. 

As exemplified below, an important observation in the presence of the antisymmetric spin-orbit coupling in Eq.~(\ref{eq:perturbation_nq_SOC}) is that the CDW modulation depends on not only the magnetic texture but also the form of the effective spin interaction. 
In other words, additional CDW modulations can appear by taking into account the antisymmetric spin-orbit coupling even for the same magnetic structure. 
This implies that one can deduce the relevant spin-orbit coupling once the patterns of the CDW and the SDW are identified in experiments. 
In the following, we demonstrate additional CDW modulations for the 2$Q$ chiral stripe state in Sec.~\ref{sec:2$Q$ chiral stripeSOC} and the 2$Q$ coplanar state in Sec.~\ref{sec:2$Q$ coplanarSOC}. 

\subsubsection{2$Q$ chiral stripe}
\label{sec:2$Q$ chiral stripeSOC}

\begin{figure}[htb!]
\begin{center}
\includegraphics[width=1.0 \hsize]{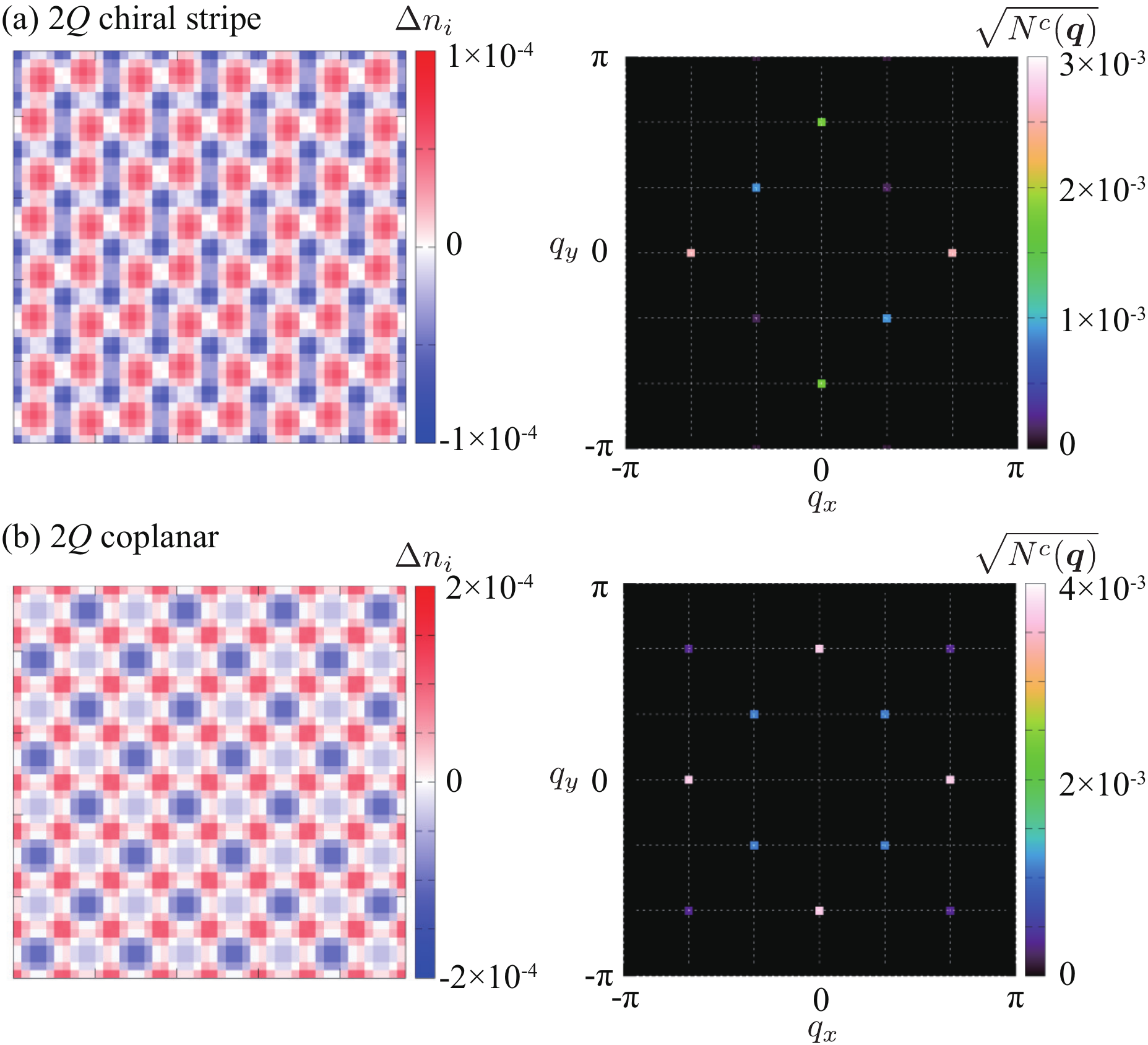} 
\caption{
\label{Fig:SOC}
Left: Real-space charge distributions measured from the average charge density in (a) the 2$Q$ chiral stripe state in Eq.~(\ref{eq:2QCS}) and (b) the 2$Q$ coplanar state in Eq.~(\ref{eq:2Qcoplanar}) at $J=0.05$, $\tilde{\alpha}=0.5$, and $\mu=3$ in the presence of the Rashba-type antisymmetric spin-orbit coupling in Eq.~(\ref{eq:ASOC}). 
Right: The square root of the charge structure factor. 
}
\end{center}
\end{figure}

Figure~\ref{Fig:SOC}(a) shows the real-space charge distribution (left panel) and the charge structure factor (right panel) in the 2$Q$ chiral stripe state, which is obtained by the direct diagonalization of $\mathcal{H}+\mathcal{H}^{\rm ASOC}$ at $J=0.05$, $\tilde{\alpha}=0.5$, and $\mu=3$. 
We consider the same spin configuration in Eq.~(\ref{eq:2QCS}) in order to show the effect of the spin-orbit coupling. 
Compared to the result in Fig.~\ref{Fig:noSOC_multiQ}(a), the local charge density shows a modulation along the $x$ direction in addition to the $y$ direction, as shown in the left panel of Fig.~\ref{Fig:SOC}(a). 
By performing the Fourier transformation, there are additional three peaks in the charge structure factor at $\pm 2\bm{Q}_1$, $\pm (\bm{Q}_1+\bm{Q}_2)$, and $\pm (\bm{Q}_1-\bm{Q}_2)$ in addition to $\pm 2\bm{Q}_2$, as shown in the right panel of Fig.~\ref{Fig:SOC}(a).  
The appearance of $n_{2\bm{Q}_1}$ is owing to $\Pi^{yy}_{(\bm{k}\bm{Q}_1\bm{Q}_1)(\sigma\sigma'\sigma'')} S^y_{\bm{Q}_1}S^y_{\bm{Q}_1} \neq \Pi^{zz}_{(\bm{k}\bm{Q}_1\bm{Q}_1)(\sigma\sigma'\sigma'')} S^z_{\bm{Q}_1}S^z_{\bm{Q}_1}$ [for instance, see Eqs.~(\ref{eq:Piyy}) and (\ref{eq:Pizz})], while that of 
$n_{\bm{Q}_1 + \bm{Q}_2}$ is owing to nonzero $\Pi^{yx}_{(\bm{k}\bm{Q}_1\bm{Q}_2)(\sigma\sigma'\sigma'')} S^y_{\bm{Q}_1}S^x_{\bm{Q}_2}$, $\Pi^{zx}_{(\bm{k}\bm{Q}_1\bm{Q}_2)(\sigma\sigma'\sigma'')} S^z_{\bm{Q}_1}S^x_{\bm{Q}_2}$ [for instance, see Eqs.~(\ref{eq:Piyx}) and (\ref{eq:Pizx})].

\subsubsection{2$Q$ coplanar}
\label{sec:2$Q$ coplanarSOC}

Figure~\ref{Fig:SOC}(b) shows the CDW in real and momentum spaces in the 2$Q$ coplanar state with the same model parameters as in Sec.~\ref{sec:2$Q$ chiral stripeSOC}. 
The spin configuration is given by Eq.~(\ref{eq:2Qcoplanar}).
Although the local charge density in the left panel of Fig.~\ref{Fig:SOC}(b) looks similar to that in Fig.~\ref{Fig:noSOC_multiQ}(b), additional components at $\bm{q}=\pm\bm{Q}_1 \pm \bm{Q}_2$ are found in the charge structure factor by introducing the antisymmetric spin-orbit coupling, as shown in the right panel of Fig.~\ref{Fig:SOC}(b). 
This additional modulation originates from nonzero $\Pi^{yx}_{(\bm{k}\bm{Q}_1\bm{Q}_2)(\sigma\sigma'\sigma'')} S^y_{\bm{Q}_1}S^x_{\bm{Q}_2}$ [for instance, see Eq.~(\ref{eq:Piyx})].

\section{Summary}
\label{sec:Summary}
To summarize, we have investigated the CDW induced by the SDW in itinerant magnets. 
We analyzed the Kondo lattice model on the basis of the analytical perturbative calculations and the numerical diagonalization. 
Our perturbative formula provides a clear correspondence between the charge and spin degrees of freedom, which is useful in identifying the CDW modulation from the SDW modulation and vice versa. 
We confirmed that the perturbative formula predicts correctly not only the wave numbers but also the amplitudes of the CDW semiquantitatively in the weak spin-charge coupling regime, for various single-$Q$ and double-$Q$ SDWs including the SkX and MAX. 
Moreover, we showed that the CDW may provide richer information than the SDW by showing that it can distinguish the MAX clearly from the coplanar state although the two states have similar spin structure factors. 
We also established the relation between the CDW and bond modulation in terms of the kinetic bond energy and the vector chirality. 
In addition, we found that the CDW is sensitively modulated by the spin-charge coupling and electron filling in the region beyond the perturbative regime; in particular, the CDW is enhanced at some commensurate electron fillings due to the tendency of gap opening in the electronic band structure. 
Furthermore, we examined the role of the spin-orbit coupling in the CDW modulation, which brings about additional CDW modulations through the effective long-range anisotropic interactions arising from the spin-orbit coupling. 

Our formula in Eq.~(\ref{eq:perturbation_nq}) is generic to any magnetic structures on any lattice structures. 
Indeed, it well explains the behavior of the CDW under the multiple-$Q$ spin textures, such as the SkX and the 2$Q$ coplanar state, in GdRu$_2$Si$_2$~\cite{Yasui2020imaging}.  
Since the concomitant CDW and SDW implies the coupling between the spin and charge degrees of freedom, the CDW observation in the materials with showing  multiple-$Q$ SDWs will indicate the importance of itinerant electrons~\cite{hayami2021topological}. 
The candidate materials are Y$_3$Co$_8$Sn$_4$ hosting the 3$Q$ vortex crystal~\cite{takagi2018multiple}, EuPtSi~\cite{kakihana2018giant,kaneko2019unique,kakihana2019unique,tabata2019magnetic,hayami2021field} and Gd$_3$Ru$_4$Al$_{12}$~\cite{Nakamura_PhysRevB.98.054410,hirschberger2019skyrmion,Hirschberger_10.1088/1367-2630/abdef9} hosting the 3$Q$ SkX, and MnSi$_{1-x}$Ge$_{x}$~\cite{tanigaki2015real,kanazawa2017noncentrosymmetric,fujishiro2019topological,Kanazawa_PhysRevLett.125.137202,Okumura_PhysRevB.101.144416,grytsiuk2020topological} and SrFeO$_3$~\cite{Ishiwata_PhysRevB.84.054427,Ishiwata_PhysRevB.101.134406,Rogge_PhysRevMaterials.3.084404,Onose_PhysRevMaterials.4.114420} hosting the $3Q$ and $4Q$ hedgehog crystals. 
Furthermore, our finding will provide a clue to understand complex charge and magnetic orderings recently found in materials, such as GdSb$_x$Te$_{2-x-\delta}$~\cite{Shiming_PhysRevB.103.134418} and EuAl$_4$~\cite{onuki2020unique,Shang_PhysRevB.103.L020405,kaneko2021charge}. 
Thus, the present results stimulate a further study of exotic spin-charge entanglement in itinerant magnets. 

It is noteworthy to mention a possibility of controlling the SDW by the CDW through their intimate relation in the spin-charge coupled systems. 
In the present Kondo lattice model, the CDW is always induced by the SDW via the effective spin-channel interactions arising from the spin-charge coupling rather than the charge-channel interaction, since the itinerant electrons have no bare Coulomb interactions. 
In this case, the optimal magnetic spin configuration is given by the effective magnetic interactions~\cite{Hayami_PhysRevB.95.224424}. 
Meanwhile, the charge-channel interaction may also contribute to the magnetic orderings, since the characteristic wave vectors of the CDW can be different in the presence of charge-charge interactions. 
Thus, an extension of the model by taking into account, e.g., Coulomb interactions and electron-phonon couplings might result in yet another stabilization mechanism of the multiple-$Q$ SDWs. 
Indeed, a periodic array of nonmagnetic impurities, which mimics a CDW, in insulating magnets gives rise to a plethora of multiple-$Q$ SDWs including the SkX~\cite{Hayami_PhysRevB.94.174420}. 
Such an interesting analysis is left for future study.

\begin{acknowledgments}
The authors thank T. Arima, T. Hanaguri, S. Seki, and Y. Yasui for fruitful discussions. 
This research was supported by JSPS KAKENHI Grants Numbers JP18K13488, JP19K03752, JP19H01834, JP19H05825, JP21H01037, and by JST PRESTO (JPMJPR20L8) and JST CREST (JP-MJCR18T2). 
Parts of the numerical calculations were performed in the supercomputing systems in ISSP, the University of Tokyo.
\end{acknowledgments}

\bibliography{ref.bib}

\begin{thebibliography}{114}%
\makeatletter
\providecommand \@ifxundefined [1]{%
 \@ifx{#1\undefined}
}%
\providecommand \@ifnum [1]{%
 \ifnum #1\expandafter \@firstoftwo
 \else \expandafter \@secondoftwo
 \fi
}%
\providecommand \@ifx [1]{%
 \ifx #1\expandafter \@firstoftwo
 \else \expandafter \@secondoftwo
 \fi
}%
\providecommand \natexlab [1]{#1}%
\providecommand \enquote  [1]{``#1''}%
\providecommand \bibnamefont  [1]{#1}%
\providecommand \bibfnamefont [1]{#1}%
\providecommand \citenamefont [1]{#1}%
\providecommand \href@noop [0]{\@secondoftwo}%
\providecommand \href [0]{\begingroup \@sanitize@url \@href}%
\providecommand \@href[1]{\@@startlink{#1}\@@href}%
\providecommand \@@href[1]{\endgroup#1\@@endlink}%
\providecommand \@sanitize@url [0]{\catcode `\\12\catcode `\$12\catcode
  `\&12\catcode `\#12\catcode `\^12\catcode `\_12\catcode `\%12\relax}%
\providecommand \@@startlink[1]{}%
\providecommand \@@endlink[0]{}%
\providecommand \url  [0]{\begingroup\@sanitize@url \@url }%
\providecommand \@url [1]{\endgroup\@href {#1}{\urlprefix }}%
\providecommand \urlprefix  [0]{URL }%
\providecommand \Eprint [0]{\href }%
\providecommand \doibase [0]{https://doi.org/}%
\providecommand \selectlanguage [0]{\@gobble}%
\providecommand \bibinfo  [0]{\@secondoftwo}%
\providecommand \bibfield  [0]{\@secondoftwo}%
\providecommand \translation [1]{[#1]}%
\providecommand \BibitemOpen [0]{}%
\providecommand \bibitemStop [0]{}%
\providecommand \bibitemNoStop [0]{.\EOS\space}%
\providecommand \EOS [0]{\spacefactor3000\relax}%
\providecommand \BibitemShut  [1]{\csname bibitem#1\endcsname}%
\let\auto@bib@innerbib\@empty
\bibitem [{\citenamefont {Stewart}(1984)}]{Stewart_RevModPhys.56.755}%
  \BibitemOpen
  \bibfield  {author} {\bibinfo {author} {\bibfnamefont {G.~R.}\ \bibnamefont
  {Stewart}},\ }\bibfield  {title} {\bibinfo {title} {Heavy-fermion systems},\
  }\href {https://doi.org/10.1103/RevModPhys.56.755} {\bibfield  {journal}
  {\bibinfo  {journal} {Rev. Mod. Phys.}\ }\textbf {\bibinfo {volume} {56}},\
  \bibinfo {pages} {755} (\bibinfo {year} {1984})}\BibitemShut {NoStop}%
\bibitem [{\citenamefont {Stewart}(2001)}]{Stewart_RevModPhys.73.797}%
  \BibitemOpen
  \bibfield  {author} {\bibinfo {author} {\bibfnamefont {G.~R.}\ \bibnamefont
  {Stewart}},\ }\bibfield  {title} {\bibinfo {title} {Non-fermi-liquid behavior
  in $d$- and $f$-electron metals},\ }\href
  {https://doi.org/10.1103/RevModPhys.73.797} {\bibfield  {journal} {\bibinfo
  {journal} {Rev. Mod. Phys.}\ }\textbf {\bibinfo {volume} {73}},\ \bibinfo
  {pages} {797} (\bibinfo {year} {2001})}\BibitemShut {NoStop}%
\bibitem [{\citenamefont {Ruderman}\ and\ \citenamefont
  {Kittel}(1954)}]{Ruderman}%
  \BibitemOpen
  \bibfield  {author} {\bibinfo {author} {\bibfnamefont {M.~A.}\ \bibnamefont
  {Ruderman}}\ and\ \bibinfo {author} {\bibfnamefont {C.}~\bibnamefont
  {Kittel}},\ }\bibfield  {title} {\bibinfo {title} {Indirect exchange coupling
  of nuclear magnetic moments by conduction electrons},\ }\href
  {https://doi.org/10.1103/PhysRev.96.99} {\bibfield  {journal} {\bibinfo
  {journal} {Phys. Rev.}\ }\textbf {\bibinfo {volume} {96}},\ \bibinfo {pages}
  {99} (\bibinfo {year} {1954})}\BibitemShut {NoStop}%
\bibitem [{\citenamefont {Kasuya}(1956)}]{Kasuya}%
  \BibitemOpen
  \bibfield  {author} {\bibinfo {author} {\bibfnamefont {T.}~\bibnamefont
  {Kasuya}},\ }\bibfield  {title} {\bibinfo {title} {A theory of metallic
  ferro- and antiferromagnetism on zener's model},\ }\href
  {https://doi.org/10.1143/PTP.16.45} {\bibfield  {journal} {\bibinfo
  {journal} {Prog. Theor. Phys.}\ }\textbf {\bibinfo {volume} {16}},\ \bibinfo
  {pages} {45} (\bibinfo {year} {1956})}\BibitemShut {NoStop}%
\bibitem [{\citenamefont {Yosida}(1957)}]{Yosida1957}%
  \BibitemOpen
  \bibfield  {author} {\bibinfo {author} {\bibfnamefont {K.}~\bibnamefont
  {Yosida}},\ }\bibfield  {title} {\bibinfo {title} {Magnetic properties of
  cu-mn alloys},\ }\href {https://doi.org/10.1103/PhysRev.106.893} {\bibfield
  {journal} {\bibinfo  {journal} {Phys. Rev.}\ }\textbf {\bibinfo {volume}
  {106}},\ \bibinfo {pages} {893} (\bibinfo {year} {1957})}\BibitemShut
  {NoStop}%
\bibitem [{\citenamefont {Shindou}\ and\ \citenamefont
  {Nagaosa}(2001)}]{Shindou_PhysRevLett.87.116801}%
  \BibitemOpen
  \bibfield  {author} {\bibinfo {author} {\bibfnamefont {R.}~\bibnamefont
  {Shindou}}\ and\ \bibinfo {author} {\bibfnamefont {N.}~\bibnamefont
  {Nagaosa}},\ }\bibfield  {title} {\bibinfo {title} {Orbital ferromagnetism
  and anomalous hall effect in antiferromagnets on the distorted fcc lattice},\
  }\href {https://doi.org/10.1103/PhysRevLett.87.116801} {\bibfield  {journal}
  {\bibinfo  {journal} {Phys. Rev. Lett.}\ }\textbf {\bibinfo {volume} {87}},\
  \bibinfo {pages} {116801} (\bibinfo {year} {2001})}\BibitemShut {NoStop}%
\bibitem [{\citenamefont {Martin}\ and\ \citenamefont
  {Batista}(2008)}]{Martin_PhysRevLett.101.156402}%
  \BibitemOpen
  \bibfield  {author} {\bibinfo {author} {\bibfnamefont {I.}~\bibnamefont
  {Martin}}\ and\ \bibinfo {author} {\bibfnamefont {C.~D.}\ \bibnamefont
  {Batista}},\ }\bibfield  {title} {\bibinfo {title} {Itinerant electron-driven
  chiral magnetic ordering and spontaneous quantum hall effect in triangular
  lattice models},\ }\href {https://doi.org/10.1103/PhysRevLett.101.156402}
  {\bibfield  {journal} {\bibinfo  {journal} {Phys. Rev. Lett.}\ }\textbf
  {\bibinfo {volume} {101}},\ \bibinfo {pages} {156402} (\bibinfo {year}
  {2008})}\BibitemShut {NoStop}%
\bibitem [{\citenamefont {Akagi}\ and\ \citenamefont
  {Motome}(2010)}]{Akagi_JPSJ.79.083711}%
  \BibitemOpen
  \bibfield  {author} {\bibinfo {author} {\bibfnamefont {Y.}~\bibnamefont
  {Akagi}}\ and\ \bibinfo {author} {\bibfnamefont {Y.}~\bibnamefont {Motome}},\
  }\bibfield  {title} {\bibinfo {title} {Spin chirality ordering and anomalous
  hall effect in the ferromagnetic kondo lattice model on a triangular
  lattice},\ }\href {https://doi.org/10.1143/JPSJ.79.083711} {\bibfield
  {journal} {\bibinfo  {journal} {J. Phys. Soc. Jpn.}\ }\textbf {\bibinfo
  {volume} {79}},\ \bibinfo {pages} {083711} (\bibinfo {year}
  {2010})}\BibitemShut {NoStop}%
\bibitem [{\citenamefont {Hayami}\ and\ \citenamefont
  {Motome}(2014)}]{Hayami_PhysRevB.90.060402}%
  \BibitemOpen
  \bibfield  {author} {\bibinfo {author} {\bibfnamefont {S.}~\bibnamefont
  {Hayami}}\ and\ \bibinfo {author} {\bibfnamefont {Y.}~\bibnamefont
  {Motome}},\ }\bibfield  {title} {\bibinfo {title} {Multiple-$q$ instability
  by $(d$-${}2)$-dimensional connections of fermi surfaces},\ }\href
  {https://doi.org/10.1103/PhysRevB.90.060402} {\bibfield  {journal} {\bibinfo
  {journal} {Phys. Rev. B}\ }\textbf {\bibinfo {volume} {90}},\ \bibinfo
  {pages} {060402(R)} (\bibinfo {year} {2014})}\BibitemShut {NoStop}%
\bibitem [{\citenamefont {Batista}\ \emph {et~al.}(2016)\citenamefont
  {Batista}, \citenamefont {Lin}, \citenamefont {Hayami},\ and\ \citenamefont
  {Kamiya}}]{batista2016frustration}%
  \BibitemOpen
  \bibfield  {author} {\bibinfo {author} {\bibfnamefont {C.~D.}\ \bibnamefont
  {Batista}}, \bibinfo {author} {\bibfnamefont {S.-Z.}\ \bibnamefont {Lin}},
  \bibinfo {author} {\bibfnamefont {S.}~\bibnamefont {Hayami}},\ and\ \bibinfo
  {author} {\bibfnamefont {Y.}~\bibnamefont {Kamiya}},\ }\bibfield  {title}
  {\bibinfo {title} {Frustration and chiral orderings in correlated electron
  systems},\ }\href@noop {} {\bibfield  {journal} {\bibinfo  {journal} {Rep.
  Prog. Phys.}\ }\textbf {\bibinfo {volume} {79}},\ \bibinfo {pages} {084504}
  (\bibinfo {year} {2016})}\BibitemShut {NoStop}%
\bibitem [{\citenamefont {Ozawa}\ \emph {et~al.}(2017)\citenamefont {Ozawa},
  \citenamefont {Hayami},\ and\ \citenamefont
  {Motome}}]{Ozawa_PhysRevLett.118.147205}%
  \BibitemOpen
  \bibfield  {author} {\bibinfo {author} {\bibfnamefont {R.}~\bibnamefont
  {Ozawa}}, \bibinfo {author} {\bibfnamefont {S.}~\bibnamefont {Hayami}},\ and\
  \bibinfo {author} {\bibfnamefont {Y.}~\bibnamefont {Motome}},\ }\bibfield
  {title} {\bibinfo {title} {Zero-field skyrmions with a high topological
  number in itinerant magnets},\ }\href
  {https://doi.org/10.1103/PhysRevLett.118.147205} {\bibfield  {journal}
  {\bibinfo  {journal} {Phys. Rev. Lett.}\ }\textbf {\bibinfo {volume} {118}},\
  \bibinfo {pages} {147205} (\bibinfo {year} {2017})}\BibitemShut {NoStop}%
\bibitem [{\citenamefont {Hayami}\ and\ \citenamefont
  {Motome}(2021{\natexlab{a}})}]{hayami2021topological}%
  \BibitemOpen
  \bibfield  {author} {\bibinfo {author} {\bibfnamefont {S.}~\bibnamefont
  {Hayami}}\ and\ \bibinfo {author} {\bibfnamefont {Y.}~\bibnamefont
  {Motome}},\ }\bibfield  {title} {\bibinfo {title} {Topological spin crystals
  by itinerant frustration},\ }\href {https://doi.org/10.1088/1361-648x/ac1a30}
  {\bibfield  {journal} {\bibinfo  {journal} {J. Phys.: Condens. Matter}\
  }\textbf {\bibinfo {volume} {33}},\ \bibinfo {pages} {443001} (\bibinfo
  {year} {2021}{\natexlab{a}})}\BibitemShut {NoStop}%
\bibitem [{\citenamefont {Baibich}\ \emph {et~al.}(1988)\citenamefont
  {Baibich}, \citenamefont {Broto}, \citenamefont {Fert}, \citenamefont
  {Van~Dau}, \citenamefont {Petroff}, \citenamefont {Etienne}, \citenamefont
  {Creuzet}, \citenamefont {Friederich},\ and\ \citenamefont
  {Chazelas}}]{baibich1988giant}%
  \BibitemOpen
  \bibfield  {author} {\bibinfo {author} {\bibfnamefont {M.~N.}\ \bibnamefont
  {Baibich}}, \bibinfo {author} {\bibfnamefont {J.}~\bibnamefont {Broto}},
  \bibinfo {author} {\bibfnamefont {A.}~\bibnamefont {Fert}}, \bibinfo {author}
  {\bibfnamefont {F.~N.}\ \bibnamefont {Van~Dau}}, \bibinfo {author}
  {\bibfnamefont {F.}~\bibnamefont {Petroff}}, \bibinfo {author} {\bibfnamefont
  {P.}~\bibnamefont {Etienne}}, \bibinfo {author} {\bibfnamefont
  {G.}~\bibnamefont {Creuzet}}, \bibinfo {author} {\bibfnamefont
  {A.}~\bibnamefont {Friederich}},\ and\ \bibinfo {author} {\bibfnamefont
  {J.}~\bibnamefont {Chazelas}},\ }\bibfield  {title} {\bibinfo {title} {Giant
  magnetoresistance of (001) fe/(001) cr magnetic superlattices},\ }\href@noop
  {} {\bibfield  {journal} {\bibinfo  {journal} {Phys. Rev. Lett.}\ }\textbf
  {\bibinfo {volume} {61}},\ \bibinfo {pages} {2472} (\bibinfo {year}
  {1988})}\BibitemShut {NoStop}%
\bibitem [{\citenamefont {Ramirez}(1997)}]{ramirez1997colossal}%
  \BibitemOpen
  \bibfield  {author} {\bibinfo {author} {\bibfnamefont {A.}~\bibnamefont
  {Ramirez}},\ }\bibfield  {title} {\bibinfo {title} {Colossal
  magnetoresistance},\ }\href@noop {} {\bibfield  {journal} {\bibinfo
  {journal} {J. Phys.: Condens. Matter}\ }\textbf {\bibinfo {volume} {9}},\
  \bibinfo {pages} {8171} (\bibinfo {year} {1997})}\BibitemShut {NoStop}%
\bibitem [{\citenamefont {Tokura}\ and\ \citenamefont
  {Tomioka}(1999)}]{tokura1999colossal}%
  \BibitemOpen
  \bibfield  {author} {\bibinfo {author} {\bibfnamefont {Y.}~\bibnamefont
  {Tokura}}\ and\ \bibinfo {author} {\bibfnamefont {Y.}~\bibnamefont
  {Tomioka}},\ }\bibfield  {title} {\bibinfo {title} {Colossal magnetoresistive
  manganites},\ }\href@noop {} {\bibfield  {journal} {\bibinfo  {journal} {J.
  Magn. Magn. Mat.}\ }\textbf {\bibinfo {volume} {200}},\ \bibinfo {pages} {1}
  (\bibinfo {year} {1999})}\BibitemShut {NoStop}%
\bibitem [{\citenamefont {Tokura}(1999)}]{Tokura200005}%
  \BibitemOpen
  \bibinfo {editor} {\bibfnamefont {Y.}~\bibnamefont {Tokura}},\ ed.,\
  \href@noop {} {\emph {\bibinfo {title} {Colossal Magnetoresistive Oxides}}}\
  (\bibinfo  {publisher} {Gordon \& Breach Science Pub.},\ \bibinfo {year}
  {1999})\BibitemShut {NoStop}%
\bibitem [{\citenamefont {Dagotto}\ \emph {et~al.}(2001)\citenamefont
  {Dagotto}, \citenamefont {Hotta},\ and\ \citenamefont
  {Moreo}}]{dagotto2001colossal}%
  \BibitemOpen
  \bibfield  {author} {\bibinfo {author} {\bibfnamefont {E.}~\bibnamefont
  {Dagotto}}, \bibinfo {author} {\bibfnamefont {T.}~\bibnamefont {Hotta}},\
  and\ \bibinfo {author} {\bibfnamefont {A.}~\bibnamefont {Moreo}},\ }\bibfield
   {title} {\bibinfo {title} {Colossal magnetoresistant materials: the key role
  of phase separation},\ }\href@noop {} {\bibfield  {journal} {\bibinfo
  {journal} {Phys. Rep.}\ }\textbf {\bibinfo {volume} {344}},\ \bibinfo {pages}
  {1} (\bibinfo {year} {2001})}\BibitemShut {NoStop}%
\bibitem [{\citenamefont {Ye}\ \emph {et~al.}(1999)\citenamefont {Ye},
  \citenamefont {Kim}, \citenamefont {Millis}, \citenamefont {Shraiman},
  \citenamefont {Majumdar},\ and\ \citenamefont {Te\ifmmode \check{s}\else
  \v{s}\fi{}anovi\ifmmode~\acute{c}\else \'{c}\fi{}}}]{Ye_PhysRevLett.83.3737}%
  \BibitemOpen
  \bibfield  {author} {\bibinfo {author} {\bibfnamefont {J.}~\bibnamefont
  {Ye}}, \bibinfo {author} {\bibfnamefont {Y.~B.}\ \bibnamefont {Kim}},
  \bibinfo {author} {\bibfnamefont {A.~J.}\ \bibnamefont {Millis}}, \bibinfo
  {author} {\bibfnamefont {B.~I.}\ \bibnamefont {Shraiman}}, \bibinfo {author}
  {\bibfnamefont {P.}~\bibnamefont {Majumdar}},\ and\ \bibinfo {author}
  {\bibfnamefont {Z.}~\bibnamefont {Te\ifmmode \check{s}\else
  \v{s}\fi{}anovi\ifmmode~\acute{c}\else \'{c}\fi{}}},\ }\bibfield  {title}
  {\bibinfo {title} {Berry phase theory of the anomalous hall effect:
  Application to colossal magnetoresistance manganites},\ }\href
  {https://doi.org/10.1103/PhysRevLett.83.3737} {\bibfield  {journal} {\bibinfo
   {journal} {Phys. Rev. Lett.}\ }\textbf {\bibinfo {volume} {83}},\ \bibinfo
  {pages} {3737} (\bibinfo {year} {1999})}\BibitemShut {NoStop}%
\bibitem [{\citenamefont {Ohgushi}\ \emph {et~al.}(2000)\citenamefont
  {Ohgushi}, \citenamefont {Murakami},\ and\ \citenamefont
  {Nagaosa}}]{Ohgushi_PhysRevB.62.R6065}%
  \BibitemOpen
  \bibfield  {author} {\bibinfo {author} {\bibfnamefont {K.}~\bibnamefont
  {Ohgushi}}, \bibinfo {author} {\bibfnamefont {S.}~\bibnamefont {Murakami}},\
  and\ \bibinfo {author} {\bibfnamefont {N.}~\bibnamefont {Nagaosa}},\
  }\bibfield  {title} {\bibinfo {title} {Spin anisotropy and quantum hall
  effect in the \textit{kagom\'e} lattice: Chiral spin state based on a
  ferromagnet},\ }\href {https://doi.org/10.1103/PhysRevB.62.R6065} {\bibfield
  {journal} {\bibinfo  {journal} {Phys. Rev. B}\ }\textbf {\bibinfo {volume}
  {62}},\ \bibinfo {pages} {R6065} (\bibinfo {year} {2000})}\BibitemShut
  {NoStop}%
\bibitem [{\citenamefont {Tatara}\ and\ \citenamefont
  {Kawamura}(2002)}]{tatara2002chirality}%
  \BibitemOpen
  \bibfield  {author} {\bibinfo {author} {\bibfnamefont {G.}~\bibnamefont
  {Tatara}}\ and\ \bibinfo {author} {\bibfnamefont {H.}~\bibnamefont
  {Kawamura}},\ }\bibfield  {title} {\bibinfo {title} {Chirality-driven
  anomalous hall effect in weak coupling regime},\ }\href@noop {} {\bibfield
  {journal} {\bibinfo  {journal} {J. Phys. Soc. Jpn.}\ }\textbf {\bibinfo
  {volume} {71}},\ \bibinfo {pages} {2613} (\bibinfo {year}
  {2002})}\BibitemShut {NoStop}%
\bibitem [{\citenamefont {Nagaosa}\ \emph {et~al.}(2010)\citenamefont
  {Nagaosa}, \citenamefont {Sinova}, \citenamefont {Onoda}, \citenamefont
  {MacDonald},\ and\ \citenamefont {Ong}}]{Nagaosa_RevModPhys.82.1539}%
  \BibitemOpen
  \bibfield  {author} {\bibinfo {author} {\bibfnamefont {N.}~\bibnamefont
  {Nagaosa}}, \bibinfo {author} {\bibfnamefont {J.}~\bibnamefont {Sinova}},
  \bibinfo {author} {\bibfnamefont {S.}~\bibnamefont {Onoda}}, \bibinfo
  {author} {\bibfnamefont {A.~H.}\ \bibnamefont {MacDonald}},\ and\ \bibinfo
  {author} {\bibfnamefont {N.~P.}\ \bibnamefont {Ong}},\ }\bibfield  {title}
  {\bibinfo {title} {Anomalous hall effect},\ }\href
  {https://doi.org/10.1103/RevModPhys.82.1539} {\bibfield  {journal} {\bibinfo
  {journal} {Rev. Mod. Phys.}\ }\textbf {\bibinfo {volume} {82}},\ \bibinfo
  {pages} {1539} (\bibinfo {year} {2010})}\BibitemShut {NoStop}%
\bibitem [{\citenamefont {Nagaosa}\ and\ \citenamefont
  {Tokura}(2013)}]{nagaosa2013topological}%
  \BibitemOpen
  \bibfield  {author} {\bibinfo {author} {\bibfnamefont {N.}~\bibnamefont
  {Nagaosa}}\ and\ \bibinfo {author} {\bibfnamefont {Y.}~\bibnamefont
  {Tokura}},\ }\bibfield  {title} {\bibinfo {title} {Topological properties and
  dynamics of magnetic skyrmions},\ }\href@noop {} {\bibfield  {journal}
  {\bibinfo  {journal} {Nat. Nanotech.}\ }\textbf {\bibinfo {volume} {8}},\
  \bibinfo {pages} {899} (\bibinfo {year} {2013})}\BibitemShut {NoStop}%
\bibitem [{\citenamefont {Lee}\ \emph {et~al.}(2009)\citenamefont {Lee},
  \citenamefont {Kang}, \citenamefont {Onose}, \citenamefont {Tokura},\ and\
  \citenamefont {Ong}}]{Lee_PhysRevLett.102.186601}%
  \BibitemOpen
  \bibfield  {author} {\bibinfo {author} {\bibfnamefont {M.}~\bibnamefont
  {Lee}}, \bibinfo {author} {\bibfnamefont {W.}~\bibnamefont {Kang}}, \bibinfo
  {author} {\bibfnamefont {Y.}~\bibnamefont {Onose}}, \bibinfo {author}
  {\bibfnamefont {Y.}~\bibnamefont {Tokura}},\ and\ \bibinfo {author}
  {\bibfnamefont {N.~P.}\ \bibnamefont {Ong}},\ }\bibfield  {title} {\bibinfo
  {title} {Unusual hall effect anomaly in mnsi under pressure},\ }\href
  {https://doi.org/10.1103/PhysRevLett.102.186601} {\bibfield  {journal}
  {\bibinfo  {journal} {Phys. Rev. Lett.}\ }\textbf {\bibinfo {volume} {102}},\
  \bibinfo {pages} {186601} (\bibinfo {year} {2009})}\BibitemShut {NoStop}%
\bibitem [{\citenamefont {Neubauer}\ \emph {et~al.}(2009)\citenamefont
  {Neubauer}, \citenamefont {Pfleiderer}, \citenamefont {Binz}, \citenamefont
  {Rosch}, \citenamefont {Ritz}, \citenamefont {Niklowitz},\ and\ \citenamefont
  {B\"oni}}]{Neubauer_PhysRevLett.102.186602}%
  \BibitemOpen
  \bibfield  {author} {\bibinfo {author} {\bibfnamefont {A.}~\bibnamefont
  {Neubauer}}, \bibinfo {author} {\bibfnamefont {C.}~\bibnamefont
  {Pfleiderer}}, \bibinfo {author} {\bibfnamefont {B.}~\bibnamefont {Binz}},
  \bibinfo {author} {\bibfnamefont {A.}~\bibnamefont {Rosch}}, \bibinfo
  {author} {\bibfnamefont {R.}~\bibnamefont {Ritz}}, \bibinfo {author}
  {\bibfnamefont {P.~G.}\ \bibnamefont {Niklowitz}},\ and\ \bibinfo {author}
  {\bibfnamefont {P.}~\bibnamefont {B\"oni}},\ }\bibfield  {title} {\bibinfo
  {title} {Topological hall effect in the $a$ phase of mnsi},\ }\href
  {https://doi.org/10.1103/PhysRevLett.102.186602} {\bibfield  {journal}
  {\bibinfo  {journal} {Phys. Rev. Lett.}\ }\textbf {\bibinfo {volume} {102}},\
  \bibinfo {pages} {186602} (\bibinfo {year} {2009})}\BibitemShut {NoStop}%
\bibitem [{\citenamefont {Hayami}\ \emph
  {et~al.}(2014{\natexlab{a}})\citenamefont {Hayami}, \citenamefont
  {Kusunose},\ and\ \citenamefont {Motome}}]{Hayami_PhysRevB.90.024432}%
  \BibitemOpen
  \bibfield  {author} {\bibinfo {author} {\bibfnamefont {S.}~\bibnamefont
  {Hayami}}, \bibinfo {author} {\bibfnamefont {H.}~\bibnamefont {Kusunose}},\
  and\ \bibinfo {author} {\bibfnamefont {Y.}~\bibnamefont {Motome}},\
  }\bibfield  {title} {\bibinfo {title} {Toroidal order in metals without local
  inversion symmetry},\ }\href {https://doi.org/10.1103/PhysRevB.90.024432}
  {\bibfield  {journal} {\bibinfo  {journal} {Phys. Rev. B}\ }\textbf {\bibinfo
  {volume} {90}},\ \bibinfo {pages} {024432} (\bibinfo {year}
  {2014}{\natexlab{a}})}\BibitemShut {NoStop}%
\bibitem [{\citenamefont {Th{\"o}le}\ and\ \citenamefont
  {Spaldin}(2018)}]{thole2018magnetoelectric}%
  \BibitemOpen
  \bibfield  {author} {\bibinfo {author} {\bibfnamefont {F.}~\bibnamefont
  {Th{\"o}le}}\ and\ \bibinfo {author} {\bibfnamefont {N.~A.}\ \bibnamefont
  {Spaldin}},\ }\bibfield  {title} {\bibinfo {title} {Magnetoelectric
  multipoles in metals},\ }\href@noop {} {\bibfield  {journal} {\bibinfo
  {journal} {Philos. Trans. R. Soc. A}\ }\textbf {\bibinfo {volume} {376}},\
  \bibinfo {pages} {20170450} (\bibinfo {year} {2018})}\BibitemShut {NoStop}%
\bibitem [{\citenamefont {Gao}\ \emph {et~al.}(2018)\citenamefont {Gao},
  \citenamefont {Vanderbilt},\ and\ \citenamefont
  {Xiao}}]{Gao_PhysRevB.97.134423}%
  \BibitemOpen
  \bibfield  {author} {\bibinfo {author} {\bibfnamefont {Y.}~\bibnamefont
  {Gao}}, \bibinfo {author} {\bibfnamefont {D.}~\bibnamefont {Vanderbilt}},\
  and\ \bibinfo {author} {\bibfnamefont {D.}~\bibnamefont {Xiao}},\ }\bibfield
  {title} {\bibinfo {title} {Microscopic theory of spin toroidization in
  periodic crystals},\ }\href {https://doi.org/10.1103/PhysRevB.97.134423}
  {\bibfield  {journal} {\bibinfo  {journal} {Phys. Rev. B}\ }\textbf {\bibinfo
  {volume} {97}},\ \bibinfo {pages} {134423} (\bibinfo {year}
  {2018})}\BibitemShut {NoStop}%
\bibitem [{\citenamefont {Shitade}\ \emph {et~al.}(2018)\citenamefont
  {Shitade}, \citenamefont {Watanabe},\ and\ \citenamefont
  {Yanase}}]{Shitade_PhysRevB.98.020407}%
  \BibitemOpen
  \bibfield  {author} {\bibinfo {author} {\bibfnamefont {A.}~\bibnamefont
  {Shitade}}, \bibinfo {author} {\bibfnamefont {H.}~\bibnamefont {Watanabe}},\
  and\ \bibinfo {author} {\bibfnamefont {Y.}~\bibnamefont {Yanase}},\
  }\bibfield  {title} {\bibinfo {title} {Theory of orbital magnetic quadrupole
  moment and magnetoelectric susceptibility},\ }\href
  {https://doi.org/10.1103/PhysRevB.98.020407} {\bibfield  {journal} {\bibinfo
  {journal} {Phys. Rev. B}\ }\textbf {\bibinfo {volume} {98}},\ \bibinfo
  {pages} {020407} (\bibinfo {year} {2018})}\BibitemShut {NoStop}%
\bibitem [{\citenamefont {Tokura}\ and\ \citenamefont
  {Nagaosa}(2018)}]{tokura2018nonreciprocal}%
  \BibitemOpen
  \bibfield  {author} {\bibinfo {author} {\bibfnamefont {Y.}~\bibnamefont
  {Tokura}}\ and\ \bibinfo {author} {\bibfnamefont {N.}~\bibnamefont
  {Nagaosa}},\ }\bibfield  {title} {\bibinfo {title} {Nonreciprocal responses
  from non-centrosymmetric quantum materials},\ }\href@noop {} {\bibfield
  {journal} {\bibinfo  {journal} {Nat. Commun.}\ }\textbf {\bibinfo {volume}
  {9}},\ \bibinfo {pages} {3740} (\bibinfo {year} {2018})}\BibitemShut
  {NoStop}%
\bibitem [{\citenamefont {Gao}\ and\ \citenamefont
  {Xiao}(2018)}]{Gao_PhysRevB.98.060402}%
  \BibitemOpen
  \bibfield  {author} {\bibinfo {author} {\bibfnamefont {Y.}~\bibnamefont
  {Gao}}\ and\ \bibinfo {author} {\bibfnamefont {D.}~\bibnamefont {Xiao}},\
  }\bibfield  {title} {\bibinfo {title} {Orbital magnetic quadrupole moment and
  nonlinear anomalous thermoelectric transport},\ }\href
  {https://doi.org/10.1103/PhysRevB.98.060402} {\bibfield  {journal} {\bibinfo
  {journal} {Phys. Rev. B}\ }\textbf {\bibinfo {volume} {98}},\ \bibinfo
  {pages} {060402} (\bibinfo {year} {2018})}\BibitemShut {NoStop}%
\bibitem [{\citenamefont {Ishizuka}\ and\ \citenamefont
  {Nagaosa}(2020)}]{ishizuka2020anomalous}%
  \BibitemOpen
  \bibfield  {author} {\bibinfo {author} {\bibfnamefont {H.}~\bibnamefont
  {Ishizuka}}\ and\ \bibinfo {author} {\bibfnamefont {N.}~\bibnamefont
  {Nagaosa}},\ }\bibfield  {title} {\bibinfo {title} {Anomalous electrical
  magnetochiral effect by chiral spin-cluster scattering},\ }\href@noop {}
  {\bibfield  {journal} {\bibinfo  {journal} {Nat. Commun.}\ }\textbf {\bibinfo
  {volume} {11}},\ \bibinfo {pages} {2986} (\bibinfo {year}
  {2020})}\BibitemShut {NoStop}%
\bibitem [{\citenamefont {Hayami}\ \emph
  {et~al.}(2020{\natexlab{a}})\citenamefont {Hayami}, \citenamefont {Yanagi},\
  and\ \citenamefont {Kusunose}}]{Hayami_PhysRevB.102.144441}%
  \BibitemOpen
  \bibfield  {author} {\bibinfo {author} {\bibfnamefont {S.}~\bibnamefont
  {Hayami}}, \bibinfo {author} {\bibfnamefont {Y.}~\bibnamefont {Yanagi}},\
  and\ \bibinfo {author} {\bibfnamefont {H.}~\bibnamefont {Kusunose}},\
  }\bibfield  {title} {\bibinfo {title} {Bottom-up design of spin-split and
  reshaped electronic band structures in antiferromagnets without spin-orbit
  coupling: Procedure on the basis of augmented multipoles},\ }\href
  {https://doi.org/10.1103/PhysRevB.102.144441} {\bibfield  {journal} {\bibinfo
   {journal} {Phys. Rev. B}\ }\textbf {\bibinfo {volume} {102}},\ \bibinfo
  {pages} {144441} (\bibinfo {year} {2020}{\natexlab{a}})}\BibitemShut
  {NoStop}%
\bibitem [{\citenamefont {Hirsch}(1984)}]{Hirsch_PhysRevB.30.5383}%
  \BibitemOpen
  \bibfield  {author} {\bibinfo {author} {\bibfnamefont {J.~E.}\ \bibnamefont
  {Hirsch}},\ }\bibfield  {title} {\bibinfo {title} {Strong-coupling expansion
  for a kondo-lattice model},\ }\href
  {https://doi.org/10.1103/PhysRevB.30.5383} {\bibfield  {journal} {\bibinfo
  {journal} {Phys. Rev. B}\ }\textbf {\bibinfo {volume} {30}},\ \bibinfo
  {pages} {5383} (\bibinfo {year} {1984})}\BibitemShut {NoStop}%
\bibitem [{\citenamefont {Huang}\ \emph {et~al.}(2019)\citenamefont {Huang},
  \citenamefont {Sheng},\ and\ \citenamefont
  {Ting}}]{Huang_PhysRevB.99.195109}%
  \BibitemOpen
  \bibfield  {author} {\bibinfo {author} {\bibfnamefont {Y.}~\bibnamefont
  {Huang}}, \bibinfo {author} {\bibfnamefont {D.~N.}\ \bibnamefont {Sheng}},\
  and\ \bibinfo {author} {\bibfnamefont {C.~S.}\ \bibnamefont {Ting}},\
  }\bibfield  {title} {\bibinfo {title} {Charge density wave in a doped kondo
  chain},\ }\href {https://doi.org/10.1103/PhysRevB.99.195109} {\bibfield
  {journal} {\bibinfo  {journal} {Phys. Rev. B}\ }\textbf {\bibinfo {volume}
  {99}},\ \bibinfo {pages} {195109} (\bibinfo {year} {2019})}\BibitemShut
  {NoStop}%
\bibitem [{\citenamefont {Misawa}\ \emph {et~al.}(2013)\citenamefont {Misawa},
  \citenamefont {Yoshitake},\ and\ \citenamefont
  {Motome}}]{Misawa_PhysRevLett.110.246401}%
  \BibitemOpen
  \bibfield  {author} {\bibinfo {author} {\bibfnamefont {T.}~\bibnamefont
  {Misawa}}, \bibinfo {author} {\bibfnamefont {J.}~\bibnamefont {Yoshitake}},\
  and\ \bibinfo {author} {\bibfnamefont {Y.}~\bibnamefont {Motome}},\
  }\bibfield  {title} {\bibinfo {title} {Charge order in a two-dimensional
  kondo lattice model},\ }\href
  {https://doi.org/10.1103/PhysRevLett.110.246401} {\bibfield  {journal}
  {\bibinfo  {journal} {Phys. Rev. Lett.}\ }\textbf {\bibinfo {volume} {110}},\
  \bibinfo {pages} {246401} (\bibinfo {year} {2013})}\BibitemShut {NoStop}%
\bibitem [{\citenamefont {Otsuki}\ \emph {et~al.}(2009)\citenamefont {Otsuki},
  \citenamefont {Kusunose},\ and\ \citenamefont
  {Kuramoto}}]{Otsuki_JPSJ.78.034719}%
  \BibitemOpen
  \bibfield  {author} {\bibinfo {author} {\bibfnamefont {J.}~\bibnamefont
  {Otsuki}}, \bibinfo {author} {\bibfnamefont {H.}~\bibnamefont {Kusunose}},\
  and\ \bibinfo {author} {\bibfnamefont {Y.}~\bibnamefont {Kuramoto}},\
  }\bibfield  {title} {\bibinfo {title} {The kondo lattice model in infinite
  dimensions: Ii. static susceptibilities and phase diagram},\ }\href
  {https://doi.org/10.1143/JPSJ.78.034719} {\bibfield  {journal} {\bibinfo
  {journal} {J. Phys. Soc. Jpn.}\ }\textbf {\bibinfo {volume} {78}},\ \bibinfo
  {pages} {034719} (\bibinfo {year} {2009})}\BibitemShut {NoStop}%
\bibitem [{\citenamefont {Peters}\ \emph {et~al.}(2013)\citenamefont {Peters},
  \citenamefont {Hoshino}, \citenamefont {Kawakami}, \citenamefont {Otsuki},\
  and\ \citenamefont {Kuramoto}}]{Peters_PhysRevB.87.165133}%
  \BibitemOpen
  \bibfield  {author} {\bibinfo {author} {\bibfnamefont {R.}~\bibnamefont
  {Peters}}, \bibinfo {author} {\bibfnamefont {S.}~\bibnamefont {Hoshino}},
  \bibinfo {author} {\bibfnamefont {N.}~\bibnamefont {Kawakami}}, \bibinfo
  {author} {\bibfnamefont {J.}~\bibnamefont {Otsuki}},\ and\ \bibinfo {author}
  {\bibfnamefont {Y.}~\bibnamefont {Kuramoto}},\ }\bibfield  {title} {\bibinfo
  {title} {Charge order in kondo lattice systems},\ }\href
  {https://doi.org/10.1103/PhysRevB.87.165133} {\bibfield  {journal} {\bibinfo
  {journal} {Phys. Rev. B}\ }\textbf {\bibinfo {volume} {87}},\ \bibinfo
  {pages} {165133} (\bibinfo {year} {2013})}\BibitemShut {NoStop}%
\bibitem [{\citenamefont {Tahvildar-Zadeh}\ \emph {et~al.}(1997)\citenamefont
  {Tahvildar-Zadeh}, \citenamefont {Jarrell},\ and\ \citenamefont
  {Freericks}}]{Zadeh_PhysRevB.55.R3332}%
  \BibitemOpen
  \bibfield  {author} {\bibinfo {author} {\bibfnamefont {A.~N.}\ \bibnamefont
  {Tahvildar-Zadeh}}, \bibinfo {author} {\bibfnamefont {M.}~\bibnamefont
  {Jarrell}},\ and\ \bibinfo {author} {\bibfnamefont {J.~K.}\ \bibnamefont
  {Freericks}},\ }\bibfield  {title} {\bibinfo {title} {Protracted screening in
  the periodic anderson model},\ }\href
  {https://doi.org/10.1103/PhysRevB.55.R3332} {\bibfield  {journal} {\bibinfo
  {journal} {Phys. Rev. B}\ }\textbf {\bibinfo {volume} {55}},\ \bibinfo
  {pages} {R3332} (\bibinfo {year} {1997})}\BibitemShut {NoStop}%
\bibitem [{\citenamefont {Majidi}\ \emph {et~al.}(2007)\citenamefont {Majidi},
  \citenamefont {Doluweera}, \citenamefont {Moritz}, \citenamefont {Kent},
  \citenamefont {Moreno},\ and\ \citenamefont {Jarrell}}]{Majidi_2007charge}%
  \BibitemOpen
  \bibfield  {author} {\bibinfo {author} {\bibfnamefont {M.}~\bibnamefont
  {Majidi}}, \bibinfo {author} {\bibfnamefont {D.}~\bibnamefont {Doluweera}},
  \bibinfo {author} {\bibfnamefont {B.}~\bibnamefont {Moritz}}, \bibinfo
  {author} {\bibfnamefont {P.}~\bibnamefont {Kent}}, \bibinfo {author}
  {\bibfnamefont {J.}~\bibnamefont {Moreno}},\ and\ \bibinfo {author}
  {\bibfnamefont {M.}~\bibnamefont {Jarrell}},\ }\bibfield  {title} {\bibinfo
  {title} {Charge density wave driven ferromagnetism in the periodic anderson
  model},\ }\href@noop {} {\bibfield  {journal} {\bibinfo  {journal}
  {arXiv:0710.5937}\ } (\bibinfo {year} {2007})}\BibitemShut {NoStop}%
\bibitem [{\citenamefont {Motome}\ \emph {et~al.}(2010)\citenamefont {Motome},
  \citenamefont {Nakamikawa}, \citenamefont {Yamaji},\ and\ \citenamefont
  {Udagawa}}]{Motome2010}%
  \BibitemOpen
  \bibfield  {author} {\bibinfo {author} {\bibfnamefont {Y.}~\bibnamefont
  {Motome}}, \bibinfo {author} {\bibfnamefont {K.}~\bibnamefont {Nakamikawa}},
  \bibinfo {author} {\bibfnamefont {Y.}~\bibnamefont {Yamaji}},\ and\ \bibinfo
  {author} {\bibfnamefont {M.}~\bibnamefont {Udagawa}},\ }\bibfield  {title}
  {\bibinfo {title} {Partial kondo screening in frustrated kondo lattice
  systems},\ }\href {https://doi.org/10.1103/PhysRevLett.105.036403} {\bibfield
   {journal} {\bibinfo  {journal} {Phys. Rev. Lett.}\ }\textbf {\bibinfo
  {volume} {105}},\ \bibinfo {pages} {036403} (\bibinfo {year}
  {2010})}\BibitemShut {NoStop}%
\bibitem [{\citenamefont {Hayami}\ \emph {et~al.}(2011)\citenamefont {Hayami},
  \citenamefont {Udagawa},\ and\ \citenamefont {Motome}}]{Hayami2011}%
  \BibitemOpen
  \bibfield  {author} {\bibinfo {author} {\bibfnamefont {S.}~\bibnamefont
  {Hayami}}, \bibinfo {author} {\bibfnamefont {M.}~\bibnamefont {Udagawa}},\
  and\ \bibinfo {author} {\bibfnamefont {Y.}~\bibnamefont {Motome}},\
  }\bibfield  {title} {\bibinfo {title} {Partial disorder in the periodic
  anderson model on a triangular lattice},\ }\href
  {https://doi.org/10.1143/JPSJ.80.073704} {\bibfield  {journal} {\bibinfo
  {journal} {J. Phys. Soc. Jpn.}\ }\textbf {\bibinfo {volume} {80}},\ \bibinfo
  {pages} {073704} (\bibinfo {year} {2011})}\BibitemShut {NoStop}%
\bibitem [{\citenamefont {Ishizuka}\ and\ \citenamefont
  {Motome}(2012{\natexlab{a}})}]{Ishizuka_PhysRevLett.108.257205}%
  \BibitemOpen
  \bibfield  {author} {\bibinfo {author} {\bibfnamefont {H.}~\bibnamefont
  {Ishizuka}}\ and\ \bibinfo {author} {\bibfnamefont {Y.}~\bibnamefont
  {Motome}},\ }\bibfield  {title} {\bibinfo {title} {Partial disorder in an
  ising-spin kondo lattice model on a triangular lattice},\ }\href
  {https://doi.org/10.1103/PhysRevLett.108.257205} {\bibfield  {journal}
  {\bibinfo  {journal} {Phys. Rev. Lett.}\ }\textbf {\bibinfo {volume} {108}},\
  \bibinfo {pages} {257205} (\bibinfo {year} {2012}{\natexlab{a}})}\BibitemShut
  {NoStop}%
\bibitem [{\citenamefont {Hayami}\ \emph {et~al.}(2012)\citenamefont {Hayami},
  \citenamefont {Udagawa},\ and\ \citenamefont {Motome}}]{Hayami2012}%
  \BibitemOpen
  \bibfield  {author} {\bibinfo {author} {\bibfnamefont {S.}~\bibnamefont
  {Hayami}}, \bibinfo {author} {\bibfnamefont {M.}~\bibnamefont {Udagawa}},\
  and\ \bibinfo {author} {\bibfnamefont {Y.}~\bibnamefont {Motome}},\
  }\bibfield  {title} {\bibinfo {title} {Partial disorder and metal--insulator
  transition in the periodic anderson model on a triangular lattice},\ }\href
  {https://doi.org/10.1143/JPSJ.81.103707} {\bibfield  {journal} {\bibinfo
  {journal} {J. Phys. Soc. Jpn.}\ }\textbf {\bibinfo {volume} {81}},\ \bibinfo
  {pages} {103707} (\bibinfo {year} {2012})}\BibitemShut {NoStop}%
\bibitem [{\citenamefont {Aulbach}\ \emph {et~al.}(2015)\citenamefont
  {Aulbach}, \citenamefont {Assaad},\ and\ \citenamefont
  {Potthoff}}]{Aulbach_PhysRevB.92.235131}%
  \BibitemOpen
  \bibfield  {author} {\bibinfo {author} {\bibfnamefont {M.~W.}\ \bibnamefont
  {Aulbach}}, \bibinfo {author} {\bibfnamefont {F.~F.}\ \bibnamefont
  {Assaad}},\ and\ \bibinfo {author} {\bibfnamefont {M.}~\bibnamefont
  {Potthoff}},\ }\bibfield  {title} {\bibinfo {title} {Dynamical mean-field
  study of partial kondo screening in the periodic anderson model on the
  triangular lattice},\ }\href {https://doi.org/10.1103/PhysRevB.92.235131}
  {\bibfield  {journal} {\bibinfo  {journal} {Phys. Rev. B}\ }\textbf {\bibinfo
  {volume} {92}},\ \bibinfo {pages} {235131} (\bibinfo {year}
  {2015})}\BibitemShut {NoStop}%
\bibitem [{\citenamefont {Ke\ss{}ler}\ and\ \citenamefont
  {Eder}(2020)}]{Keler_PhysRevB.102.235125}%
  \BibitemOpen
  \bibfield  {author} {\bibinfo {author} {\bibfnamefont {M.}~\bibnamefont
  {Ke\ss{}ler}}\ and\ \bibinfo {author} {\bibfnamefont {R.}~\bibnamefont
  {Eder}},\ }\bibfield  {title} {\bibinfo {title} {Magnetic phases of the
  triangular kondo lattice},\ }\href
  {https://doi.org/10.1103/PhysRevB.102.235125} {\bibfield  {journal} {\bibinfo
   {journal} {Phys. Rev. B}\ }\textbf {\bibinfo {volume} {102}},\ \bibinfo
  {pages} {235125} (\bibinfo {year} {2020})}\BibitemShut {NoStop}%
\bibitem [{\citenamefont {Ishizuka}\ and\ \citenamefont
  {Motome}(2012{\natexlab{b}})}]{Ishizuka_PhysRevLett.109.237207}%
  \BibitemOpen
  \bibfield  {author} {\bibinfo {author} {\bibfnamefont {H.}~\bibnamefont
  {Ishizuka}}\ and\ \bibinfo {author} {\bibfnamefont {Y.}~\bibnamefont
  {Motome}},\ }\bibfield  {title} {\bibinfo {title} {Dirac half-metal in a
  triangular ferrimagnet},\ }\href
  {https://doi.org/10.1103/PhysRevLett.109.237207} {\bibfield  {journal}
  {\bibinfo  {journal} {Phys. Rev. Lett.}\ }\textbf {\bibinfo {volume} {109}},\
  \bibinfo {pages} {237207} (\bibinfo {year} {2012}{\natexlab{b}})}\BibitemShut
  {NoStop}%
\bibitem [{\citenamefont {Akagi}\ and\ \citenamefont
  {Motome}(2015)}]{Akagi_PhysRevB.91.155132}%
  \BibitemOpen
  \bibfield  {author} {\bibinfo {author} {\bibfnamefont {Y.}~\bibnamefont
  {Akagi}}\ and\ \bibinfo {author} {\bibfnamefont {Y.}~\bibnamefont {Motome}},\
  }\bibfield  {title} {\bibinfo {title} {Spontaneous formation of kagome
  network and dirac half-semimetal on a triangular lattice},\ }\href
  {https://doi.org/10.1103/PhysRevB.91.155132} {\bibfield  {journal} {\bibinfo
  {journal} {Phys. Rev. B}\ }\textbf {\bibinfo {volume} {91}},\ \bibinfo
  {pages} {155132} (\bibinfo {year} {2015})}\BibitemShut {NoStop}%
\bibitem [{\citenamefont {Hayami}\ \emph
  {et~al.}(2014{\natexlab{b}})\citenamefont {Hayami}, \citenamefont {Misawa},
  \citenamefont {Yamaji},\ and\ \citenamefont
  {Motome}}]{Hayami_PhysRevB.89.085124}%
  \BibitemOpen
  \bibfield  {author} {\bibinfo {author} {\bibfnamefont {S.}~\bibnamefont
  {Hayami}}, \bibinfo {author} {\bibfnamefont {T.}~\bibnamefont {Misawa}},
  \bibinfo {author} {\bibfnamefont {Y.}~\bibnamefont {Yamaji}},\ and\ \bibinfo
  {author} {\bibfnamefont {Y.}~\bibnamefont {Motome}},\ }\bibfield  {title}
  {\bibinfo {title} {Three-dimensional dirac electrons on a cubic lattice with
  noncoplanar multiple-$q$ order},\ }\href
  {https://doi.org/10.1103/PhysRevB.89.085124} {\bibfield  {journal} {\bibinfo
  {journal} {Phys. Rev. B}\ }\textbf {\bibinfo {volume} {89}},\ \bibinfo
  {pages} {085124} (\bibinfo {year} {2014}{\natexlab{b}})}\BibitemShut
  {NoStop}%
\bibitem [{\citenamefont {Hayami}\ \emph
  {et~al.}(2014{\natexlab{c}})\citenamefont {Hayami}, \citenamefont {Misawa},\
  and\ \citenamefont {Motome}}]{hayami2014charge}%
  \BibitemOpen
  \bibfield  {author} {\bibinfo {author} {\bibfnamefont {S.}~\bibnamefont
  {Hayami}}, \bibinfo {author} {\bibfnamefont {T.}~\bibnamefont {Misawa}},\
  and\ \bibinfo {author} {\bibfnamefont {Y.}~\bibnamefont {Motome}},\
  }\bibfield  {title} {\bibinfo {title} {Charge order with a noncoplanar
  triple-q magnetic order on a cubic lattice},\ }\href@noop {} {\bibfield
  {journal} {\bibinfo  {journal} {JPS Conf. Proc.}\ }\textbf {\bibinfo {volume}
  {3}},\ \bibinfo {pages} {016016} (\bibinfo {year}
  {2014}{\natexlab{c}})}\BibitemShut {NoStop}%
\bibitem [{\citenamefont {Heinze}\ \emph {et~al.}(2011)\citenamefont {Heinze},
  \citenamefont {von Bergmann}, \citenamefont {Menzel}, \citenamefont {Brede},
  \citenamefont {Kubetzka}, \citenamefont {Wiesendanger}, \citenamefont
  {Bihlmayer},\ and\ \citenamefont {Bl{\"u}gel}}]{heinze2011spontaneous}%
  \BibitemOpen
  \bibfield  {author} {\bibinfo {author} {\bibfnamefont {S.}~\bibnamefont
  {Heinze}}, \bibinfo {author} {\bibfnamefont {K.}~\bibnamefont {von
  Bergmann}}, \bibinfo {author} {\bibfnamefont {M.}~\bibnamefont {Menzel}},
  \bibinfo {author} {\bibfnamefont {J.}~\bibnamefont {Brede}}, \bibinfo
  {author} {\bibfnamefont {A.}~\bibnamefont {Kubetzka}}, \bibinfo {author}
  {\bibfnamefont {R.}~\bibnamefont {Wiesendanger}}, \bibinfo {author}
  {\bibfnamefont {G.}~\bibnamefont {Bihlmayer}},\ and\ \bibinfo {author}
  {\bibfnamefont {S.}~\bibnamefont {Bl{\"u}gel}},\ }\bibfield  {title}
  {\bibinfo {title} {Spontaneous atomic-scale magnetic skyrmion lattice in two
  dimensions},\ }\href@noop {} {\bibfield  {journal} {\bibinfo  {journal} {Nat.
  Phys.}\ }\textbf {\bibinfo {volume} {7}},\ \bibinfo {pages} {713} (\bibinfo
  {year} {2011})}\BibitemShut {NoStop}%
\bibitem [{\citenamefont {von Bergmann}\ \emph {et~al.}(2012)\citenamefont {von
  Bergmann}, \citenamefont {Menzel}, \citenamefont {Serrate}, \citenamefont
  {Yoshida}, \citenamefont {Schr\"oder}, \citenamefont {Ferriani},
  \citenamefont {Kubetzka}, \citenamefont {Wiesendanger},\ and\ \citenamefont
  {Heinze}}]{Bergmann_PhysRevB.86.134422}%
  \BibitemOpen
  \bibfield  {author} {\bibinfo {author} {\bibfnamefont {K.}~\bibnamefont {von
  Bergmann}}, \bibinfo {author} {\bibfnamefont {M.}~\bibnamefont {Menzel}},
  \bibinfo {author} {\bibfnamefont {D.}~\bibnamefont {Serrate}}, \bibinfo
  {author} {\bibfnamefont {Y.}~\bibnamefont {Yoshida}}, \bibinfo {author}
  {\bibfnamefont {S.}~\bibnamefont {Schr\"oder}}, \bibinfo {author}
  {\bibfnamefont {P.}~\bibnamefont {Ferriani}}, \bibinfo {author}
  {\bibfnamefont {A.}~\bibnamefont {Kubetzka}}, \bibinfo {author}
  {\bibfnamefont {R.}~\bibnamefont {Wiesendanger}},\ and\ \bibinfo {author}
  {\bibfnamefont {S.}~\bibnamefont {Heinze}},\ }\bibfield  {title} {\bibinfo
  {title} {Tunneling anisotropic magnetoresistance on the atomic scale},\
  }\href {https://doi.org/10.1103/PhysRevB.86.134422} {\bibfield  {journal}
  {\bibinfo  {journal} {Phys. Rev. B}\ }\textbf {\bibinfo {volume} {86}},\
  \bibinfo {pages} {134422} (\bibinfo {year} {2012})}\BibitemShut {NoStop}%
\bibitem [{\citenamefont {von Bergmann}\ \emph {et~al.}(2015)\citenamefont {von
  Bergmann}, \citenamefont {Menzel}, \citenamefont {Kubetzka},\ and\
  \citenamefont {Wiesendanger}}]{von2015influence}%
  \BibitemOpen
  \bibfield  {author} {\bibinfo {author} {\bibfnamefont {K.}~\bibnamefont {von
  Bergmann}}, \bibinfo {author} {\bibfnamefont {M.}~\bibnamefont {Menzel}},
  \bibinfo {author} {\bibfnamefont {A.}~\bibnamefont {Kubetzka}},\ and\
  \bibinfo {author} {\bibfnamefont {R.}~\bibnamefont {Wiesendanger}},\
  }\bibfield  {title} {\bibinfo {title} {Influence of the local atom
  configuration on a hexagonal skyrmion lattice},\ }\href@noop {} {\bibfield
  {journal} {\bibinfo  {journal} {Nano Lett.}\ }\textbf {\bibinfo {volume}
  {15}},\ \bibinfo {pages} {3280} (\bibinfo {year} {2015})}\BibitemShut
  {NoStop}%
\bibitem [{\citenamefont {Hanneken}\ \emph {et~al.}(2015)\citenamefont
  {Hanneken}, \citenamefont {Otte}, \citenamefont {Kubetzka}, \citenamefont
  {Dup{\'e}}, \citenamefont {Romming}, \citenamefont {Von~Bergmann},
  \citenamefont {Wiesendanger},\ and\ \citenamefont
  {Heinze}}]{hanneken2015electrical}%
  \BibitemOpen
  \bibfield  {author} {\bibinfo {author} {\bibfnamefont {C.}~\bibnamefont
  {Hanneken}}, \bibinfo {author} {\bibfnamefont {F.}~\bibnamefont {Otte}},
  \bibinfo {author} {\bibfnamefont {A.}~\bibnamefont {Kubetzka}}, \bibinfo
  {author} {\bibfnamefont {B.}~\bibnamefont {Dup{\'e}}}, \bibinfo {author}
  {\bibfnamefont {N.}~\bibnamefont {Romming}}, \bibinfo {author} {\bibfnamefont
  {K.}~\bibnamefont {Von~Bergmann}}, \bibinfo {author} {\bibfnamefont
  {R.}~\bibnamefont {Wiesendanger}},\ and\ \bibinfo {author} {\bibfnamefont
  {S.}~\bibnamefont {Heinze}},\ }\bibfield  {title} {\bibinfo {title}
  {Electrical detection of magnetic skyrmions by tunnelling non-collinear
  magnetoresistance},\ }\href@noop {} {\bibfield  {journal} {\bibinfo
  {journal} {Nat. Nanotech.}\ }\textbf {\bibinfo {volume} {10}},\ \bibinfo
  {pages} {1039} (\bibinfo {year} {2015})}\BibitemShut {NoStop}%
\bibitem [{\citenamefont {Kubetzka}\ \emph {et~al.}(2017)\citenamefont
  {Kubetzka}, \citenamefont {Hanneken}, \citenamefont {Wiesendanger},\ and\
  \citenamefont {von Bergmann}}]{Kubetzka_PhysRevB.95.104433}%
  \BibitemOpen
  \bibfield  {author} {\bibinfo {author} {\bibfnamefont {A.}~\bibnamefont
  {Kubetzka}}, \bibinfo {author} {\bibfnamefont {C.}~\bibnamefont {Hanneken}},
  \bibinfo {author} {\bibfnamefont {R.}~\bibnamefont {Wiesendanger}},\ and\
  \bibinfo {author} {\bibfnamefont {K.}~\bibnamefont {von Bergmann}},\
  }\bibfield  {title} {\bibinfo {title} {Impact of the skyrmion spin texture on
  magnetoresistance},\ }\href {https://doi.org/10.1103/PhysRevB.95.104433}
  {\bibfield  {journal} {\bibinfo  {journal} {Phys. Rev. B}\ }\textbf {\bibinfo
  {volume} {95}},\ \bibinfo {pages} {104433} (\bibinfo {year}
  {2017})}\BibitemShut {NoStop}%
\bibitem [{\citenamefont {Kathyat}\ \emph {et~al.}(2021)\citenamefont
  {Kathyat}, \citenamefont {Mukherjee},\ and\ \citenamefont
  {Kumar}}]{Kathyat_PhysRevB.103.035111}%
  \BibitemOpen
  \bibfield  {author} {\bibinfo {author} {\bibfnamefont {D.~S.}\ \bibnamefont
  {Kathyat}}, \bibinfo {author} {\bibfnamefont {A.}~\bibnamefont {Mukherjee}},\
  and\ \bibinfo {author} {\bibfnamefont {S.}~\bibnamefont {Kumar}},\ }\bibfield
   {title} {\bibinfo {title} {Electronic mechanism for nanoscale skyrmions and
  topological metals},\ }\href {https://doi.org/10.1103/PhysRevB.103.035111}
  {\bibfield  {journal} {\bibinfo  {journal} {Phys. Rev. B}\ }\textbf {\bibinfo
  {volume} {103}},\ \bibinfo {pages} {035111} (\bibinfo {year}
  {2021})}\BibitemShut {NoStop}%
\bibitem [{\citenamefont {Khanh}\ \emph {et~al.}(2020)\citenamefont {Khanh},
  \citenamefont {Nakajima}, \citenamefont {Yu}, \citenamefont {Gao},
  \citenamefont {Shibata}, \citenamefont {Hirschberger}, \citenamefont
  {Yamasaki}, \citenamefont {Sagayama}, \citenamefont {Nakao}, \citenamefont
  {Peng}, \citenamefont {Nakajima}, \citenamefont {Takagi}, \citenamefont
  {Arima}, \citenamefont {Tokura},\ and\ \citenamefont
  {Seki}}]{khanh2020nanometric}%
  \BibitemOpen
  \bibfield  {author} {\bibinfo {author} {\bibfnamefont {N.~D.}\ \bibnamefont
  {Khanh}}, \bibinfo {author} {\bibfnamefont {T.}~\bibnamefont {Nakajima}},
  \bibinfo {author} {\bibfnamefont {X.}~\bibnamefont {Yu}}, \bibinfo {author}
  {\bibfnamefont {S.}~\bibnamefont {Gao}}, \bibinfo {author} {\bibfnamefont
  {K.}~\bibnamefont {Shibata}}, \bibinfo {author} {\bibfnamefont
  {M.}~\bibnamefont {Hirschberger}}, \bibinfo {author} {\bibfnamefont
  {Y.}~\bibnamefont {Yamasaki}}, \bibinfo {author} {\bibfnamefont
  {H.}~\bibnamefont {Sagayama}}, \bibinfo {author} {\bibfnamefont
  {H.}~\bibnamefont {Nakao}}, \bibinfo {author} {\bibfnamefont
  {L.}~\bibnamefont {Peng}}, \bibinfo {author} {\bibfnamefont {K.}~\bibnamefont
  {Nakajima}}, \bibinfo {author} {\bibfnamefont {R.}~\bibnamefont {Takagi}},
  \bibinfo {author} {\bibfnamefont {T.-h.}\ \bibnamefont {Arima}}, \bibinfo
  {author} {\bibfnamefont {Y.}~\bibnamefont {Tokura}},\ and\ \bibinfo {author}
  {\bibfnamefont {S.}~\bibnamefont {Seki}},\ }\bibfield  {title} {\bibinfo
  {title} {Nanometric square skyrmion lattice in a centrosymmetric tetragonal
  magnet},\ }\href@noop {} {\bibfield  {journal} {\bibinfo  {journal} {Nat.
  Nanotech.}\ }\textbf {\bibinfo {volume} {15}},\ \bibinfo {pages} {444}
  (\bibinfo {year} {2020})}\BibitemShut {NoStop}%
\bibitem [{\citenamefont {Yasui}\ \emph {et~al.}(2020)\citenamefont {Yasui},
  \citenamefont {Butler}, \citenamefont {Khanh}, \citenamefont {Hayami},
  \citenamefont {Nomoto}, \citenamefont {Hanaguri}, \citenamefont {Motome},
  \citenamefont {Arita}, \citenamefont {h.~Arima}, \citenamefont {Tokura},\
  and\ \citenamefont {Seki}}]{Yasui2020imaging}%
  \BibitemOpen
  \bibfield  {author} {\bibinfo {author} {\bibfnamefont {Y.}~\bibnamefont
  {Yasui}}, \bibinfo {author} {\bibfnamefont {C.~J.}\ \bibnamefont {Butler}},
  \bibinfo {author} {\bibfnamefont {N.~D.}\ \bibnamefont {Khanh}}, \bibinfo
  {author} {\bibfnamefont {S.}~\bibnamefont {Hayami}}, \bibinfo {author}
  {\bibfnamefont {T.}~\bibnamefont {Nomoto}}, \bibinfo {author} {\bibfnamefont
  {T.}~\bibnamefont {Hanaguri}}, \bibinfo {author} {\bibfnamefont
  {Y.}~\bibnamefont {Motome}}, \bibinfo {author} {\bibfnamefont
  {R.}~\bibnamefont {Arita}}, \bibinfo {author} {\bibfnamefont
  {T.}~\bibnamefont {h.~Arima}}, \bibinfo {author} {\bibfnamefont
  {Y.}~\bibnamefont {Tokura}},\ and\ \bibinfo {author} {\bibfnamefont
  {S.}~\bibnamefont {Seki}},\ }\bibfield  {title} {\bibinfo {title} {Imaging
  the coupling between itinerant electrons and localised moments in the
  centrosymmetric skyrmion magnet gdru$_2$si$_2$},\ }\href@noop {} {\bibfield
  {journal} {\bibinfo  {journal} {Nat. Commun.}\ }\textbf {\bibinfo {volume}
  {11}},\ \bibinfo {pages} {5925} (\bibinfo {year} {2020})}\BibitemShut
  {NoStop}%
\bibitem [{\citenamefont {Hayami}\ and\ \citenamefont
  {Motome}(2021{\natexlab{b}})}]{Hayami_PhysRevB.103.024439}%
  \BibitemOpen
  \bibfield  {author} {\bibinfo {author} {\bibfnamefont {S.}~\bibnamefont
  {Hayami}}\ and\ \bibinfo {author} {\bibfnamefont {Y.}~\bibnamefont
  {Motome}},\ }\bibfield  {title} {\bibinfo {title} {Square skyrmion crystal in
  centrosymmetric itinerant magnets},\ }\href
  {https://doi.org/10.1103/PhysRevB.103.024439} {\bibfield  {journal} {\bibinfo
   {journal} {Phys. Rev. B}\ }\textbf {\bibinfo {volume} {103}},\ \bibinfo
  {pages} {024439} (\bibinfo {year} {2021}{\natexlab{b}})}\BibitemShut
  {NoStop}%
\bibitem [{\citenamefont {Hayami}\ and\ \citenamefont
  {Motome}(2019)}]{Hayami_PhysRevB.99.094420}%
  \BibitemOpen
  \bibfield  {author} {\bibinfo {author} {\bibfnamefont {S.}~\bibnamefont
  {Hayami}}\ and\ \bibinfo {author} {\bibfnamefont {Y.}~\bibnamefont
  {Motome}},\ }\bibfield  {title} {\bibinfo {title} {Effect of magnetic
  anisotropy on skyrmions with a high topological number in itinerant
  magnets},\ }\href {https://doi.org/10.1103/PhysRevB.99.094420} {\bibfield
  {journal} {\bibinfo  {journal} {Phys. Rev. B}\ }\textbf {\bibinfo {volume}
  {99}},\ \bibinfo {pages} {094420} (\bibinfo {year} {2019})}\BibitemShut
  {NoStop}%
\bibitem [{\citenamefont {Mohanta}\ \emph {et~al.}(2019)\citenamefont
  {Mohanta}, \citenamefont {Dagotto},\ and\ \citenamefont
  {Okamoto}}]{Mohanta_PhysRevB.100.064429}%
  \BibitemOpen
  \bibfield  {author} {\bibinfo {author} {\bibfnamefont {N.}~\bibnamefont
  {Mohanta}}, \bibinfo {author} {\bibfnamefont {E.}~\bibnamefont {Dagotto}},\
  and\ \bibinfo {author} {\bibfnamefont {S.}~\bibnamefont {Okamoto}},\
  }\bibfield  {title} {\bibinfo {title} {Topological hall effect and emergent
  skyrmion crystal at manganite-iridate oxide interfaces},\ }\href
  {https://doi.org/10.1103/PhysRevB.100.064429} {\bibfield  {journal} {\bibinfo
   {journal} {Phys. Rev. B}\ }\textbf {\bibinfo {volume} {100}},\ \bibinfo
  {pages} {064429} (\bibinfo {year} {2019})}\BibitemShut {NoStop}%
\bibitem [{\citenamefont {Wang}\ \emph {et~al.}(2020)\citenamefont {Wang},
  \citenamefont {Su}, \citenamefont {Lin},\ and\ \citenamefont
  {Batista}}]{Wang_PhysRevLett.124.207201}%
  \BibitemOpen
  \bibfield  {author} {\bibinfo {author} {\bibfnamefont {Z.}~\bibnamefont
  {Wang}}, \bibinfo {author} {\bibfnamefont {Y.}~\bibnamefont {Su}}, \bibinfo
  {author} {\bibfnamefont {S.-Z.}\ \bibnamefont {Lin}},\ and\ \bibinfo {author}
  {\bibfnamefont {C.~D.}\ \bibnamefont {Batista}},\ }\bibfield  {title}
  {\bibinfo {title} {Skyrmion crystal from rkky interaction mediated by 2d
  electron gas},\ }\href {https://doi.org/10.1103/PhysRevLett.124.207201}
  {\bibfield  {journal} {\bibinfo  {journal} {Phys. Rev. Lett.}\ }\textbf
  {\bibinfo {volume} {124}},\ \bibinfo {pages} {207201} (\bibinfo {year}
  {2020})}\BibitemShut {NoStop}%
\bibitem [{\citenamefont {Solenov}\ \emph {et~al.}(2012)\citenamefont
  {Solenov}, \citenamefont {Mozyrsky},\ and\ \citenamefont
  {Martin}}]{Solenov_PhysRevLett.108.096403}%
  \BibitemOpen
  \bibfield  {author} {\bibinfo {author} {\bibfnamefont {D.}~\bibnamefont
  {Solenov}}, \bibinfo {author} {\bibfnamefont {D.}~\bibnamefont {Mozyrsky}},\
  and\ \bibinfo {author} {\bibfnamefont {I.}~\bibnamefont {Martin}},\
  }\bibfield  {title} {\bibinfo {title} {Chirality waves in two-dimensional
  magnets},\ }\href {https://doi.org/10.1103/PhysRevLett.108.096403} {\bibfield
   {journal} {\bibinfo  {journal} {Phys. Rev. Lett.}\ }\textbf {\bibinfo
  {volume} {108}},\ \bibinfo {pages} {096403} (\bibinfo {year}
  {2012})}\BibitemShut {NoStop}%
\bibitem [{\citenamefont {Hayami}\ \emph
  {et~al.}(2016{\natexlab{a}})\citenamefont {Hayami}, \citenamefont {Ozawa},\
  and\ \citenamefont {Motome}}]{Hayami_PhysRevB.94.024424}%
  \BibitemOpen
  \bibfield  {author} {\bibinfo {author} {\bibfnamefont {S.}~\bibnamefont
  {Hayami}}, \bibinfo {author} {\bibfnamefont {R.}~\bibnamefont {Ozawa}},\ and\
  \bibinfo {author} {\bibfnamefont {Y.}~\bibnamefont {Motome}},\ }\bibfield
  {title} {\bibinfo {title} {Engineering chiral density waves and topological
  band structures by multiple-$q$ superpositions of collinear up-up-down-down
  orders},\ }\href {https://doi.org/10.1103/PhysRevB.94.024424} {\bibfield
  {journal} {\bibinfo  {journal} {Phys. Rev. B}\ }\textbf {\bibinfo {volume}
  {94}},\ \bibinfo {pages} {024424} (\bibinfo {year}
  {2016}{\natexlab{a}})}\BibitemShut {NoStop}%
\bibitem [{\citenamefont {Ozawa}\ \emph {et~al.}(2016)\citenamefont {Ozawa},
  \citenamefont {Hayami}, \citenamefont {Barros}, \citenamefont {Chern},
  \citenamefont {Motome},\ and\ \citenamefont
  {Batista}}]{Ozawa_doi:10.7566/JPSJ.85.103703}%
  \BibitemOpen
  \bibfield  {author} {\bibinfo {author} {\bibfnamefont {R.}~\bibnamefont
  {Ozawa}}, \bibinfo {author} {\bibfnamefont {S.}~\bibnamefont {Hayami}},
  \bibinfo {author} {\bibfnamefont {K.}~\bibnamefont {Barros}}, \bibinfo
  {author} {\bibfnamefont {G.-W.}\ \bibnamefont {Chern}}, \bibinfo {author}
  {\bibfnamefont {Y.}~\bibnamefont {Motome}},\ and\ \bibinfo {author}
  {\bibfnamefont {C.~D.}\ \bibnamefont {Batista}},\ }\bibfield  {title}
  {\bibinfo {title} {Vortex crystals with chiral stripes in itinerant
  magnets},\ }\href {https://doi.org/10.7566/JPSJ.85.103703} {\bibfield
  {journal} {\bibinfo  {journal} {J. Phys. Soc. Jpn.}\ }\textbf {\bibinfo
  {volume} {85}},\ \bibinfo {pages} {103703} (\bibinfo {year}
  {2016})}\BibitemShut {NoStop}%
\bibitem [{\citenamefont {Hayami}\ \emph
  {et~al.}(2020{\natexlab{b}})\citenamefont {Hayami}, \citenamefont {Okubo},\
  and\ \citenamefont {Motome}}]{hayami2020phase}%
  \BibitemOpen
  \bibfield  {author} {\bibinfo {author} {\bibfnamefont {S.}~\bibnamefont
  {Hayami}}, \bibinfo {author} {\bibfnamefont {T.}~\bibnamefont {Okubo}},\ and\
  \bibinfo {author} {\bibfnamefont {Y.}~\bibnamefont {Motome}},\ }\bibfield
  {title} {\bibinfo {title} {Phase shift in skyrmion crystals},\ }\href@noop {}
  {\bibfield  {journal} {\bibinfo  {journal} {arXiv:2005.03168}\ } (\bibinfo
  {year} {2020}{\natexlab{b}})}\BibitemShut {NoStop}%
\bibitem [{\citenamefont {Hayami}\ \emph {et~al.}(2017)\citenamefont {Hayami},
  \citenamefont {Ozawa},\ and\ \citenamefont
  {Motome}}]{Hayami_PhysRevB.95.224424}%
  \BibitemOpen
  \bibfield  {author} {\bibinfo {author} {\bibfnamefont {S.}~\bibnamefont
  {Hayami}}, \bibinfo {author} {\bibfnamefont {R.}~\bibnamefont {Ozawa}},\ and\
  \bibinfo {author} {\bibfnamefont {Y.}~\bibnamefont {Motome}},\ }\bibfield
  {title} {\bibinfo {title} {Effective bilinear-biquadratic model for
  noncoplanar ordering in itinerant magnets},\ }\href
  {https://doi.org/10.1103/PhysRevB.95.224424} {\bibfield  {journal} {\bibinfo
  {journal} {Phys. Rev. B}\ }\textbf {\bibinfo {volume} {95}},\ \bibinfo
  {pages} {224424} (\bibinfo {year} {2017})}\BibitemShut {NoStop}%
\bibitem [{\citenamefont {Hayami}(2020)}]{hayami2020multiple}%
  \BibitemOpen
  \bibfield  {author} {\bibinfo {author} {\bibfnamefont {S.}~\bibnamefont
  {Hayami}},\ }\bibfield  {title} {\bibinfo {title} {Multiple-q magnetism by
  anisotropic bilinear-biquadratic interactions in momentum space},\
  }\href@noop {} {\bibfield  {journal} {\bibinfo  {journal} {J. Mag. Mag.
  Mater.}\ }\textbf {\bibinfo {volume} {513}},\ \bibinfo {pages} {167181}
  (\bibinfo {year} {2020})}\BibitemShut {NoStop}%
\bibitem [{\citenamefont {Hayami}\ and\ \citenamefont
  {Motome}(2021{\natexlab{c}})}]{Hayami_PhysRevB.103.054422}%
  \BibitemOpen
  \bibfield  {author} {\bibinfo {author} {\bibfnamefont {S.}~\bibnamefont
  {Hayami}}\ and\ \bibinfo {author} {\bibfnamefont {Y.}~\bibnamefont
  {Motome}},\ }\bibfield  {title} {\bibinfo {title} {Noncoplanar multiple-$q$
  spin textures by itinerant frustration: Effects of single-ion anisotropy and
  bond-dependent anisotropy},\ }\href
  {https://doi.org/10.1103/PhysRevB.103.054422} {\bibfield  {journal} {\bibinfo
   {journal} {Phys. Rev. B}\ }\textbf {\bibinfo {volume} {103}},\ \bibinfo
  {pages} {054422} (\bibinfo {year} {2021}{\natexlab{c}})}\BibitemShut
  {NoStop}%
\bibitem [{\citenamefont {Okumura}\ \emph {et~al.}(2020)\citenamefont
  {Okumura}, \citenamefont {Hayami}, \citenamefont {Kato},\ and\ \citenamefont
  {Motome}}]{Okumura_PhysRevB.101.144416}%
  \BibitemOpen
  \bibfield  {author} {\bibinfo {author} {\bibfnamefont {S.}~\bibnamefont
  {Okumura}}, \bibinfo {author} {\bibfnamefont {S.}~\bibnamefont {Hayami}},
  \bibinfo {author} {\bibfnamefont {Y.}~\bibnamefont {Kato}},\ and\ \bibinfo
  {author} {\bibfnamefont {Y.}~\bibnamefont {Motome}},\ }\bibfield  {title}
  {\bibinfo {title} {Magnetic hedgehog lattices in noncentrosymmetric metals},\
  }\href {https://doi.org/10.1103/PhysRevB.101.144416} {\bibfield  {journal}
  {\bibinfo  {journal} {Phys. Rev. B}\ }\textbf {\bibinfo {volume} {101}},\
  \bibinfo {pages} {144416} (\bibinfo {year} {2020})}\BibitemShut {NoStop}%
\bibitem [{\citenamefont {Yambe}\ and\ \citenamefont
  {Hayami}(2020)}]{yambe2020double}%
  \BibitemOpen
  \bibfield  {author} {\bibinfo {author} {\bibfnamefont {R.}~\bibnamefont
  {Yambe}}\ and\ \bibinfo {author} {\bibfnamefont {S.}~\bibnamefont {Hayami}},\
  }\bibfield  {title} {\bibinfo {title} {Double-q chiral stripe in the d--p
  model with strong spin--charge coupling},\ }\href@noop {} {\bibfield
  {journal} {\bibinfo  {journal} {J. Phys. Soc. Jpn.}\ }\textbf {\bibinfo
  {volume} {89}},\ \bibinfo {pages} {013702} (\bibinfo {year}
  {2020})}\BibitemShut {NoStop}%
\bibitem [{\citenamefont {Hayami}\ and\ \citenamefont
  {Motome}(2018{\natexlab{a}})}]{hayami2018multiple}%
  \BibitemOpen
  \bibfield  {author} {\bibinfo {author} {\bibfnamefont {S.}~\bibnamefont
  {Hayami}}\ and\ \bibinfo {author} {\bibfnamefont {Y.}~\bibnamefont
  {Motome}},\ }\bibfield  {title} {\bibinfo {title} {Multiple-q magnetic states
  in spin-orbit coupled metals},\ }\href@noop {} {\bibfield  {journal}
  {\bibinfo  {journal} {IEEE Transactions on Magnetics}\ }\textbf {\bibinfo
  {volume} {55}},\ \bibinfo {pages} {1} (\bibinfo {year}
  {2018}{\natexlab{a}})}\BibitemShut {NoStop}%
\bibitem [{\citenamefont {Utesov}(2021)}]{Utesov_PhysRevB.103.064414}%
  \BibitemOpen
  \bibfield  {author} {\bibinfo {author} {\bibfnamefont {O.~I.}\ \bibnamefont
  {Utesov}},\ }\bibfield  {title} {\bibinfo {title} {Thermodynamically stable
  skyrmion lattice in a tetragonal frustrated antiferromagnet with dipolar
  interaction},\ }\href {https://doi.org/10.1103/PhysRevB.103.064414}
  {\bibfield  {journal} {\bibinfo  {journal} {Phys. Rev. B}\ }\textbf {\bibinfo
  {volume} {103}},\ \bibinfo {pages} {064414} (\bibinfo {year}
  {2021})}\BibitemShut {NoStop}%
\bibitem [{\citenamefont {Wang}\ \emph {et~al.}(2021)\citenamefont {Wang},
  \citenamefont {Su}, \citenamefont {Lin},\ and\ \citenamefont
  {Batista}}]{Wang_PhysRevB.103.104408}%
  \BibitemOpen
  \bibfield  {author} {\bibinfo {author} {\bibfnamefont {Z.}~\bibnamefont
  {Wang}}, \bibinfo {author} {\bibfnamefont {Y.}~\bibnamefont {Su}}, \bibinfo
  {author} {\bibfnamefont {S.-Z.}\ \bibnamefont {Lin}},\ and\ \bibinfo {author}
  {\bibfnamefont {C.~D.}\ \bibnamefont {Batista}},\ }\bibfield  {title}
  {\bibinfo {title} {Meron, skyrmion, and vortex crystals in centrosymmetric
  tetragonal magnets},\ }\href {https://doi.org/10.1103/PhysRevB.103.104408}
  {\bibfield  {journal} {\bibinfo  {journal} {Phys. Rev. B}\ }\textbf {\bibinfo
  {volume} {103}},\ \bibinfo {pages} {104408} (\bibinfo {year}
  {2021})}\BibitemShut {NoStop}%
\bibitem [{\citenamefont {Brey}\ \emph {et~al.}(1996)\citenamefont {Brey},
  \citenamefont {Fertig}, \citenamefont {C{\^o}t{\'e}},\ and\ \citenamefont
  {MacDonald}}]{brey1996skyrme}%
  \BibitemOpen
  \bibfield  {author} {\bibinfo {author} {\bibfnamefont {L.}~\bibnamefont
  {Brey}}, \bibinfo {author} {\bibfnamefont {H.}~\bibnamefont {Fertig}},
  \bibinfo {author} {\bibfnamefont {R.}~\bibnamefont {C{\^o}t{\'e}}},\ and\
  \bibinfo {author} {\bibfnamefont {A.}~\bibnamefont {MacDonald}},\ }\bibfield
  {title} {\bibinfo {title} {Skyrme and meron crystals in quantum hall
  ferromagnets},\ }\href@noop {} {\bibfield  {journal} {\bibinfo  {journal}
  {Physica Scripta}\ }\textbf {\bibinfo {volume} {1996}},\ \bibinfo {pages}
  {154} (\bibinfo {year} {1996})}\BibitemShut {NoStop}%
\bibitem [{\citenamefont {Yu}\ \emph {et~al.}(2018)\citenamefont {Yu},
  \citenamefont {Koshibae}, \citenamefont {Tokunaga}, \citenamefont {Shibata},
  \citenamefont {Taguchi}, \citenamefont {Nagaosa},\ and\ \citenamefont
  {Tokura}}]{yu2018transformation}%
  \BibitemOpen
  \bibfield  {author} {\bibinfo {author} {\bibfnamefont {X.~Z.}\ \bibnamefont
  {Yu}}, \bibinfo {author} {\bibfnamefont {W.}~\bibnamefont {Koshibae}},
  \bibinfo {author} {\bibfnamefont {Y.}~\bibnamefont {Tokunaga}}, \bibinfo
  {author} {\bibfnamefont {K.}~\bibnamefont {Shibata}}, \bibinfo {author}
  {\bibfnamefont {Y.}~\bibnamefont {Taguchi}}, \bibinfo {author} {\bibfnamefont
  {N.}~\bibnamefont {Nagaosa}},\ and\ \bibinfo {author} {\bibfnamefont
  {Y.}~\bibnamefont {Tokura}},\ }\bibfield  {title} {\bibinfo {title}
  {Transformation between meron and skyrmion topological spin textures in a
  chiral magnet},\ }\href@noop {} {\bibfield  {journal} {\bibinfo  {journal}
  {Nature}\ }\textbf {\bibinfo {volume} {564}},\ \bibinfo {pages} {95}
  (\bibinfo {year} {2018})}\BibitemShut {NoStop}%
\bibitem [{\citenamefont {Kurumaji}\ \emph {et~al.}(2019)\citenamefont
  {Kurumaji}, \citenamefont {Nakajima}, \citenamefont {Hirschberger},
  \citenamefont {Kikkawa}, \citenamefont {Yamasaki}, \citenamefont {Sagayama},
  \citenamefont {Nakao}, \citenamefont {Taguchi}, \citenamefont {Arima},\ and\
  \citenamefont {Tokura}}]{kurumaji2019skyrmion}%
  \BibitemOpen
  \bibfield  {author} {\bibinfo {author} {\bibfnamefont {T.}~\bibnamefont
  {Kurumaji}}, \bibinfo {author} {\bibfnamefont {T.}~\bibnamefont {Nakajima}},
  \bibinfo {author} {\bibfnamefont {M.}~\bibnamefont {Hirschberger}}, \bibinfo
  {author} {\bibfnamefont {A.}~\bibnamefont {Kikkawa}}, \bibinfo {author}
  {\bibfnamefont {Y.}~\bibnamefont {Yamasaki}}, \bibinfo {author}
  {\bibfnamefont {H.}~\bibnamefont {Sagayama}}, \bibinfo {author}
  {\bibfnamefont {H.}~\bibnamefont {Nakao}}, \bibinfo {author} {\bibfnamefont
  {Y.}~\bibnamefont {Taguchi}}, \bibinfo {author} {\bibfnamefont {T.-h.}\
  \bibnamefont {Arima}},\ and\ \bibinfo {author} {\bibfnamefont
  {Y.}~\bibnamefont {Tokura}},\ }\bibfield  {title} {\bibinfo {title} {Skyrmion
  lattice with a giant topological hall effect in a frustrated
  triangular-lattice magnet},\ }\href@noop {} {\bibfield  {journal} {\bibinfo
  {journal} {Science}\ }\textbf {\bibinfo {volume} {365}},\ \bibinfo {pages}
  {914} (\bibinfo {year} {2019})}\BibitemShut {NoStop}%
\bibitem [{\citenamefont {Lin}\ \emph {et~al.}(2015)\citenamefont {Lin},
  \citenamefont {Saxena},\ and\ \citenamefont
  {Batista}}]{Lin_PhysRevB.91.224407}%
  \BibitemOpen
  \bibfield  {author} {\bibinfo {author} {\bibfnamefont {S.-Z.}\ \bibnamefont
  {Lin}}, \bibinfo {author} {\bibfnamefont {A.}~\bibnamefont {Saxena}},\ and\
  \bibinfo {author} {\bibfnamefont {C.~D.}\ \bibnamefont {Batista}},\
  }\bibfield  {title} {\bibinfo {title} {Skyrmion fractionalization and merons
  in chiral magnets with easy-plane anisotropy},\ }\href
  {https://doi.org/10.1103/PhysRevB.91.224407} {\bibfield  {journal} {\bibinfo
  {journal} {Phys. Rev. B}\ }\textbf {\bibinfo {volume} {91}},\ \bibinfo
  {pages} {224407} (\bibinfo {year} {2015})}\BibitemShut {NoStop}%
\bibitem [{\citenamefont {Hayami}\ and\ \citenamefont
  {Motome}(2018{\natexlab{b}})}]{Hayami_PhysRevLett.121.137202}%
  \BibitemOpen
  \bibfield  {author} {\bibinfo {author} {\bibfnamefont {S.}~\bibnamefont
  {Hayami}}\ and\ \bibinfo {author} {\bibfnamefont {Y.}~\bibnamefont
  {Motome}},\ }\bibfield  {title} {\bibinfo {title} {N\'eel- and bloch-type
  magnetic vortices in rashba metals},\ }\href
  {https://doi.org/10.1103/PhysRevLett.121.137202} {\bibfield  {journal}
  {\bibinfo  {journal} {Phys. Rev. Lett.}\ }\textbf {\bibinfo {volume} {121}},\
  \bibinfo {pages} {137202} (\bibinfo {year} {2018}{\natexlab{b}})}\BibitemShut
  {NoStop}%
\bibitem [{\citenamefont {Hayami}\ and\ \citenamefont
  {Yambe}(2021{\natexlab{a}})}]{hayami2021meron}%
  \BibitemOpen
  \bibfield  {author} {\bibinfo {author} {\bibfnamefont {S.}~\bibnamefont
  {Hayami}}\ and\ \bibinfo {author} {\bibfnamefont {R.}~\bibnamefont {Yambe}},\
  }\bibfield  {title} {\bibinfo {title} {Meron-antimeron crystals in
  noncentrosymmetric itinerant magnets on a triangular lattice},\ }\href
  {https://doi.org/10.1103/PhysRevB.104.094425} {\bibfield  {journal} {\bibinfo
   {journal} {Phys. Rev. B}\ }\textbf {\bibinfo {volume} {104}},\ \bibinfo
  {pages} {094425} (\bibinfo {year} {2021}{\natexlab{a}})}\BibitemShut
  {NoStop}%
\bibitem [{\citenamefont {Marcus}\ \emph {et~al.}(2018)\citenamefont {Marcus},
  \citenamefont {Kim}, \citenamefont {Tutmaher}, \citenamefont
  {Rodriguez-Rivera}, \citenamefont {Birk}, \citenamefont {Niedermeyer},
  \citenamefont {Lee}, \citenamefont {Fisk},\ and\ \citenamefont
  {Broholm}}]{Marcus_PhysRevLett.120.097201}%
  \BibitemOpen
  \bibfield  {author} {\bibinfo {author} {\bibfnamefont {G.~G.}\ \bibnamefont
  {Marcus}}, \bibinfo {author} {\bibfnamefont {D.-J.}\ \bibnamefont {Kim}},
  \bibinfo {author} {\bibfnamefont {J.~A.}\ \bibnamefont {Tutmaher}}, \bibinfo
  {author} {\bibfnamefont {J.~A.}\ \bibnamefont {Rodriguez-Rivera}}, \bibinfo
  {author} {\bibfnamefont {J.~O.}\ \bibnamefont {Birk}}, \bibinfo {author}
  {\bibfnamefont {C.}~\bibnamefont {Niedermeyer}}, \bibinfo {author}
  {\bibfnamefont {H.}~\bibnamefont {Lee}}, \bibinfo {author} {\bibfnamefont
  {Z.}~\bibnamefont {Fisk}},\ and\ \bibinfo {author} {\bibfnamefont {C.~L.}\
  \bibnamefont {Broholm}},\ }\bibfield  {title} {\bibinfo {title} {Multi-$q$
  mesoscale magnetism in ${\mathrm{ceausb}}_{2}$},\ }\href
  {https://doi.org/10.1103/PhysRevLett.120.097201} {\bibfield  {journal}
  {\bibinfo  {journal} {Phys. Rev. Lett.}\ }\textbf {\bibinfo {volume} {120}},\
  \bibinfo {pages} {097201} (\bibinfo {year} {2018})}\BibitemShut {NoStop}%
\bibitem [{\citenamefont {Seo}\ \emph {et~al.}(2020)\citenamefont {Seo},
  \citenamefont {Wang}, \citenamefont {Thomas}, \citenamefont {Rahn},
  \citenamefont {Carmo}, \citenamefont {Ronning}, \citenamefont {Bauer},
  \citenamefont {dos Reis}, \citenamefont {Janoschek}, \citenamefont
  {Thompson}, \citenamefont {Fernandes},\ and\ \citenamefont
  {Rosa}}]{Seo_PhysRevX.10.011035}%
  \BibitemOpen
  \bibfield  {author} {\bibinfo {author} {\bibfnamefont {S.}~\bibnamefont
  {Seo}}, \bibinfo {author} {\bibfnamefont {X.}~\bibnamefont {Wang}}, \bibinfo
  {author} {\bibfnamefont {S.~M.}\ \bibnamefont {Thomas}}, \bibinfo {author}
  {\bibfnamefont {M.~C.}\ \bibnamefont {Rahn}}, \bibinfo {author}
  {\bibfnamefont {D.}~\bibnamefont {Carmo}}, \bibinfo {author} {\bibfnamefont
  {F.}~\bibnamefont {Ronning}}, \bibinfo {author} {\bibfnamefont {E.~D.}\
  \bibnamefont {Bauer}}, \bibinfo {author} {\bibfnamefont {R.~D.}\ \bibnamefont
  {dos Reis}}, \bibinfo {author} {\bibfnamefont {M.}~\bibnamefont {Janoschek}},
  \bibinfo {author} {\bibfnamefont {J.~D.}\ \bibnamefont {Thompson}}, \bibinfo
  {author} {\bibfnamefont {R.~M.}\ \bibnamefont {Fernandes}},\ and\ \bibinfo
  {author} {\bibfnamefont {P.~F.~S.}\ \bibnamefont {Rosa}},\ }\bibfield
  {title} {\bibinfo {title} {Nematic state in ${\mathrm{ceausb}}_{2}$},\ }\href
  {https://doi.org/10.1103/PhysRevX.10.011035} {\bibfield  {journal} {\bibinfo
  {journal} {Phys. Rev. X}\ }\textbf {\bibinfo {volume} {10}},\ \bibinfo
  {pages} {011035} (\bibinfo {year} {2020})}\BibitemShut {NoStop}%
\bibitem [{\citenamefont {Seo}\ \emph {et~al.}(2021)\citenamefont {Seo},
  \citenamefont {Hayami}, \citenamefont {Su}, \citenamefont {Thomas},
  \citenamefont {Ronning}, \citenamefont {Bauer}, \citenamefont {Thompson},
  \citenamefont {Lin},\ and\ \citenamefont {Rosa}}]{seo2021spin}%
  \BibitemOpen
  \bibfield  {author} {\bibinfo {author} {\bibfnamefont {S.}~\bibnamefont
  {Seo}}, \bibinfo {author} {\bibfnamefont {S.}~\bibnamefont {Hayami}},
  \bibinfo {author} {\bibfnamefont {Y.}~\bibnamefont {Su}}, \bibinfo {author}
  {\bibfnamefont {S.~M.}\ \bibnamefont {Thomas}}, \bibinfo {author}
  {\bibfnamefont {F.}~\bibnamefont {Ronning}}, \bibinfo {author} {\bibfnamefont
  {E.~D.}\ \bibnamefont {Bauer}}, \bibinfo {author} {\bibfnamefont {J.~D.}\
  \bibnamefont {Thompson}}, \bibinfo {author} {\bibfnamefont {S.-Z.}\
  \bibnamefont {Lin}},\ and\ \bibinfo {author} {\bibfnamefont {P.~F.}\
  \bibnamefont {Rosa}},\ }\bibfield  {title} {\bibinfo {title}
  {Spin-texture-driven electrical transport in multi-q antiferromagnets},\
  }\href@noop {} {\bibfield  {journal} {\bibinfo  {journal} {Commun. Phys.}\
  }\textbf {\bibinfo {volume} {4}},\ \bibinfo {pages} {58} (\bibinfo {year}
  {2021})}\BibitemShut {NoStop}%
\bibitem [{\citenamefont {Katsura}\ \emph {et~al.}(2005)\citenamefont
  {Katsura}, \citenamefont {Nagaosa},\ and\ \citenamefont
  {Balatsky}}]{Katsura_PhysRevLett.95.057205}%
  \BibitemOpen
  \bibfield  {author} {\bibinfo {author} {\bibfnamefont {H.}~\bibnamefont
  {Katsura}}, \bibinfo {author} {\bibfnamefont {N.}~\bibnamefont {Nagaosa}},\
  and\ \bibinfo {author} {\bibfnamefont {A.~V.}\ \bibnamefont {Balatsky}},\
  }\bibfield  {title} {\bibinfo {title} {Spin current and magnetoelectric
  effect in noncollinear magnets},\ }\href
  {https://doi.org/10.1103/PhysRevLett.95.057205} {\bibfield  {journal}
  {\bibinfo  {journal} {Phys. Rev. Lett.}\ }\textbf {\bibinfo {volume} {95}},\
  \bibinfo {pages} {057205} (\bibinfo {year} {2005})}\BibitemShut {NoStop}%
\bibitem [{\citenamefont {Mostovoy}(2006)}]{Mostovoy_PhysRevLett.96.067601}%
  \BibitemOpen
  \bibfield  {author} {\bibinfo {author} {\bibfnamefont {M.}~\bibnamefont
  {Mostovoy}},\ }\bibfield  {title} {\bibinfo {title} {Ferroelectricity in
  spiral magnets},\ }\href {https://doi.org/10.1103/PhysRevLett.96.067601}
  {\bibfield  {journal} {\bibinfo  {journal} {Phys. Rev. Lett.}\ }\textbf
  {\bibinfo {volume} {96}},\ \bibinfo {pages} {067601} (\bibinfo {year}
  {2006})}\BibitemShut {NoStop}%
\bibitem [{\citenamefont {Tokura}\ \emph {et~al.}(2014)\citenamefont {Tokura},
  \citenamefont {Seki},\ and\ \citenamefont
  {Nagaosa}}]{tokura2014multiferroics}%
  \BibitemOpen
  \bibfield  {author} {\bibinfo {author} {\bibfnamefont {Y.}~\bibnamefont
  {Tokura}}, \bibinfo {author} {\bibfnamefont {S.}~\bibnamefont {Seki}},\ and\
  \bibinfo {author} {\bibfnamefont {N.}~\bibnamefont {Nagaosa}},\ }\bibfield
  {title} {\bibinfo {title} {Multiferroics of spin origin},\ }\href@noop {}
  {\bibfield  {journal} {\bibinfo  {journal} {Rep. Prog. Phys.}\ }\textbf
  {\bibinfo {volume} {77}},\ \bibinfo {pages} {076501} (\bibinfo {year}
  {2014})}\BibitemShut {NoStop}%
\bibitem [{\citenamefont {Rashba}(1960)}]{rashba1960properties}%
  \BibitemOpen
  \bibfield  {author} {\bibinfo {author} {\bibfnamefont {E.~I.}\ \bibnamefont
  {Rashba}},\ }\bibfield  {title} {\bibinfo {title} {Properties of
  semiconductors with an extremum loop. 1. cyclotron and combinational
  resonance in a magnetic field perpendicular to the plane of the loop},\
  }\href@noop {} {\bibfield  {journal} {\bibinfo  {journal} {Sov. Phys. Solid
  State}\ }\textbf {\bibinfo {volume} {2}},\ \bibinfo {pages} {1109} (\bibinfo
  {year} {1960})}\BibitemShut {NoStop}%
\bibitem [{\citenamefont {Bychkov}\ and\ \citenamefont
  {Rashba}(1984)}]{bychkov1984oscillatory}%
  \BibitemOpen
  \bibfield  {author} {\bibinfo {author} {\bibfnamefont {Y.~A.}\ \bibnamefont
  {Bychkov}}\ and\ \bibinfo {author} {\bibfnamefont {E.~I.}\ \bibnamefont
  {Rashba}},\ }\bibfield  {title} {\bibinfo {title} {Oscillatory effects and
  the magnetic susceptibility of carriers in inversion layers},\ }\href@noop {}
  {\bibfield  {journal} {\bibinfo  {journal} {J. Phys. C: Solid State Phys.}\
  }\textbf {\bibinfo {volume} {17}},\ \bibinfo {pages} {6039} (\bibinfo {year}
  {1984})}\BibitemShut {NoStop}%
\bibitem [{\citenamefont {Lounis}(2020)}]{lounis2020multiple}%
  \BibitemOpen
  \bibfield  {author} {\bibinfo {author} {\bibfnamefont {S.}~\bibnamefont
  {Lounis}},\ }\bibfield  {title} {\bibinfo {title} {Multiple-scattering
  approach for multi-spin chiral magnetic interactions: application to the
  one-and two-dimensional rashba electron gas},\ }\href@noop {} {\bibfield
  {journal} {\bibinfo  {journal} {New J. Phys.}\ }\textbf {\bibinfo {volume}
  {22}},\ \bibinfo {pages} {103003} (\bibinfo {year} {2020})}\BibitemShut
  {NoStop}%
\bibitem [{\citenamefont {Shibuya}\ \emph {et~al.}(2016)\citenamefont
  {Shibuya}, \citenamefont {Matsuura},\ and\ \citenamefont
  {Ogata}}]{shibuya2016magnetic}%
  \BibitemOpen
  \bibfield  {author} {\bibinfo {author} {\bibfnamefont {T.}~\bibnamefont
  {Shibuya}}, \bibinfo {author} {\bibfnamefont {H.}~\bibnamefont {Matsuura}},\
  and\ \bibinfo {author} {\bibfnamefont {M.}~\bibnamefont {Ogata}},\ }\bibfield
   {title} {\bibinfo {title} {Magnetic chirality induced from
  ruderman--kittel--kasuya--yosida interaction at an interface of a
  ferromagnet/heavy metal heterostructure},\ }\href@noop {} {\bibfield
  {journal} {\bibinfo  {journal} {J. Phys. Soc. Jpn.}\ }\textbf {\bibinfo
  {volume} {85}},\ \bibinfo {pages} {114701} (\bibinfo {year}
  {2016})}\BibitemShut {NoStop}%
\bibitem [{\citenamefont {Yambe}\ and\ \citenamefont
  {Hayami}(2021)}]{yambe2021skyrmion}%
  \BibitemOpen
  \bibfield  {author} {\bibinfo {author} {\bibfnamefont {R.}~\bibnamefont
  {Yambe}}\ and\ \bibinfo {author} {\bibfnamefont {S.}~\bibnamefont {Hayami}},\
  }\bibfield  {title} {\bibinfo {title} {Skyrmion crystals in centrosymmetric
  itinerant magnets without horizontal mirror plane},\ }\href@noop {}
  {\bibfield  {journal} {\bibinfo  {journal} {Sci. Rep.}\ }\textbf {\bibinfo
  {volume} {11}},\ \bibinfo {pages} {11184} (\bibinfo {year}
  {2021})}\BibitemShut {NoStop}%
\bibitem [{\citenamefont {Okada}\ \emph {et~al.}(2018)\citenamefont {Okada},
  \citenamefont {Kato},\ and\ \citenamefont
  {Motome}}]{Okada_PhysRevB.98.224406}%
  \BibitemOpen
  \bibfield  {author} {\bibinfo {author} {\bibfnamefont {K.~N.}\ \bibnamefont
  {Okada}}, \bibinfo {author} {\bibfnamefont {Y.}~\bibnamefont {Kato}},\ and\
  \bibinfo {author} {\bibfnamefont {Y.}~\bibnamefont {Motome}},\ }\bibfield
  {title} {\bibinfo {title} {Multiple-$q$ magnetic orders in rashba-dresselhaus
  metals},\ }\href {https://doi.org/10.1103/PhysRevB.98.224406} {\bibfield
  {journal} {\bibinfo  {journal} {Phys. Rev. B}\ }\textbf {\bibinfo {volume}
  {98}},\ \bibinfo {pages} {224406} (\bibinfo {year} {2018})}\BibitemShut
  {NoStop}%
\bibitem [{\citenamefont {Takagi}\ \emph {et~al.}(2018)\citenamefont {Takagi},
  \citenamefont {White}, \citenamefont {Hayami}, \citenamefont {Arita},
  \citenamefont {Honecker}, \citenamefont {R{\o}nnow}, \citenamefont {Tokura},\
  and\ \citenamefont {Seki}}]{takagi2018multiple}%
  \BibitemOpen
  \bibfield  {author} {\bibinfo {author} {\bibfnamefont {R.}~\bibnamefont
  {Takagi}}, \bibinfo {author} {\bibfnamefont {J.}~\bibnamefont {White}},
  \bibinfo {author} {\bibfnamefont {S.}~\bibnamefont {Hayami}}, \bibinfo
  {author} {\bibfnamefont {R.}~\bibnamefont {Arita}}, \bibinfo {author}
  {\bibfnamefont {D.}~\bibnamefont {Honecker}}, \bibinfo {author}
  {\bibfnamefont {H.}~\bibnamefont {R{\o}nnow}}, \bibinfo {author}
  {\bibfnamefont {Y.}~\bibnamefont {Tokura}},\ and\ \bibinfo {author}
  {\bibfnamefont {S.}~\bibnamefont {Seki}},\ }\bibfield  {title} {\bibinfo
  {title} {Multiple-q noncollinear magnetism in an itinerant hexagonal
  magnet},\ }\href@noop {} {\bibfield  {journal} {\bibinfo  {journal} {Science
  advances}\ }\textbf {\bibinfo {volume} {4}},\ \bibinfo {pages} {eaau3402}
  (\bibinfo {year} {2018})}\BibitemShut {NoStop}%
\bibitem [{\citenamefont {Kakihana}\ \emph {et~al.}(2018)\citenamefont
  {Kakihana}, \citenamefont {Aoki}, \citenamefont {Nakamura}, \citenamefont
  {Honda}, \citenamefont {Nakashima}, \citenamefont {Amako}, \citenamefont
  {Nakamura}, \citenamefont {Sakakibara}, \citenamefont {Hedo}, \citenamefont
  {Nakama},\ and\ \citenamefont {Onuki}}]{kakihana2018giant}%
  \BibitemOpen
  \bibfield  {author} {\bibinfo {author} {\bibfnamefont {M.}~\bibnamefont
  {Kakihana}}, \bibinfo {author} {\bibfnamefont {D.}~\bibnamefont {Aoki}},
  \bibinfo {author} {\bibfnamefont {A.}~\bibnamefont {Nakamura}}, \bibinfo
  {author} {\bibfnamefont {F.}~\bibnamefont {Honda}}, \bibinfo {author}
  {\bibfnamefont {M.}~\bibnamefont {Nakashima}}, \bibinfo {author}
  {\bibfnamefont {Y.}~\bibnamefont {Amako}}, \bibinfo {author} {\bibfnamefont
  {S.}~\bibnamefont {Nakamura}}, \bibinfo {author} {\bibfnamefont
  {T.}~\bibnamefont {Sakakibara}}, \bibinfo {author} {\bibfnamefont
  {M.}~\bibnamefont {Hedo}}, \bibinfo {author} {\bibfnamefont {T.}~\bibnamefont
  {Nakama}},\ and\ \bibinfo {author} {\bibfnamefont {Y.}~\bibnamefont
  {Onuki}},\ }\bibfield  {title} {\bibinfo {title} {Giant hall resistivity and
  magnetoresistance in cubic chiral antiferromagnet euptsi},\ }\href@noop {}
  {\bibfield  {journal} {\bibinfo  {journal} {J. Phys. Soc. Jpn.}\ }\textbf
  {\bibinfo {volume} {87}},\ \bibinfo {pages} {023701} (\bibinfo {year}
  {2018})}\BibitemShut {NoStop}%
\bibitem [{\citenamefont {Kaneko}\ \emph {et~al.}(2019)\citenamefont {Kaneko},
  \citenamefont {Frontzek}, \citenamefont {Matsuda}, \citenamefont {Nakao},
  \citenamefont {Munakata}, \citenamefont {Ohhara}, \citenamefont {Kakihana},
  \citenamefont {Haga}, \citenamefont {Hedo}, \citenamefont {Nakama},\ and\
  \citenamefont {Onuki}}]{kaneko2019unique}%
  \BibitemOpen
  \bibfield  {author} {\bibinfo {author} {\bibfnamefont {K.}~\bibnamefont
  {Kaneko}}, \bibinfo {author} {\bibfnamefont {M.~D.}\ \bibnamefont
  {Frontzek}}, \bibinfo {author} {\bibfnamefont {M.}~\bibnamefont {Matsuda}},
  \bibinfo {author} {\bibfnamefont {A.}~\bibnamefont {Nakao}}, \bibinfo
  {author} {\bibfnamefont {K.}~\bibnamefont {Munakata}}, \bibinfo {author}
  {\bibfnamefont {T.}~\bibnamefont {Ohhara}}, \bibinfo {author} {\bibfnamefont
  {M.}~\bibnamefont {Kakihana}}, \bibinfo {author} {\bibfnamefont
  {Y.}~\bibnamefont {Haga}}, \bibinfo {author} {\bibfnamefont {M.}~\bibnamefont
  {Hedo}}, \bibinfo {author} {\bibfnamefont {T.}~\bibnamefont {Nakama}},\ and\
  \bibinfo {author} {\bibfnamefont {Y.}~\bibnamefont {Onuki}},\ }\bibfield
  {title} {\bibinfo {title} {Unique helical magnetic order and field-induced
  phase in trillium lattice antiferromagnet euptsi},\ }\href@noop {} {\bibfield
   {journal} {\bibinfo  {journal} {J. Phys. Soc. Jpn.}\ }\textbf {\bibinfo
  {volume} {88}},\ \bibinfo {pages} {013702} (\bibinfo {year}
  {2019})}\BibitemShut {NoStop}%
\bibitem [{\citenamefont {Kakihana}\ \emph {et~al.}(2019)\citenamefont
  {Kakihana}, \citenamefont {Aoki}, \citenamefont {Nakamura}, \citenamefont
  {Honda}, \citenamefont {Nakashima}, \citenamefont {Amako}, \citenamefont
  {Takeuchi}, \citenamefont {Harima}, \citenamefont {Hedo}, \citenamefont
  {Nakama},\ and\ \citenamefont {Onuki}}]{kakihana2019unique}%
  \BibitemOpen
  \bibfield  {author} {\bibinfo {author} {\bibfnamefont {M.}~\bibnamefont
  {Kakihana}}, \bibinfo {author} {\bibfnamefont {D.}~\bibnamefont {Aoki}},
  \bibinfo {author} {\bibfnamefont {A.}~\bibnamefont {Nakamura}}, \bibinfo
  {author} {\bibfnamefont {F.}~\bibnamefont {Honda}}, \bibinfo {author}
  {\bibfnamefont {M.}~\bibnamefont {Nakashima}}, \bibinfo {author}
  {\bibfnamefont {Y.}~\bibnamefont {Amako}}, \bibinfo {author} {\bibfnamefont
  {T.}~\bibnamefont {Takeuchi}}, \bibinfo {author} {\bibfnamefont
  {H.}~\bibnamefont {Harima}}, \bibinfo {author} {\bibfnamefont
  {M.}~\bibnamefont {Hedo}}, \bibinfo {author} {\bibfnamefont {T.}~\bibnamefont
  {Nakama}},\ and\ \bibinfo {author} {\bibfnamefont {Y.}~\bibnamefont
  {Onuki}},\ }\bibfield  {title} {\bibinfo {title} {Unique magnetic phases in
  the skyrmion lattice and fermi surface properties in cubic chiral
  antiferromagnet euptsi},\ }\href@noop {} {\bibfield  {journal} {\bibinfo
  {journal} {J. Phys. Soc. Jpn.}\ }\textbf {\bibinfo {volume} {88}},\ \bibinfo
  {pages} {094705} (\bibinfo {year} {2019})}\BibitemShut {NoStop}%
\bibitem [{\citenamefont {Tabata}\ \emph {et~al.}(2019)\citenamefont {Tabata},
  \citenamefont {Matsumura}, \citenamefont {Nakao}, \citenamefont {Michimura},
  \citenamefont {Kakihana}, \citenamefont {Inami}, \citenamefont {Kaneko},
  \citenamefont {Hedo}, \citenamefont {Nakama},\ and\ \citenamefont
  {{\=O}nuki}}]{tabata2019magnetic}%
  \BibitemOpen
  \bibfield  {author} {\bibinfo {author} {\bibfnamefont {C.}~\bibnamefont
  {Tabata}}, \bibinfo {author} {\bibfnamefont {T.}~\bibnamefont {Matsumura}},
  \bibinfo {author} {\bibfnamefont {H.}~\bibnamefont {Nakao}}, \bibinfo
  {author} {\bibfnamefont {S.}~\bibnamefont {Michimura}}, \bibinfo {author}
  {\bibfnamefont {M.}~\bibnamefont {Kakihana}}, \bibinfo {author}
  {\bibfnamefont {T.}~\bibnamefont {Inami}}, \bibinfo {author} {\bibfnamefont
  {K.}~\bibnamefont {Kaneko}}, \bibinfo {author} {\bibfnamefont
  {M.}~\bibnamefont {Hedo}}, \bibinfo {author} {\bibfnamefont {T.}~\bibnamefont
  {Nakama}},\ and\ \bibinfo {author} {\bibfnamefont {Y.}~\bibnamefont
  {{\=O}nuki}},\ }\bibfield  {title} {\bibinfo {title} {Magnetic field induced
  triple-q magnetic order in trillium lattice antiferromagnet euptsi studied by
  resonant x-ray scattering},\ }\href@noop {} {\bibfield  {journal} {\bibinfo
  {journal} {J. Phys. Soc. Jpn.}\ }\textbf {\bibinfo {volume} {88}},\ \bibinfo
  {pages} {093704} (\bibinfo {year} {2019})}\BibitemShut {NoStop}%
\bibitem [{\citenamefont {Hayami}\ and\ \citenamefont
  {Yambe}(2021{\natexlab{b}})}]{hayami2021field}%
  \BibitemOpen
  \bibfield  {author} {\bibinfo {author} {\bibfnamefont {S.}~\bibnamefont
  {Hayami}}\ and\ \bibinfo {author} {\bibfnamefont {R.}~\bibnamefont {Yambe}},\
  }\bibfield  {title} {\bibinfo {title} {Field-direction sensitive skyrmion
  crystals in cubic chiral systems: Implication to 4 f-electron compound
  euptsi},\ }\href@noop {} {\bibfield  {journal} {\bibinfo  {journal} {J. Phys.
  Soc. Jpn.}\ }\textbf {\bibinfo {volume} {90}},\ \bibinfo {pages} {073705}
  (\bibinfo {year} {2021}{\natexlab{b}})}\BibitemShut {NoStop}%
\bibitem [{\citenamefont {Nakamura}\ \emph {et~al.}(2018)\citenamefont
  {Nakamura}, \citenamefont {Kabeya}, \citenamefont {Kobayashi}, \citenamefont
  {Araki}, \citenamefont {Katoh},\ and\ \citenamefont
  {Ochiai}}]{Nakamura_PhysRevB.98.054410}%
  \BibitemOpen
  \bibfield  {author} {\bibinfo {author} {\bibfnamefont {S.}~\bibnamefont
  {Nakamura}}, \bibinfo {author} {\bibfnamefont {N.}~\bibnamefont {Kabeya}},
  \bibinfo {author} {\bibfnamefont {M.}~\bibnamefont {Kobayashi}}, \bibinfo
  {author} {\bibfnamefont {K.}~\bibnamefont {Araki}}, \bibinfo {author}
  {\bibfnamefont {K.}~\bibnamefont {Katoh}},\ and\ \bibinfo {author}
  {\bibfnamefont {A.}~\bibnamefont {Ochiai}},\ }\bibfield  {title} {\bibinfo
  {title} {Spin trimer formation in the metallic compound
  ${\mathrm{gd}}_{3}{\mathrm{ru}}_{4}{\mathrm{al}}_{12}$ with a distorted
  kagome lattice structure},\ }\href
  {https://doi.org/10.1103/PhysRevB.98.054410} {\bibfield  {journal} {\bibinfo
  {journal} {Phys. Rev. B}\ }\textbf {\bibinfo {volume} {98}},\ \bibinfo
  {pages} {054410} (\bibinfo {year} {2018})}\BibitemShut {NoStop}%
\bibitem [{\citenamefont {Hirschberger}\ \emph {et~al.}(2019)\citenamefont
  {Hirschberger}, \citenamefont {Nakajima}, \citenamefont {Gao}, \citenamefont
  {Peng}, \citenamefont {Kikkawa}, \citenamefont {Kurumaji}, \citenamefont
  {Kriener}, \citenamefont {Yamasaki}, \citenamefont {Sagayama}, \citenamefont
  {Nakao}, \citenamefont {Ohishi}, \citenamefont {Kakurai}, \citenamefont
  {Taguchi}, \citenamefont {Yu}, \citenamefont {Arima},\ and\ \citenamefont
  {Tokura}}]{hirschberger2019skyrmion}%
  \BibitemOpen
  \bibfield  {author} {\bibinfo {author} {\bibfnamefont {M.}~\bibnamefont
  {Hirschberger}}, \bibinfo {author} {\bibfnamefont {T.}~\bibnamefont
  {Nakajima}}, \bibinfo {author} {\bibfnamefont {S.}~\bibnamefont {Gao}},
  \bibinfo {author} {\bibfnamefont {L.}~\bibnamefont {Peng}}, \bibinfo {author}
  {\bibfnamefont {A.}~\bibnamefont {Kikkawa}}, \bibinfo {author} {\bibfnamefont
  {T.}~\bibnamefont {Kurumaji}}, \bibinfo {author} {\bibfnamefont
  {M.}~\bibnamefont {Kriener}}, \bibinfo {author} {\bibfnamefont
  {Y.}~\bibnamefont {Yamasaki}}, \bibinfo {author} {\bibfnamefont
  {H.}~\bibnamefont {Sagayama}}, \bibinfo {author} {\bibfnamefont
  {H.}~\bibnamefont {Nakao}}, \bibinfo {author} {\bibfnamefont
  {K.}~\bibnamefont {Ohishi}}, \bibinfo {author} {\bibfnamefont
  {K.}~\bibnamefont {Kakurai}}, \bibinfo {author} {\bibfnamefont
  {Y.}~\bibnamefont {Taguchi}}, \bibinfo {author} {\bibfnamefont
  {X.}~\bibnamefont {Yu}}, \bibinfo {author} {\bibfnamefont {T.-h.}\
  \bibnamefont {Arima}},\ and\ \bibinfo {author} {\bibfnamefont
  {Y.}~\bibnamefont {Tokura}},\ }\bibfield  {title} {\bibinfo {title} {Skyrmion
  phase and competing magnetic orders on a breathing kagome lattice},\
  }\href@noop {} {\bibfield  {journal} {\bibinfo  {journal} {Nat. Commun.}\
  }\textbf {\bibinfo {volume} {10}},\ \bibinfo {pages} {5831} (\bibinfo {year}
  {2019})}\BibitemShut {NoStop}%
\bibitem [{\citenamefont {Hirschberger}\ \emph {et~al.}(2021)\citenamefont
  {Hirschberger}, \citenamefont {Hayami},\ and\ \citenamefont
  {Tokura}}]{Hirschberger_10.1088/1367-2630/abdef9}%
  \BibitemOpen
  \bibfield  {author} {\bibinfo {author} {\bibfnamefont {M.}~\bibnamefont
  {Hirschberger}}, \bibinfo {author} {\bibfnamefont {S.}~\bibnamefont
  {Hayami}},\ and\ \bibinfo {author} {\bibfnamefont {Y.}~\bibnamefont
  {Tokura}},\ }\bibfield  {title} {\bibinfo {title} {Nanometric skyrmion
  lattice from anisotropic exchange interactions in a centrosymmetric host},\
  }\href@noop {} {\bibfield  {journal} {\bibinfo  {journal} {New J. Phys.}\
  }\textbf {\bibinfo {volume} {23}},\ \bibinfo {pages} {023039} (\bibinfo
  {year} {2021})}\BibitemShut {NoStop}%
\bibitem [{\citenamefont {Tanigaki}\ \emph {et~al.}(2015)\citenamefont
  {Tanigaki}, \citenamefont {Shibata}, \citenamefont {Kanazawa}, \citenamefont
  {Yu}, \citenamefont {Onose}, \citenamefont {Park}, \citenamefont {Shindo},\
  and\ \citenamefont {Tokura}}]{tanigaki2015real}%
  \BibitemOpen
  \bibfield  {author} {\bibinfo {author} {\bibfnamefont {T.}~\bibnamefont
  {Tanigaki}}, \bibinfo {author} {\bibfnamefont {K.}~\bibnamefont {Shibata}},
  \bibinfo {author} {\bibfnamefont {N.}~\bibnamefont {Kanazawa}}, \bibinfo
  {author} {\bibfnamefont {X.}~\bibnamefont {Yu}}, \bibinfo {author}
  {\bibfnamefont {Y.}~\bibnamefont {Onose}}, \bibinfo {author} {\bibfnamefont
  {H.~S.}\ \bibnamefont {Park}}, \bibinfo {author} {\bibfnamefont
  {D.}~\bibnamefont {Shindo}},\ and\ \bibinfo {author} {\bibfnamefont
  {Y.}~\bibnamefont {Tokura}},\ }\bibfield  {title} {\bibinfo {title}
  {Real-space observation of short-period cubic lattice of skyrmions in mnge},\
  }\href@noop {} {\bibfield  {journal} {\bibinfo  {journal} {Nano Lett.}\
  }\textbf {\bibinfo {volume} {15}},\ \bibinfo {pages} {5438} (\bibinfo {year}
  {2015})}\BibitemShut {NoStop}%
\bibitem [{\citenamefont {Kanazawa}\ \emph {et~al.}(2017)\citenamefont
  {Kanazawa}, \citenamefont {Seki},\ and\ \citenamefont
  {Tokura}}]{kanazawa2017noncentrosymmetric}%
  \BibitemOpen
  \bibfield  {author} {\bibinfo {author} {\bibfnamefont {N.}~\bibnamefont
  {Kanazawa}}, \bibinfo {author} {\bibfnamefont {S.}~\bibnamefont {Seki}},\
  and\ \bibinfo {author} {\bibfnamefont {Y.}~\bibnamefont {Tokura}},\
  }\bibfield  {title} {\bibinfo {title} {Noncentrosymmetric magnets hosting
  magnetic skyrmions},\ }\href@noop {} {\bibfield  {journal} {\bibinfo
  {journal} {Adv. Mater.}\ }\textbf {\bibinfo {volume} {29}},\ \bibinfo {pages}
  {1603227} (\bibinfo {year} {2017})}\BibitemShut {NoStop}%
\bibitem [{\citenamefont {Fujishiro}\ \emph {et~al.}(2019)\citenamefont
  {Fujishiro}, \citenamefont {Kanazawa}, \citenamefont {Nakajima},
  \citenamefont {Yu}, \citenamefont {Ohishi}, \citenamefont {Kawamura},
  \citenamefont {Kakurai}, \citenamefont {Arima}, \citenamefont {Mitamura},
  \citenamefont {Miyake}, \citenamefont {Akiba}, \citenamefont {Tokunaga},
  \citenamefont {Matsuo}, \citenamefont {Kindo}, \citenamefont {Koretsune},
  \citenamefont {Arita},\ and\ \citenamefont
  {Tokura}}]{fujishiro2019topological}%
  \BibitemOpen
  \bibfield  {author} {\bibinfo {author} {\bibfnamefont {Y.}~\bibnamefont
  {Fujishiro}}, \bibinfo {author} {\bibfnamefont {N.}~\bibnamefont {Kanazawa}},
  \bibinfo {author} {\bibfnamefont {T.}~\bibnamefont {Nakajima}}, \bibinfo
  {author} {\bibfnamefont {X.~Z.}\ \bibnamefont {Yu}}, \bibinfo {author}
  {\bibfnamefont {K.}~\bibnamefont {Ohishi}}, \bibinfo {author} {\bibfnamefont
  {Y.}~\bibnamefont {Kawamura}}, \bibinfo {author} {\bibfnamefont
  {K.}~\bibnamefont {Kakurai}}, \bibinfo {author} {\bibfnamefont
  {T.}~\bibnamefont {Arima}}, \bibinfo {author} {\bibfnamefont
  {H.}~\bibnamefont {Mitamura}}, \bibinfo {author} {\bibfnamefont
  {A.}~\bibnamefont {Miyake}}, \bibinfo {author} {\bibfnamefont
  {K.}~\bibnamefont {Akiba}}, \bibinfo {author} {\bibfnamefont
  {M.}~\bibnamefont {Tokunaga}}, \bibinfo {author} {\bibfnamefont
  {A.}~\bibnamefont {Matsuo}}, \bibinfo {author} {\bibfnamefont
  {K.}~\bibnamefont {Kindo}}, \bibinfo {author} {\bibfnamefont
  {T.}~\bibnamefont {Koretsune}}, \bibinfo {author} {\bibfnamefont
  {R.}~\bibnamefont {Arita}},\ and\ \bibinfo {author} {\bibfnamefont
  {Y.}~\bibnamefont {Tokura}},\ }\bibfield  {title} {\bibinfo {title}
  {Topological transitions among skyrmion-and hedgehog-lattice states in cubic
  chiral magnets},\ }\href@noop {} {\bibfield  {journal} {\bibinfo  {journal}
  {Nat. Commun.}\ }\textbf {\bibinfo {volume} {10}},\ \bibinfo {pages} {1059}
  (\bibinfo {year} {2019})}\BibitemShut {NoStop}%
\bibitem [{\citenamefont {Kanazawa}\ \emph {et~al.}(2020)\citenamefont
  {Kanazawa}, \citenamefont {Kitaori}, \citenamefont {White}, \citenamefont
  {Ukleev}, \citenamefont {R\o{}nnow}, \citenamefont {Tsukazaki}, \citenamefont
  {Ichikawa}, \citenamefont {Kawasaki},\ and\ \citenamefont
  {Tokura}}]{Kanazawa_PhysRevLett.125.137202}%
  \BibitemOpen
  \bibfield  {author} {\bibinfo {author} {\bibfnamefont {N.}~\bibnamefont
  {Kanazawa}}, \bibinfo {author} {\bibfnamefont {A.}~\bibnamefont {Kitaori}},
  \bibinfo {author} {\bibfnamefont {J.~S.}\ \bibnamefont {White}}, \bibinfo
  {author} {\bibfnamefont {V.}~\bibnamefont {Ukleev}}, \bibinfo {author}
  {\bibfnamefont {H.~M.}\ \bibnamefont {R\o{}nnow}}, \bibinfo {author}
  {\bibfnamefont {A.}~\bibnamefont {Tsukazaki}}, \bibinfo {author}
  {\bibfnamefont {M.}~\bibnamefont {Ichikawa}}, \bibinfo {author}
  {\bibfnamefont {M.}~\bibnamefont {Kawasaki}},\ and\ \bibinfo {author}
  {\bibfnamefont {Y.}~\bibnamefont {Tokura}},\ }\bibfield  {title} {\bibinfo
  {title} {Direct observation of the statics and dynamics of emergent magnetic
  monopoles in a chiral magnet},\ }\href
  {https://doi.org/10.1103/PhysRevLett.125.137202} {\bibfield  {journal}
  {\bibinfo  {journal} {Phys. Rev. Lett.}\ }\textbf {\bibinfo {volume} {125}},\
  \bibinfo {pages} {137202} (\bibinfo {year} {2020})}\BibitemShut {NoStop}%
\bibitem [{\citenamefont {Grytsiuk}\ \emph {et~al.}(2020)\citenamefont
  {Grytsiuk}, \citenamefont {Hanke}, \citenamefont {Hoffmann}, \citenamefont
  {Bouaziz}, \citenamefont {Gomonay}, \citenamefont {Bihlmayer}, \citenamefont
  {Lounis}, \citenamefont {Mokrousov},\ and\ \citenamefont
  {Bl{\"u}gel}}]{grytsiuk2020topological}%
  \BibitemOpen
  \bibfield  {author} {\bibinfo {author} {\bibfnamefont {S.}~\bibnamefont
  {Grytsiuk}}, \bibinfo {author} {\bibfnamefont {J.-P.}\ \bibnamefont {Hanke}},
  \bibinfo {author} {\bibfnamefont {M.}~\bibnamefont {Hoffmann}}, \bibinfo
  {author} {\bibfnamefont {J.}~\bibnamefont {Bouaziz}}, \bibinfo {author}
  {\bibfnamefont {O.}~\bibnamefont {Gomonay}}, \bibinfo {author} {\bibfnamefont
  {G.}~\bibnamefont {Bihlmayer}}, \bibinfo {author} {\bibfnamefont
  {S.}~\bibnamefont {Lounis}}, \bibinfo {author} {\bibfnamefont
  {Y.}~\bibnamefont {Mokrousov}},\ and\ \bibinfo {author} {\bibfnamefont
  {S.}~\bibnamefont {Bl{\"u}gel}},\ }\bibfield  {title} {\bibinfo {title}
  {Topological--chiral magnetic interactions driven by emergent orbital
  magnetism},\ }\href@noop {} {\bibfield  {journal} {\bibinfo  {journal} {Nat.
  Commun.}\ }\textbf {\bibinfo {volume} {11}},\ \bibinfo {pages} {511}
  (\bibinfo {year} {2020})}\BibitemShut {NoStop}%
\bibitem [{\citenamefont {Ishiwata}\ \emph {et~al.}(2011)\citenamefont
  {Ishiwata}, \citenamefont {Tokunaga}, \citenamefont {Kaneko}, \citenamefont
  {Okuyama}, \citenamefont {Tokunaga}, \citenamefont {Wakimoto}, \citenamefont
  {Kakurai}, \citenamefont {Arima}, \citenamefont {Taguchi},\ and\
  \citenamefont {Tokura}}]{Ishiwata_PhysRevB.84.054427}%
  \BibitemOpen
  \bibfield  {author} {\bibinfo {author} {\bibfnamefont {S.}~\bibnamefont
  {Ishiwata}}, \bibinfo {author} {\bibfnamefont {M.}~\bibnamefont {Tokunaga}},
  \bibinfo {author} {\bibfnamefont {Y.}~\bibnamefont {Kaneko}}, \bibinfo
  {author} {\bibfnamefont {D.}~\bibnamefont {Okuyama}}, \bibinfo {author}
  {\bibfnamefont {Y.}~\bibnamefont {Tokunaga}}, \bibinfo {author}
  {\bibfnamefont {S.}~\bibnamefont {Wakimoto}}, \bibinfo {author}
  {\bibfnamefont {K.}~\bibnamefont {Kakurai}}, \bibinfo {author} {\bibfnamefont
  {T.}~\bibnamefont {Arima}}, \bibinfo {author} {\bibfnamefont
  {Y.}~\bibnamefont {Taguchi}},\ and\ \bibinfo {author} {\bibfnamefont
  {Y.}~\bibnamefont {Tokura}},\ }\bibfield  {title} {\bibinfo {title}
  {Versatile helimagnetic phases under magnetic fields in cubic perovskite
  ${SrFeO}_{3}$},\ }\href {https://doi.org/10.1103/PhysRevB.84.054427}
  {\bibfield  {journal} {\bibinfo  {journal} {Phys. Rev. B}\ }\textbf {\bibinfo
  {volume} {84}},\ \bibinfo {pages} {054427} (\bibinfo {year}
  {2011})}\BibitemShut {NoStop}%
\bibitem [{\citenamefont {Ishiwata}\ \emph {et~al.}(2020)\citenamefont
  {Ishiwata}, \citenamefont {Nakajima}, \citenamefont {Kim}, \citenamefont
  {Inosov}, \citenamefont {Kanazawa}, \citenamefont {White}, \citenamefont
  {Gavilano}, \citenamefont {Georgii}, \citenamefont {Seemann}, \citenamefont
  {Brandl}, \citenamefont {Manuel}, \citenamefont {Khalyavin}, \citenamefont
  {Seki}, \citenamefont {Tokunaga}, \citenamefont {Kinoshita}, \citenamefont
  {Long}, \citenamefont {Kaneko}, \citenamefont {Taguchi}, \citenamefont
  {Arima}, \citenamefont {Keimer},\ and\ \citenamefont
  {Tokura}}]{Ishiwata_PhysRevB.101.134406}%
  \BibitemOpen
  \bibfield  {author} {\bibinfo {author} {\bibfnamefont {S.}~\bibnamefont
  {Ishiwata}}, \bibinfo {author} {\bibfnamefont {T.}~\bibnamefont {Nakajima}},
  \bibinfo {author} {\bibfnamefont {J.-H.}\ \bibnamefont {Kim}}, \bibinfo
  {author} {\bibfnamefont {D.~S.}\ \bibnamefont {Inosov}}, \bibinfo {author}
  {\bibfnamefont {N.}~\bibnamefont {Kanazawa}}, \bibinfo {author}
  {\bibfnamefont {J.~S.}\ \bibnamefont {White}}, \bibinfo {author}
  {\bibfnamefont {J.~L.}\ \bibnamefont {Gavilano}}, \bibinfo {author}
  {\bibfnamefont {R.}~\bibnamefont {Georgii}}, \bibinfo {author} {\bibfnamefont
  {K.~M.}\ \bibnamefont {Seemann}}, \bibinfo {author} {\bibfnamefont
  {G.}~\bibnamefont {Brandl}}, \bibinfo {author} {\bibfnamefont
  {P.}~\bibnamefont {Manuel}}, \bibinfo {author} {\bibfnamefont {D.~D.}\
  \bibnamefont {Khalyavin}}, \bibinfo {author} {\bibfnamefont {S.}~\bibnamefont
  {Seki}}, \bibinfo {author} {\bibfnamefont {Y.}~\bibnamefont {Tokunaga}},
  \bibinfo {author} {\bibfnamefont {M.}~\bibnamefont {Kinoshita}}, \bibinfo
  {author} {\bibfnamefont {Y.~W.}\ \bibnamefont {Long}}, \bibinfo {author}
  {\bibfnamefont {Y.}~\bibnamefont {Kaneko}}, \bibinfo {author} {\bibfnamefont
  {Y.}~\bibnamefont {Taguchi}}, \bibinfo {author} {\bibfnamefont
  {T.}~\bibnamefont {Arima}}, \bibinfo {author} {\bibfnamefont
  {B.}~\bibnamefont {Keimer}},\ and\ \bibinfo {author} {\bibfnamefont
  {Y.}~\bibnamefont {Tokura}},\ }\bibfield  {title} {\bibinfo {title} {Emergent
  topological spin structures in the centrosymmetric cubic perovskite
  ${\mathrm{srfeo}}_{3}$},\ }\href
  {https://doi.org/10.1103/PhysRevB.101.134406} {\bibfield  {journal} {\bibinfo
   {journal} {Phys. Rev. B}\ }\textbf {\bibinfo {volume} {101}},\ \bibinfo
  {pages} {134406} (\bibinfo {year} {2020})}\BibitemShut {NoStop}%
\bibitem [{\citenamefont {Rogge}\ \emph {et~al.}(2019)\citenamefont {Rogge},
  \citenamefont {Green}, \citenamefont {Sutarto},\ and\ \citenamefont
  {May}}]{Rogge_PhysRevMaterials.3.084404}%
  \BibitemOpen
  \bibfield  {author} {\bibinfo {author} {\bibfnamefont {P.~C.}\ \bibnamefont
  {Rogge}}, \bibinfo {author} {\bibfnamefont {R.~J.}\ \bibnamefont {Green}},
  \bibinfo {author} {\bibfnamefont {R.}~\bibnamefont {Sutarto}},\ and\ \bibinfo
  {author} {\bibfnamefont {S.~J.}\ \bibnamefont {May}},\ }\bibfield  {title}
  {\bibinfo {title} {Itinerancy-dependent noncollinear spin textures in
  ${\mathrm{srfeo}}_{3}, {\mathrm{cafeo}}_{3}$, and
  ${\mathrm{cafeo}}_{3}/{\mathrm{srfeo}}_{3}$ heterostructures probed via
  resonant x-ray scattering},\ }\href
  {https://doi.org/10.1103/PhysRevMaterials.3.084404} {\bibfield  {journal}
  {\bibinfo  {journal} {Phys. Rev. Materials}\ }\textbf {\bibinfo {volume}
  {3}},\ \bibinfo {pages} {084404} (\bibinfo {year} {2019})}\BibitemShut
  {NoStop}%
\bibitem [{\citenamefont {Onose}\ \emph {et~al.}(2020)\citenamefont {Onose},
  \citenamefont {Takahashi}, \citenamefont {Sagayama}, \citenamefont
  {Yamasaki},\ and\ \citenamefont
  {Ishiwata}}]{Onose_PhysRevMaterials.4.114420}%
  \BibitemOpen
  \bibfield  {author} {\bibinfo {author} {\bibfnamefont {M.}~\bibnamefont
  {Onose}}, \bibinfo {author} {\bibfnamefont {H.}~\bibnamefont {Takahashi}},
  \bibinfo {author} {\bibfnamefont {H.}~\bibnamefont {Sagayama}}, \bibinfo
  {author} {\bibfnamefont {Y.}~\bibnamefont {Yamasaki}},\ and\ \bibinfo
  {author} {\bibfnamefont {S.}~\bibnamefont {Ishiwata}},\ }\bibfield  {title}
  {\bibinfo {title} {Complete phase diagram of
  ${\mathrm{sr}}_{1--x}{\mathrm{la}}_{x}\mathrm{Fe}{\mathrm{o}}_{3}$ with
  versatile magnetic and charge ordering},\ }\href
  {https://doi.org/10.1103/PhysRevMaterials.4.114420} {\bibfield  {journal}
  {\bibinfo  {journal} {Phys. Rev. Materials}\ }\textbf {\bibinfo {volume}
  {4}},\ \bibinfo {pages} {114420} (\bibinfo {year} {2020})}\BibitemShut
  {NoStop}%
\bibitem [{\citenamefont {Lei}\ \emph {et~al.}(2021)\citenamefont {Lei},
  \citenamefont {Saltzman},\ and\ \citenamefont
  {Schoop}}]{Shiming_PhysRevB.103.134418}%
  \BibitemOpen
  \bibfield  {author} {\bibinfo {author} {\bibfnamefont {S.}~\bibnamefont
  {Lei}}, \bibinfo {author} {\bibfnamefont {A.}~\bibnamefont {Saltzman}},\ and\
  \bibinfo {author} {\bibfnamefont {L.~M.}\ \bibnamefont {Schoop}},\ }\bibfield
   {title} {\bibinfo {title} {Complex magnetic phases enriched by charge
  density waves in the topological semimetals
  ${\mathrm{gdsb}}_{x}{\mathrm{te}}_{2\ensuremath{-}x\ensuremath{-}\ensuremath{\delta}}$},\
  }\href {https://doi.org/10.1103/PhysRevB.103.134418} {\bibfield  {journal}
  {\bibinfo  {journal} {Phys. Rev. B}\ }\textbf {\bibinfo {volume} {103}},\
  \bibinfo {pages} {134418} (\bibinfo {year} {2021})}\BibitemShut {NoStop}%
\bibitem [{\citenamefont {{\=O}nuki}\ \emph {et~al.}(2020)\citenamefont
  {{\=O}nuki}, \citenamefont {Hedo},\ and\ \citenamefont
  {Honda}}]{onuki2020unique}%
  \BibitemOpen
  \bibfield  {author} {\bibinfo {author} {\bibfnamefont {Y.}~\bibnamefont
  {{\=O}nuki}}, \bibinfo {author} {\bibfnamefont {M.}~\bibnamefont {Hedo}},\
  and\ \bibinfo {author} {\bibfnamefont {F.}~\bibnamefont {Honda}},\ }\bibfield
   {title} {\bibinfo {title} {Unique electronic states of eu-based compounds},\
  }\href@noop {} {\bibfield  {journal} {\bibinfo  {journal} {J. Phys. Soc.
  Jpn.}\ }\textbf {\bibinfo {volume} {89}},\ \bibinfo {pages} {102001}
  (\bibinfo {year} {2020})}\BibitemShut {NoStop}%
\bibitem [{\citenamefont {Shang}\ \emph {et~al.}(2021)\citenamefont {Shang},
  \citenamefont {Xu}, \citenamefont {Gawryluk}, \citenamefont {Ma},
  \citenamefont {Shiroka}, \citenamefont {Shi},\ and\ \citenamefont
  {Pomjakushina}}]{Shang_PhysRevB.103.L020405}%
  \BibitemOpen
  \bibfield  {author} {\bibinfo {author} {\bibfnamefont {T.}~\bibnamefont
  {Shang}}, \bibinfo {author} {\bibfnamefont {Y.}~\bibnamefont {Xu}}, \bibinfo
  {author} {\bibfnamefont {D.~J.}\ \bibnamefont {Gawryluk}}, \bibinfo {author}
  {\bibfnamefont {J.~Z.}\ \bibnamefont {Ma}}, \bibinfo {author} {\bibfnamefont
  {T.}~\bibnamefont {Shiroka}}, \bibinfo {author} {\bibfnamefont
  {M.}~\bibnamefont {Shi}},\ and\ \bibinfo {author} {\bibfnamefont
  {E.}~\bibnamefont {Pomjakushina}},\ }\bibfield  {title} {\bibinfo {title}
  {Anomalous hall resistivity and possible topological hall effect in the
  ${\mathrm{eual}}_{4}$ antiferromagnet},\ }\href
  {https://doi.org/10.1103/PhysRevB.103.L020405} {\bibfield  {journal}
  {\bibinfo  {journal} {Phys. Rev. B}\ }\textbf {\bibinfo {volume} {103}},\
  \bibinfo {pages} {L020405} (\bibinfo {year} {2021})}\BibitemShut {NoStop}%
\bibitem [{\citenamefont {Kaneko}\ \emph {et~al.}(2021)\citenamefont {Kaneko},
  \citenamefont {Kawasaki}, \citenamefont {Nakamura}, \citenamefont {Munakata},
  \citenamefont {Nakao}, \citenamefont {Hanashima}, \citenamefont {Kiyanagi},
  \citenamefont {Ohhara}, \citenamefont {Hedo}, \citenamefont {Nakama},\ and\
  \citenamefont {Onuki}}]{kaneko2021charge}%
  \BibitemOpen
  \bibfield  {author} {\bibinfo {author} {\bibfnamefont {K.}~\bibnamefont
  {Kaneko}}, \bibinfo {author} {\bibfnamefont {T.}~\bibnamefont {Kawasaki}},
  \bibinfo {author} {\bibfnamefont {A.}~\bibnamefont {Nakamura}}, \bibinfo
  {author} {\bibfnamefont {K.}~\bibnamefont {Munakata}}, \bibinfo {author}
  {\bibfnamefont {A.}~\bibnamefont {Nakao}}, \bibinfo {author} {\bibfnamefont
  {T.}~\bibnamefont {Hanashima}}, \bibinfo {author} {\bibfnamefont
  {R.}~\bibnamefont {Kiyanagi}}, \bibinfo {author} {\bibfnamefont
  {T.}~\bibnamefont {Ohhara}}, \bibinfo {author} {\bibfnamefont
  {M.}~\bibnamefont {Hedo}}, \bibinfo {author} {\bibfnamefont {T.}~\bibnamefont
  {Nakama}},\ and\ \bibinfo {author} {\bibfnamefont {Y.}~\bibnamefont
  {Onuki}},\ }\bibfield  {title} {\bibinfo {title} {Charge-density-wave order
  and multiple magnetic transitions in divalent europium compound eual4},\
  }\href@noop {} {\bibfield  {journal} {\bibinfo  {journal} {J. Phys. Soc.
  Jpn.}\ }\textbf {\bibinfo {volume} {90}},\ \bibinfo {pages} {064704}
  (\bibinfo {year} {2021})}\BibitemShut {NoStop}%
\bibitem [{\citenamefont {Hayami}\ \emph
  {et~al.}(2016{\natexlab{b}})\citenamefont {Hayami}, \citenamefont {Lin},
  \citenamefont {Kamiya},\ and\ \citenamefont
  {Batista}}]{Hayami_PhysRevB.94.174420}%
  \BibitemOpen
  \bibfield  {author} {\bibinfo {author} {\bibfnamefont {S.}~\bibnamefont
  {Hayami}}, \bibinfo {author} {\bibfnamefont {S.-Z.}\ \bibnamefont {Lin}},
  \bibinfo {author} {\bibfnamefont {Y.}~\bibnamefont {Kamiya}},\ and\ \bibinfo
  {author} {\bibfnamefont {C.~D.}\ \bibnamefont {Batista}},\ }\bibfield
  {title} {\bibinfo {title} {Vortices, skyrmions, and chirality waves in
  frustrated mott insulators with a quenched periodic array of impurities},\
  }\href {https://doi.org/10.1103/PhysRevB.94.174420} {\bibfield  {journal}
  {\bibinfo  {journal} {Phys. Rev. B}\ }\textbf {\bibinfo {volume} {94}},\
  \bibinfo {pages} {174420} (\bibinfo {year} {2016}{\natexlab{b}})}\BibitemShut
  {NoStop}%
\end{thebibliography}%

\end{document}